\documentstyle[10pt,epsf,epsfig,dp_delphititle]{dp_delphi}
\makeindex
\def\DpPaperGroup{PH-EP}
\def\DpPaperRef{2007-026}
\def\DpDate{12 July 2007}
\def\DpAuthors{DELPHI Collaboration}
\def\DpSubmit{(Accepted by Eur. Phys. J. C)}
\def\DpTitle{{Measurement of the Mass and Width of the W Boson
in {\boldmath $ {e^+e^-} $} Collisions at { \boldmath $ \sqrt{s}  =   161 - 209 ~\mathrm{GeV}$ }}}
\def\DpComment{}
\def\DpEMail{}

\begin{document}
\makeatletter
\newcount\@tempcntc
\def\@citex[#1]#2{\if@filesw\immediate\write\@auxout{\string\citation{#2}}\fi
  \@tempcnta\z@\@tempcntb\m@ne\def\@citea{}\@cite{\@for\@citeb:=#2\do
    {\@ifundefined
       {b@\@citeb}{\@citeo\@tempcntb\m@ne\@citea\def\@citea{,}{\bf ?}\@warning
       {Citation `\@citeb' on page \thepage \space undefined}}%
    {\setbox\z@\hbox{\global\@tempcntc0\csname b@\@citeb\endcsname\relax}%
     \ifnum\@tempcntc=\z@ \@citeo\@tempcntb\m@ne
       \@citea\def\@citea{,}\hbox{\csname b@\@citeb\endcsname}%
     \else
      \advance\@tempcntb\@ne
      \ifnum\@tempcntb=\@tempcntc
      \else\advance\@tempcntb\m@ne\@citeo
      \@tempcnta\@tempcntc\@tempcntb\@tempcntc\fi\fi}}\@citeo}{#1}}
\def\@citeo{\ifnum\@tempcnta>\@tempcntb\else\@citea\def\@citea{,}%
  \ifnum\@tempcnta=\@tempcntb\the\@tempcnta\else
   {\advance\@tempcnta\@ne\ifnum\@tempcnta=\@tempcntb \else \def\@citea{--}\fi
    \advance\@tempcnta\m@ne\the\@tempcnta\@citea\the\@tempcntb}\fi\fi}
 
\makeatother

\begin{titlepage}
\pagenumbering{roman}

\CERNpreprint{\DpPaperGroup}{\DpPaperRef}   
\date{{\small\DpDate}}                      
\title{\DpTitle}                            
\address{\DpAuthors}                        

\begin{shortabs}                            
\noindent
%
\noindent

\newcommand{\approxgt}{\raisebox{-.5ex}{\stackrel{>}{\sim}}}
\newcommand{\approxlt}{\raisebox{-.5ex}{\stackrel{<}{\sim}}}
\newcommand{\mean}[1]{{\left\langle #1 \right\rangle}}
\newcommand{\mod}[1]{{\left|#1\right|}}
\newcommand{\abs}[1]{{\left\|#1\right\|}}
\newcommand{\s}{{\sim}}
\newcommand{\ra}{{\rightarrow}}
\newcommand{\lra}{{\leftrightarrow}}
\newcommand{\longra}{{\longrightarrow}}
\newcommand{\mstrut}{\rule{0cm}{5ex}}
\newcommand{\ov}[1]{\overline{#1}}
\newcommand{\tc}[1]{{\mbox{\tiny #1}}}
\newcommand{\Real}{{\mathcal{R}e}}
\newcommand{\Imag}{{\mathcal{I}m}}


\newcommand{\bmW}{{\mathbf{W}}}
\newcommand{\bmF}{{\mathbf{F}}}
\newcommand{\bmX}{{\mathbf{X}}}
\newcommand{\bmI}{{\mathbf{I}}}
\newcommand{\bmalpha}{{\boldmath{\alpha}}}
\newcommand{\bmtheta}{{\mathbf{\theta}}}
\newcommand{\bmsigma}{{\mathbf{\sigma}}}
\newcommand{\bmlambda}{{\mathbf{\lambda}}}
\newcommand{\bmtau}{{\mathbf{\tau}}}
\newcommand{\bmdag}{{\mathbf{\dagger}}}
\newcommand{\Delx}{{\Delta x}}
\newcommand{\Dely}{{\Delta y}}
\newcommand{\rphi}{{r-\phi}}
\newcommand{\sz}{{s-z}}
\newcommand{\xy}{{x-y}}
\newcommand{\zo}{{z_0}}
\newcommand{\modzo}{{\left| z_0 \right|}}
\newcommand{\meanzo}{{\left\langle z_0 \right\rangle}}
\newcommand{\bphi}{{\bar\phi}}
\newcommand{\bthe}{{\bar\theta}} 

\newcommand{\chisq}{{{\chi}^{2}}}
\newcommand{\chindof}{{\chisq / \mathrm{dof}}}

\newcommand{\grpsuth}{{\mathrm{SU}(3)}}
\newcommand{\grpsutw}{{\mathrm{SU}(2)}}
\newcommand{\grpuone}{{\mathrm{U}(1)}}

\newcommand{\qq}{{q{\bar q}}}
\newcommand{\ppbar}{{p{\bar p}}}
\newcommand{\qqgam}{{q{\bar q}(\gamma)}}
\newcommand{\ff}{{f{\bar f}}}
\newcommand{\lplm}{{\l^{+}\l^{-}}}
\newcommand{\bb}{{b{\bar b}}}
\newcommand{\cc}{{c{\bar c}}}
\newcommand{\ee}{{e^{+}e^{-}}}
\newcommand{\mumu}{{\mu^{+}\mu^{-}}}
\newcommand{\tautau}{{\tau^{+}\tau^{-}}}
\newcommand{\eemm}{{\ee \ra \mumu}}
\newcommand{\eemmg}{{\ee \ra \mumu \gamma}}
\newcommand{\eett}{{\ee \ra \tautau}}
\newcommand{\eeee}{{\ee \ra \ee}}
\newcommand{\eell}{{\ee \ra \lplm}}
\newcommand{\nl}{{\nu_{\ell}}}
\newcommand{\nbl}{{{\overline{\nu}}_{\ell}}}
\newcommand{\nmu}{{\nu_{\mu}}}
\newcommand{\nbmu}{{{\overline{\nu}}_{\mu}}}
\newcommand{\nel}{{\nu_{e}}}
\newcommand{\nbel}{{{\overline{\nu}}_{e}}}
\newcommand{\ntau}{{\nu_{\tau}}}
\newcommand{\nbtau}{{{\overline{\nu}}_{\tau}}}
\newcommand{\ppos}{{\pi^{+}}}
\newcommand{\pneg}{{\pi^{-}}}
\newcommand{\pzer}{{\pi^{0}}}
\newcommand{\ppm}{{\pi^{\pm }}}
\newcommand{\Zo}{{\mathrm{Z}^{0}}}
\newcommand{\Zostar}{{\mathrm{Z}^{0\ast}}}
\newcommand{\Wpm}{{\mathrm{W}^\pm}}
\newcommand{\Wp}{{\mathrm{W}^{+}}}
\newcommand{\Wm}{{\mathrm{W}^{-}}}
\newcommand{\W}{{\mathrm{W}}}
\newcommand{\bfW}{{\mathbf{W}}}
\newcommand{\Z}{{\mathrm{Z}}}
\newcommand{\ZZ}{{\mathrm{ZZ}}}
\newcommand{\Zee}{{\mathrm{Z \ee}}}
\newcommand{\Wen}{{\mathrm{W e \nel}}}
\newcommand{\Zgam}{{\mathrm{Z/\gamma}}}
\newcommand{\Ho}{{\mathrm{H}}}
\newcommand{\WW}{{\mathrm{ W^{+} W^{-} }}}
\newcommand{\bfWW}{{\mathbf{ W^{+} W^{-} }}}

\newcommand{\ffb}{{f{\bar f'}}}
\newcommand{\qqb}{{q{\bar q'}}}
\newcommand{\fbf}{{{\bar f}f'}}
\newcommand{\qbq}{{{\bar q}q'}}
\newcommand{\len}{{\ell \nbl}}
\newcommand{\mn}{{\mu \nbmu}}
\newcommand{\en}{{e \nbel}}
\newcommand{\tn}{{\tau \nbtau}}
\newcommand{\nmn}{{{\overline{\mu}} \nmu}}
\newcommand{\nen}{{{\overline{e}} \nel}}
\newcommand{\ntn}{{{\overline{\tau}} \ntau}}

\newcommand{\ffff}{{\mathrm{ \ffb \fbf}}}
\newcommand{\qqqq}{{\mathrm{ \qqb \qbq}}} 
\newcommand{\lnqq}{{\mathrm{ \len  \qqb}}}
\newcommand{\nqq}{{\mathrm{ \nbl  \qqb}}}
\newcommand{\enqq}{{\mathrm{ \en  \qqb}}}
\newcommand{\mnqq}{{\mathrm{ \mn  \qqb}}}
\newcommand{\tnqq}{{\mathrm{ \tn  \qqb}}}
\newcommand{\lnln}{{\ell \nbl {\overline{\ell}} \nl}}
\newcommand{\enen}{{\en \nen }}  
\newcommand{\mnmn}{{\mn \nmn}}
\newcommand{\tntn}{{\tn \ntn}}
\newcommand{\enmn}{{\en \nmn}}
\newcommand{\entn}{{\en \ntn}}
\newcommand{\mntn}{{\mn \ntn}}
\newcommand{\qqll}{{\mathrm{ \qqb \lplm}}} 

\newcommand{\mw}{{\mathrm{M_W}}}
\newcommand{\mwone}{{\mathrm{m_{\W 1}}}}
\newcommand{\mwtwo}{{\mathrm{m_{\W 2}}}}
\newcommand{\smw}{{\mathrm{m_W}}}
\newcommand{\mwav}{{\mathrm{\bar{m}_W}}}

\newcommand{\mz}{{\mathrm{M_Z}}}
\newcommand{\MZrp}[1]{{\mathrm{M_{\mbox{\tiny{Z}}}\!\!^{#1}}}}
\newcommand{\gz}{{\Gamma_{\mathrm{Z}}}}
\newcommand{\tz}{{\tau_{\mathrm{Z}}}}

\newcommand{\gw}{{\Gamma_{\mathrm{W}}}}
\newcommand{\mh}{{\mathrm{M_H}}}
\newcommand{\mt}{{\mathrm{m_t}}}
\newcommand{\ssqtw}{{\sin^{2}\!\theta_{\mathrm{W}}}}
\newcommand{\csqtw}{{\cos^{2}\!\theta_{\mathrm{W}}}}
\newcommand{\stw}{{\sin\theta_{\mathrm{W}}}}
\newcommand{\ctw}{{\cos\theta_{\mathrm{W}}}}
\newcommand{\ssqtwef}{{\sin}^{2}\theta_{\mathrm{W}}^{\mathrm{eff}}}
\newcommand{\ssqtfef}{{{\sin}^{2}\theta^{\mathrm{eff}}_{f}}}
\newcommand{\ssqtlef}{{{\sin}^{2}\theta^{\mathrm{eff}}_{l}}}
\newcommand{\csqtwef}{{{\cos}^{2}\theta_{\mathrm{W}}^{\mathrm{eff}}}}
\newcommand{\stwef}{\sin\theta_{\mathrm{W}}^{\mathrm{eff}}}
\newcommand{\ctwef}{\cos\theta_{\mathrm{W}}^{\mathrm{eff}}}
\newcommand{\gv}{{g_{\mbox{\tiny V}}}}
\newcommand{\ga}{{g_{\mbox{\tiny A}}}}
\newcommand{\gvel}{{g_{\mbox{\tiny V}}^{e}}}
\newcommand{\gael}{{g_{\mbox{\tiny A}}^{e}}}
\newcommand{\gvmu}{{g_{\mbox{\tiny V}}^{\mu}}}
\newcommand{\gamu}{{g_{\mbox{\tiny A}}^{\mu}}}
\newcommand{\gvf}{{g_{\mbox{\tiny V}}^{f}}}
\newcommand{\gaf}{{g_{\mbox{\tiny A}}^{f}}}
\newcommand{\gvl}{{g_{\mbox{\tiny V}}^{l}}}
\newcommand{\gal}{{g_{\mbox{\tiny A}}^{l}}}
\newcommand{\ghvf}{{\hat{g}_{\mbox{\tiny V}}^{f}}}
\newcommand{\ghaf}{{\hat{g}_{\mbox{\tiny A}}^{f}}}
\newcommand{\gvh}{{\hat{g}_{\mbox{\tiny V}}}}
\newcommand{\gah}{{\hat{g}_{\mbox{\tiny A}}}}
\newcommand{\thw}{{\theta_{\mbox{\mathrm{W}}}}}
\newcommand{\GF}{{G_{\mbox{\tiny F}}}}
\newcommand{\Vub}{{\mathrm{V}_{ub}}}
\newcommand{\Vcb}{{\mathrm{V}_{cb}=}}

\newcommand{\mgen}{{\mathrm{M_{gen}}}}
\newcommand{\ordalph}{{\mathcal{O}(\alpha)}}
\newcommand{\ordalsq}{{\mathcal{O}(\alpha^{2})}}
\newcommand{\ordalcb}{{\mathcal{O}(\alpha^{3})}}
\newcommand{\sigtot}{{\sigma_{\mbox{\tiny TOT}}}}
\newcommand{\sigf}{{\sigma_{\mbox{\tiny F}}}}
\newcommand{\sigb}{{\sigma_{\mbox{\tiny B}}}}
\newcommand{\dsigf}{{\delta\sigma_{\mbox{\tiny F}}}}
\newcommand{\dsigb}{{\delta\sigma_{\mbox{\tiny B}}}}
\newcommand{\dsfbi}{{\delta\sigma_{\mbox{\tiny fb}}^{\mbox{\tiny int}}}}
\newcommand{\AFB}{{A_{\mbox{\tiny FB}}}}
\newcommand{\Afbmm}{{A_{\mbox{\tiny FB}}^{\mbox{\tiny $\mu\mu$}}}}
\newcommand{\Afbtt}{{A_{\mbox{\tiny FB}}^{\mbox{\tiny $\tau\tau$}}}}
\newcommand{\APOL}{{A_{\mbox{\tiny POL}}}}
\newcommand{\pwff}{{\Gamma_{ff}}}
\newcommand{\pwee}{{\Gamma_{ee}}}
\newcommand{\pwmm}{{\Gamma_{\mu\mu}}}
\newcommand{\Af}{{\mathcal{A}_{f}}}
\newcommand{\Ael}{{\mathcal{A}_{e}}}
\newcommand{\Amu}{{\mathcal{A}_{\mu}}}
\newcommand{\dafbint}{{\delta A_{\mbox{\tiny FB}}^{\mbox{\tiny int}}}}
\newcommand{\vcs}{{\left| V_{cs} \right|} }

\newcommand{\Ephot}{{E_{\gamma}}}
\newcommand{\Ebeam}{{E_{\mbox{\tiny BEAM}}}}
\newcommand{\sqs}{{\protect\sqrt{s}}}
\newcommand{\sprime}{{\protect\sqrt{s^{\prime}}}}
\newcommand{\pT}{{\mathrm{p_T}}}
\newcommand{\mmu}{{m_{\mu}}}
\newcommand{\mb}{{m_{b}}}
\newcommand{\thacop}{{\theta_{\mbox{\tiny acop}}}}
\newcommand{\thacol}{{\theta_{\mbox{\tiny acol}}}}
\newcommand{\prad}{{p_{\mbox{\tiny rad}}}}
\newcommand{\lambdabar}{{\lambda \! \! \! \! {\raisebox{+.5ex}{$-$}}}}

\newcommand{\rprg}{{r_{\mbox{\tiny p}}}}
\newcommand{\zprg}{{z_{\mbox{\tiny p}}}}
\newcommand{\thprg}{{\theta_{\mbox{\tiny p}}}}
\newcommand{\phprg}{{\phi_{\mbox{\tiny p}}}}

\newcommand{\mrad}{{\mathrm{mrad}}}
\newcommand{\rad}{{\mathrm{rad}}}
\newcommand{\dgr}{{^\circ}}
\newcommand{\TeV}{{\mathrm{TeV}}}
\newcommand{\GeV}{{\mathrm{GeV}}}
\newcommand{\GeVm}{{\mathrm{GeV/}{\it c}{^2}}}
\newcommand{\GeVp}{{\mathrm{GeV/}{\it c}}}
\newcommand{\MeV}{{\mathrm{MeV}}}
\newcommand{\MeVp}{{\mathrm{MeV/}{\it c}}}
\newcommand{\MeVm}{{\mathrm{MeV/}{\it c}{^2}}}
\newcommand{\KeV}{{\mathrm{KeV}}}
\newcommand{\eV}{{\mathrm{eV}}}
\newcommand{\um}{{\mathrm{m}}}
\newcommand{\umm}{{\mathrm{mm}}}
\newcommand{\uum}{{\mu{\mathrm m}}}
\newcommand{\ucm}{{\mathrm{cm}}}
\newcommand{\ufm}{{\mathrm{fm}}}
\newcommand{\umicrom}{{\mathrm{\mu m}}}
\newcommand{\us}{{\mathrm{s}}}
\newcommand{\ums}{{\mathrm{ms}}}
\newcommand{\uus}{{\mu{\mathrm{s}}}}
\newcommand{\uns}{{\mathrm{ns}}}
\newcommand{\ups}{{\mathrm{ps}}}
\newcommand{\uub}{{\mu{\mathrm{b}}}}
\newcommand{\unb}{{\mathrm{nb}}}
\newcommand{\upb}{{\mathrm{pb}}}
\newcommand{\ipb}{{\mathrm{pb^{-1}}}}
\newcommand{\ifb}{{\mathrm{fb^{-1}}}}
\newcommand{\inb}{{\mathrm{nb^{-1}}}}

\newcommand{\eg}{\mbox{\itshape e.g.}}
\newcommand{\ie}{\mbox{\itshape i.e.}}
\newcommand{\etal}{{\slshape et al\/}\ }
\newcommand{\etc}{\mbox{\itshape etc}}
\newcommand{\cf}{\mbox{\itshape cf.}}
\newcommand{\ffp}{\mbox{\itshape ff}}
\newcommand{\ALEPH}{\mbox{\scshape Aleph}}
\newcommand{\DELPHI}{\mbox{\scshape Delphi}}
\newcommand{\OPAL}{\mbox{\scshape Opal}}
\newcommand{\LTHREE}{\mbox{\scshape L3}}
\newcommand{\CERN}{\mbox{\scshape Cern}}
\newcommand{\LEP}{\mbox{\scshape Lep}}
\newcommand{\LEPONE}{\mbox{\scshape Lep1}}
\newcommand{\LEPTWO}{\mbox{\scshape Lep2}}
\newcommand{\CDF}{\mbox{\scshape Cdf}}
\newcommand{\DO}{\mbox{\scshape D0}}
\newcommand{\SLD}{\mbox{\scshape Sld}}
\newcommand{\CLEO}{\mbox{\scshape Cleo}}
\newcommand{\UAONE}{\mbox{\scshape Ua1}}
\newcommand{\UATWO}{\mbox{\scshape Ua2}}
\newcommand{\TEVATRON}{\mbox{\scshape Tevatron}}
\newcommand{\LHC}{\mbox{\scshape LHC}}
\newcommand{\KORALZ}{\mbox{\ttfamily KORALZ}}
\newcommand{\KORALW}{\mbox{\ttfamily KORALW}}
\newcommand{\ZFITTER}{\mbox{\ttfamily ZFITTER}}
\newcommand{\GENTLE}{\mbox{\ttfamily GENTLE}}
\newcommand{\DELANA}{\mbox{\ttfamily DELANA}}
\newcommand{\DELSIM}{\mbox{\ttfamily DELSIM}}
\newcommand{\DYMU}{\mbox{\ttfamily DYMU3}}
\newcommand{\TANAGRA}{\mbox{\ttfamily TANAGRA}}
\newcommand{\ZEBRA}{\mbox{\ttfamily ZEBRA}}
\newcommand{\PAW}{\mbox{\ttfamily PAW}}
\newcommand{\WWANA}{\mbox{\ttfamily WWANA}}
\newcommand{\FASTSIM}{\mbox{\ttfamily FASTSIM}}
\newcommand{\PYTHIA}{\mbox{\ttfamily PYTHIA}}
\newcommand{\JETSET}{\mbox{\ttfamily JETSET}}
\newcommand{\TWOGAM}{\mbox{\ttfamily TWOGAM}}
\newcommand{\LUBOEI}{\mbox{\ttfamily LUBOEI}}
\newcommand{\SKI}{\mbox{\ttfamily SK-I}}
\newcommand{\SKII}{\mbox{\ttfamily SK-II}}
\newcommand{\ARIADNE}{\mbox{\ttfamily ARIADNE}}
\newcommand{\ARII}{\mbox{\ttfamily AR-II}}
\newcommand{\VNI}{\mbox{\ttfamily VNI}}
\newcommand{\HERWIG}{\mbox{\ttfamily HERWIG}}
\newcommand{\EXCALIBUR}{\mbox{\ttfamily EXCALIBUR}}
\newcommand{\WPHACT}{\mbox{\ttfamily WPHACT}}
\newcommand{\RACOONWW}{\mbox{\ttfamily RACOONWW}}
\newcommand{\YFSWW}{\mbox{\ttfamily YFSWW}}
\newcommand{\QEDPS}{\mbox{\ttfamily QEDPS}}
\newcommand{\CCTHREE}{\mbox{\ttfamily CC03}}
\newcommand{\NCEIGHT}{\mbox{\ttfamily NC08}}
\newcommand{\LUCLUS}{\mbox{\ttfamily LUCLUS}}
\newcommand{\DJOIN}{\mbox{$\mathrm{d_{join}}$}}
\newcommand{\JADE}{\mbox{\ttfamily JADE}}
\newcommand{\DURHAM}{\mbox{\ttfamily DURHAM}}
\newcommand{\CAMBRIDGE}{\mbox{\ttfamily CAMBRIDGE}}
\newcommand{\CAMJET}{\mbox{\ttfamily CAMJET}}
\newcommand{\DICLUS}{\mbox{\ttfamily DICLUS}}
\newcommand{\CONE}{\mbox{\ttfamily CONE}}
\newcommand{\PUFITC}{\mbox{\ttfamily PUFITC+}}
\newcommand{\PHDST}{\mbox{\ttfamily PHDST}}
\newcommand{\SKELANA}{\mbox{\ttfamily SKELANA}}
\newcommand{\DAFNE}{\mbox{\ttfamily DAFNE}}
\newcommand{\CERNLIB}{\mbox{\ttfamily CERNLIB}}
\newcommand{\MINUIT}{\mbox{\ttfamily MINUIT}}
\newcommand{\REMCLU}{\mbox{\ttfamily REMCLU}}
\newcommand{\DST}{\mbox{\ttfamily DST}}
\newcommand{\XSDST}{\mbox{\ttfamily XShortDST}}
\newcommand{\FDST}{\mbox{\ttfamily FullDST}}

\newcommand{\spot}{\mbox{$\bullet \;$}}

\newcommand{\eps}{{\epsilon}}
\newcommand{\erreps}{{\sigma_{\epsilon}}}
\newcommand{\errepsp}{{\sigma_{\epsilon}^{+}}}
\newcommand{\errepsm}{{\sigma_{\epsilon}^{-}}}
\newcommand{\Lmb}{{\Lambda}}
\newcommand{\lmb}{{\lambda}}
\newcommand{\Lmbsq}{{\Lambda^{2}}}
\newcommand{\Lmbisq}{{1/\Lambda^{2}}}
\newcommand{\Lmbpm}{{\Lambda^{\pm}}}
\newcommand{\Lmbp}{{\Lambda^{+}}}
\newcommand{\Lmbm}{{\Lambda^{-}}}
\newcommand{\gc}{{g_{c}}}
\newcommand{\etaij}{{\eta_{ij}}}
\newcommand{\IJ}{{\mathrm{IJ}}}
\newcommand{\IJpm}[1]{{\mathrm{IJ}^{(#1)}}}

\newcommand{\gev}{{\mathrm{GeV}}}
\newcommand{\wpm}{{\mathrm{W}^\pm}}
\newcommand{\wm}{{\mathrm{W}^{-}}}
\newcommand{\w}{{\mathrm{W}}}
\newcommand{\z}{{\mathrm{Z}}}
\newcommand{\ww}{{\mathrm{ W^{+} W^{-} }}}
\newcommand{\tev}{{\mathrm{TeV}}}
\newcommand{\lep}{\mbox{\scshape Lep}}
\newcommand{\lepone}{\mbox{\scshape Lep1}}
\newcommand{\leptwo}{\mbox{\scshape Lep2}}
\newcommand{\delphi}{\mbox{\scshape Delphi}}
\newcommand{\excalibur}{\mbox{\ttfamily EXCALIBUR}}
\newcommand{\wwana}{\mbox{\ttfamily WWANA}}
\newcommand{\fastsim}{\mbox{\ttfamily FASTSIM}}
\newcommand{\phdst}{\mbox{\ttfamily PHDST}}
\newcommand{\skelana}{\mbox{\ttfamily SKELANA}}
\newcommand{\dafne}{\mbox{\ttfamily DAFNE}}
\newcommand{\delana}{\mbox{\ttfamily DELANA}}
\newcommand{\zebra}{\mbox{\ttfamily ZEBRA}}
\newcommand{\delsim}{\mbox{\ttfamily DELSIM}}
\newcommand{\pythia}{\mbox{\ttfamily PYTHIA}}
\newcommand{\jetset}{\mbox{\ttfamily JETSET}}
\newcommand{\ariadne}{\mbox{\ttfamily ARIADNE}}
\newcommand{\cdf}{\mbox{\scshape CDF}}
\newcommand{\gentle}{\mbox{\ttfamily GENTLE}}
A measurement of the $\W$ boson mass and width has been performed by the \DELPHI\
collaboration using the data collected during the full LEP2 programme (1996-2000). The data sample has an integrated luminosity of 660~$\ipb$ and was collected over a range of centre-of-mass energies from 161 to 209 $\GeV$.

Results are obtained by applying the method of direct reconstruction of the mass of the $\W$ from its decay products in both the $\WW\ \rightarrow\ \lnqq$ and $\WW\ \rightarrow\ \qqqq$ channels. The $\W$ mass result for the combined data set is 
\begin{eqnarray*}
   \mw & = & 80.336 \pm 0.055(\rm Stat.) \pm 0.028 (\rm Syst.) \pm 0.025 (\rm FSI) \pm 0.009 (\rm LEP)~\GeVm,
 \end{eqnarray*}

where FSI represents the uncertainty due to final state interaction effects in the $\qqqq$ channel, and LEP represents that arising from the knowledge of the collision energy of the accelerator. The combined value for the $\W$ width is
\begin{eqnarray*}
   \gw & = &  2.404 \pm 0.140(\rm Stat.) \pm 0.077 (\rm Syst.) \pm 0.065 (\rm FSI)~\GeVm.
 \end{eqnarray*}

These results supersede all values previously published by the DELPHI collaboration.

\vspace{1cm}

\begin{center}
{\it This paper is dedicated to the memory of Carlo Caso.} 
\end{center}
\end{shortabs}

\vfill

\begin{center}
\DpSubmit \ \\          
\DpComment \ \\
\DpEMail \ \\
\end{center}

\vfill
\clearpage

\headsep 10.0pt

\addtolength{\textheight}{10mm}
\addtolength{\footskip}{-5mm}
\begingroup
%
\newcommand{\DpName}[2]{\hbox{#1$^{\ref{#2}}$},\hfill}
\newcommand{\DpNameTwo}[3]{\hbox{#1$^{\ref{#2},\ref{#3}}$},\hfill}
\newcommand{\DpNameThree}[4]{\hbox{#1$^{\ref{#2},\ref{#3},\ref{#4}}$},\hfill}
\newskip\Bigfill \Bigfill = 0pt plus 1000fill
\newcommand{\DpNameLast}[2]{\hbox{#1$^{\ref{#2}}$}\hspace{\Bigfill}}

%
\footnotesize
\noindent
\DpName{J.Abdallah}{LPNHE}
\DpName{P.Abreu}{LIP}
\DpName{W.Adam}{VIENNA}
\DpName{P.Adzic}{DEMOKRITOS}
\DpName{T.Albrecht}{KARLSRUHE}
\DpName{R.Alemany-Fernandez}{CERN}
\DpName{T.Allmendinger}{KARLSRUHE}
\DpName{P.P.Allport}{LIVERPOOL}
\DpName{U.Amaldi}{MILANO2}
\DpName{N.Amapane}{TORINO}
% ...
\normalsize
\endgroup
\newpage

\titlefoot{Department of Physics and Astronomy, Iowa State
     University, Ames IA 50011-3160, USA
    \label{AMES}}
\titlefoot{IIHE, ULB-VUB,
     Pleinlaan 2, B-1050 Brussels, Belgium
    \label{BRUSSELS}}
\titlefoot{Physics Laboratory, University of Athens, Solonos Str.
     104, GR-10680 Athens, Greece
    \label{ATHENS}}
\titlefoot{Department of Physics, University of Bergen,
     All\'egaten 55, NO-5007 Bergen, Norway
    \label{BERGEN}}
\titlefoot{Dipartimento di Fisica, Universit\`a di Bologna and INFN,
     Via Irnerio 46, IT-40126 Bologna, Italy
    \label{BOLOGNA}}
\titlefoot{Centro Brasileiro de Pesquisas F\'{\i}sicas, rua Xavier Sigaud 150,
     BR-22290 Rio de Janeiro, Brazil
    \label{BRASIL-CBPF}}
\titlefoot{Inst. de F\'{\i}sica, Univ. Estadual do Rio de Janeiro,
     rua S\~{a}o Francisco Xavier 524, Rio de Janeiro, Brazil
    \label{BRASIL-IFUERJ}}
\titlefoot{Coll\`ege de France, Lab. de Physique Corpusculaire, IN2P3-CNRS,
     FR-75231 Paris Cedex 05, France
    \label{CDF}}
\titlefoot{CERN, CH-1211 Geneva 23, Switzerland
    \label{CERN}}
\titlefoot{Institut de Recherches Subatomiques, IN2P3 - CNRS/ULP - BP20,
     FR-67037 Strasbourg Cedex, France
    \label{CRN}}
\titlefoot{Now at DESY-Zeuthen, Platanenallee 6, D-15735 Zeuthen, Germany
    \label{DESY}}
\titlefoot{Institute of Nuclear Physics, N.C.S.R. Demokritos,
     P.O. Box 60228, GR-15310 Athens, Greece
    \label{DEMOKRITOS}}
\titlefoot{FZU, Inst. of Phys. of the C.A.S. High Energy Physics Division,
     Na Slovance 2, CZ-182 21, Praha 8, Czech Republic
    \label{FZU}}
\titlefoot{Dipartimento di Fisica, Universit\`a di Genova and INFN,
     Via Dodecaneso 33, IT-16146 Genova, Italy
    \label{GENOVA}}
\titlefoot{Institut des Sciences Nucl\'eaires, IN2P3-CNRS, Universit\'e
     de Grenoble 1, FR-38026 Grenoble Cedex, France
    \label{GRENOBLE}}
\titlefoot{Helsinki Institute of Physics and Department of Physical Sciences,
     P.O. Box 64, FIN-00014 University of Helsinki, 
     \indent~~Finland
    \label{HELSINKI}}
\titlefoot{Joint Institute for Nuclear Research, Dubna, Head Post
     Office, P.O. Box 79, RU-101 000 Moscow, Russian Federation
    \label{JINR}}
\titlefoot{Institut f\"ur Experimentelle Kernphysik,
     Universit\"at Karlsruhe, Postfach 6980, DE-76128 Karlsruhe,
     Germany
    \label{KARLSRUHE}}
\titlefoot{Institute of Nuclear Physics PAN,Ul. Radzikowskiego 152,
     PL-31142 Krakow, Poland
    \label{KRAKOW1}}
\titlefoot{Faculty of Physics and Nuclear Techniques, University of Mining
     and Metallurgy, PL-30055 Krakow, Poland
    \label{KRAKOW2}}
\titlefoot{Universit\'e de Paris-Sud, Lab. de l'Acc\'el\'erateur
     Lin\'eaire, IN2P3-CNRS, B\^{a}t. 200, FR-91405 Orsay Cedex, France
    \label{LAL}}
\titlefoot{School of Physics and Chemistry, University of Lancaster,
     Lancaster LA1 4YB, UK
    \label{LANCASTER}}
\titlefoot{LIP, IST, FCUL - Av. Elias Garcia, 14-$1^{o}$,
     PT-1000 Lisboa Codex, Portugal
    \label{LIP}}
\titlefoot{Department of Physics, University of Liverpool, P.O.
     Box 147, Liverpool L69 3BX, UK
    \label{LIVERPOOL}}
\titlefoot{Dept. of Physics and Astronomy, Kelvin Building,
     University of Glasgow, Glasgow G12 8QQ, UK
    \label{GLASGOW}}
\titlefoot{LPNHE, IN2P3-CNRS, Univ.~Paris VI et VII, Tour 33 (RdC),
     4 place Jussieu, FR-75252 Paris Cedex 05, France
    \label{LPNHE}}
\titlefoot{Department of Physics, University of Lund,
     S\"olvegatan 14, SE-223 63 Lund, Sweden
    \label{LUND}}
\titlefoot{Universit\'e Claude Bernard de Lyon, IPNL, IN2P3-CNRS,
     FR-69622 Villeurbanne Cedex, France
    \label{LYON}}
\titlefoot{Dipartimento di Fisica, Universit\`a di Milano and INFN-MILANO,
     Via Celoria 16, IT-20133 Milan, Italy
    \label{MILANO}}
\titlefoot{Dipartimento di Fisica, Univ. di Milano-Bicocca and
     INFN-MILANO, Piazza della Scienza 3, IT-20126 Milan, Italy
    \label{MILANO2}}
\titlefoot{IPNP of MFF, Charles Univ., Areal MFF,
     V Holesovickach 2, CZ-180 00, Praha 8, Czech Republic
    \label{NC}}
\titlefoot{NIKHEF, Postbus 41882, NL-1009 DB
     Amsterdam, The Netherlands
    \label{NIKHEF}}
\titlefoot{National Technical University, Physics Department,
     Zografou Campus, GR-15773 Athens, Greece
    \label{NTU-ATHENS}}
\titlefoot{Physics Department, University of Oslo, Blindern,
     NO-0316 Oslo, Norway
    \label{OSLO}}
\titlefoot{Dpto. Fisica, Univ. Oviedo, Avda. Calvo Sotelo
     s/n, ES-33007 Oviedo, Spain
    \label{OVIEDO}}
\titlefoot{Department of Physics, University of Oxford,
     Keble Road, Oxford OX1 3RH, UK
    \label{OXFORD}}
\titlefoot{Dipartimento di Fisica, Universit\`a di Padova and
     INFN, Via Marzolo 8, IT-35131 Padua, Italy
    \label{PADOVA}}
\titlefoot{Rutherford Appleton Laboratory, Chilton, Didcot
     OX11 OQX, UK
    \label{RAL}}
\titlefoot{Dipartimento di Fisica, Universit\`a di Roma II and
     INFN, Tor Vergata, IT-00173 Rome, Italy
    \label{ROMA2}}
\titlefoot{Dipartimento di Fisica, Universit\`a di Roma III and
     INFN, Via della Vasca Navale 84, IT-00146 Rome, Italy
    \label{ROMA3}}
\titlefoot{DAPNIA/Service de Physique des Particules,
     CEA-Saclay, FR-91191 Gif-sur-Yvette Cedex, France
    \label{SACLAY}}
\titlefoot{Instituto de Fisica de Cantabria (CSIC-UC), Avda.
     los Castros s/n, ES-39006 Santander, Spain
    \label{SANTANDER}}
\titlefoot{Inst. for High Energy Physics, Serpukov
     P.O. Box 35, Protvino, (Moscow Region), Russian Federation
    \label{SERPUKHOV}}
\titlefoot{J. Stefan Institute, Jamova 39, SI-1000 Ljubljana, Slovenia
    \label{SLOVENIJA1}}
\titlefoot{Laboratory for Astroparticle Physics,
     University of Nova Gorica, Kostanjeviska 16a, SI-5000 Nova Gorica, Slovenia
    \label{SLOVENIJA2}}
\titlefoot{Department of Physics, University of Ljubljana,
     SI-1000 Ljubljana, Slovenia
    \label{SLOVENIJA3}}
\titlefoot{Fysikum, Stockholm University,
     Box 6730, SE-113 85 Stockholm, Sweden
    \label{STOCKHOLM}}
\titlefoot{Dipartimento di Fisica Sperimentale, Universit\`a di
     Torino and INFN, Via P. Giuria 1, IT-10125 Turin, Italy
    \label{TORINO}}
\titlefoot{INFN,Sezione di Torino and Dipartimento di Fisica Teorica,
     Universit\`a di Torino, Via Giuria 1,
     IT-10125 Turin, Italy
    \label{TORINOTH}}
\titlefoot{Dipartimento di Fisica, Universit\`a di Trieste and
     INFN, Via A. Valerio 2, IT-34127 Trieste, Italy
    \label{TRIESTE}}
\titlefoot{Istituto di Fisica, Universit\`a di Udine and INFN,
     IT-33100 Udine, Italy
    \label{UDINE}}
\titlefoot{Univ. Federal do Rio de Janeiro, C.P. 68528
     Cidade Univ., Ilha do Fund\~ao
     BR-21945-970 Rio de Janeiro, Brazil
    \label{UFRJ}}
\titlefoot{Department of Radiation Sciences, University of
     Uppsala, P.O. Box 535, SE-751 21 Uppsala, Sweden
    \label{UPPSALA}}
\titlefoot{IFIC, Valencia-CSIC, and D.F.A.M.N., U. de Valencia,
     Avda. Dr. Moliner 50, ES-46100 Burjassot (Valencia), Spain
    \label{VALENCIA}}
\titlefoot{Institut f\"ur Hochenergiephysik, \"Osterr. Akad.
     d. Wissensch., Nikolsdorfergasse 18, AT-1050 Vienna, Austria
    \label{VIENNA}}
\titlefoot{Inst. Nuclear Studies and University of Warsaw, Ul.
     Hoza 69, PL-00681 Warsaw, Poland
    \label{WARSZAWA}}
\titlefoot{Now at University of Warwick, Coventry CV4 7AL, UK
    \label{WARWICK}}
\titlefoot{Fachbereich Physik, University of Wuppertal, Postfach
     100 127, DE-42097 Wuppertal, Germany \\
\noindent
{$^\dagger$~deceased}
    \label{WUPPERTAL}}
\addtolength{\textheight}{-10mm}
\addtolength{\footskip}{5mm}
\clearpage

\headsep 30.0pt
\end{titlepage}

%
\pagenumbering{arabic}                              
\setcounter{footnote}{0}                            %
\large
%
%
%


\newcommand{\approxgt}{\raisebox{-.5ex}{\stackrel{>}{\sim}}}
\newcommand{\approxlt}{\raisebox{-.5ex}{\stackrel{<}{\sim}}}
\newcommand{\mean}[1]{{\left\langle #1 \right\rangle}}
\newcommand{\mod}[1]{{\left|#1\right|}}
\newcommand{\abs}[1]{{\left\|#1\right\|}}
\newcommand{\s}{{\sim}}
\newcommand{\ra}{{\rightarrow}}
\newcommand{\lra}{{\leftrightarrow}}
\newcommand{\longra}{{\longrightarrow}}
\newcommand{\mstrut}{\rule{0cm}{5ex}}
\newcommand{\ov}[1]{\overline{#1}}
\newcommand{\tc}[1]{{\mbox{\tiny #1}}}
\newcommand{\Real}{{\mathcal{R}e}}
\newcommand{\Imag}{{\mathcal{I}m}}


\newcommand{\bmW}{{\mathbf{W}}}
\newcommand{\bmF}{{\mathbf{F}}}
\newcommand{\bmX}{{\mathbf{X}}}
\newcommand{\bmI}{{\mathbf{I}}}
\newcommand{\bmalpha}{{\boldmath{\alpha}}}
\newcommand{\bmtheta}{{\mathbf{\theta}}}
\newcommand{\bmsigma}{{\mathbf{\sigma}}}
\newcommand{\bmlambda}{{\mathbf{\lambda}}}
\newcommand{\bmtau}{{\mathbf{\tau}}}
\newcommand{\bmdag}{{\mathbf{\dagger}}}
\newcommand{\Delx}{{\Delta x}}
\newcommand{\Dely}{{\Delta y}}
\newcommand{\rphi}{{r-\phi}}
\newcommand{\sz}{{s-z}}
\newcommand{\xy}{{x-y}}
\newcommand{\zo}{{z_0}}
\newcommand{\modzo}{{\left| z_0 \right|}}
\newcommand{\meanzo}{{\left\langle z_0 \right\rangle}}
\newcommand{\bphi}{{\bar\phi}}
\newcommand{\bthe}{{\bar\theta}} 

\newcommand{\chisq}{{{\chi}^{2}}}
\newcommand{\chindof}{{\chisq / \mathrm{dof}}}

\newcommand{\grpsuth}{{\mathrm{SU}(3)}}
\newcommand{\grpsutw}{{\mathrm{SU}(2)}}
\newcommand{\grpuone}{{\mathrm{U}(1)}}

\newcommand{\qq}{{q{\bar q}}}
\newcommand{\ppbar}{{p{\bar p}}}
\newcommand{\qqgam}{{q{\bar q}(\gamma)}}
\newcommand{\ff}{{f{\bar f}}}
\newcommand{\lplm}{{\l^{+}\l^{-}}}
\newcommand{\bb}{{b{\bar b}}}
\newcommand{\cc}{{c{\bar c}}}
\newcommand{\ee}{{e^{+}e^{-}}}
\newcommand{\mumu}{{\mu^{+}\mu^{-}}}
\newcommand{\tautau}{{\tau^{+}\tau^{-}}}
\newcommand{\eemm}{{\ee \ra \mumu}}
\newcommand{\eemmg}{{\ee \ra \mumu \gamma}}
\newcommand{\eett}{{\ee \ra \tautau}}
\newcommand{\eeee}{{\ee \ra \ee}}
\newcommand{\eell}{{\ee \ra \lplm}}
\newcommand{\nl}{{\nu_{\ell}}}
\newcommand{\nbl}{{{\overline{\nu}}_{\ell}}}
\newcommand{\nmu}{{\nu_{\mu}}}
\newcommand{\nbmu}{{{\overline{\nu}}_{\mu}}}
\newcommand{\nel}{{\nu_{e}}}
\newcommand{\nbel}{{{\overline{\nu}}_{e}}}
\newcommand{\ntau}{{\nu_{\tau}}}
\newcommand{\nbtau}{{{\overline{\nu}}_{\tau}}}
\newcommand{\ppos}{{\pi^{+}}}
\newcommand{\pneg}{{\pi^{-}}}
\newcommand{\pzer}{{\pi^{0}}}
\newcommand{\ppm}{{\pi^{\pm }}}
\newcommand{\Zo}{{\mathrm{Z}^{0}}}
\newcommand{\Zostar}{{\mathrm{Z}^{0\ast}}}
\newcommand{\Wpm}{{\mathrm{W}^\pm}}
\newcommand{\Wp}{{\mathrm{W}^{+}}}
\newcommand{\Wm}{{\mathrm{W}^{-}}}
\newcommand{\W}{{\mathrm{W}}}
\newcommand{\bfW}{{\mathbf{W}}}
\newcommand{\Z}{{\mathrm{Z}}}
\newcommand{\ZZ}{{\mathrm{ZZ}}}
\newcommand{\Zee}{{\mathrm{Z \ee}}}
\newcommand{\Wen}{{\mathrm{W e \nel}}}
\newcommand{\Zgam}{{\mathrm{Z/\gamma}}}
\newcommand{\Ho}{{\mathrm{H}}}
\newcommand{\WW}{{\mathrm{ W^{+} W^{-} }}}
\newcommand{\bfWW}{{\mathbf{ W^{+} W^{-} }}}

\newcommand{\ffb}{{f{\bar f'}}}
\newcommand{\qqb}{{q{\bar q'}}}
\newcommand{\fbf}{{{\bar f}f'}}
\newcommand{\qbq}{{{\bar q}q'}}
\newcommand{\len}{{\ell \nbl}}
\newcommand{\mn}{{\mu \nbmu}}
\newcommand{\en}{{e \nbel}}
\newcommand{\tn}{{\tau \nbtau}}
\newcommand{\nmn}{{{\overline{\mu}} \nmu}}
\newcommand{\nen}{{{\overline{e}} \nel}}
\newcommand{\ntn}{{{\overline{\tau}} \ntau}}

\newcommand{\ffff}{{\mathrm{ \ffb \fbf}}}
\newcommand{\qqqq}{{\mathrm{ \qqb \qbq}}} 
\newcommand{\lnqq}{{\mathrm{ \len  \qqb}}}
\newcommand{\nqq}{{\mathrm{ \nbl  \qqb}}}
\newcommand{\enqq}{{\mathrm{ \en  \qqb}}}
\newcommand{\mnqq}{{\mathrm{ \mn  \qqb}}}
\newcommand{\tnqq}{{\mathrm{ \tn  \qqb}}}
\newcommand{\lnln}{{\ell \nbl {\overline{\ell}} \nl}}
\newcommand{\enen}{{\en \nen }}  
\newcommand{\mnmn}{{\mn \nmn}}
\newcommand{\tntn}{{\tn \ntn}}
\newcommand{\enmn}{{\en \nmn}}
\newcommand{\entn}{{\en \ntn}}
\newcommand{\mntn}{{\mn \ntn}}
\newcommand{\qqll}{{\mathrm{ \qqb \lplm}}} 

\newcommand{\mw}{{\mathrm{M_W}}}
\newcommand{\mwone}{{\mathrm{m_{\W 1}}}}
\newcommand{\mwtwo}{{\mathrm{m_{\W 2}}}}
\newcommand{\smw}{{\mathrm{m_W}}}
\newcommand{\mwav}{{\mathrm{\bar{m}_W}}}

\newcommand{\mz}{{\mathrm{M_Z}}}
\newcommand{\MZrp}[1]{{\mathrm{M_{\mbox{\tiny{Z}}}\!\!^{#1}}}}
\newcommand{\gz}{{\Gamma_{\mathrm{Z}}}}
\newcommand{\tz}{{\tau_{\mathrm{Z}}}}

\newcommand{\gw}{{\Gamma_{\mathrm{W}}}}
\newcommand{\mh}{{\mathrm{M_H}}}
\newcommand{\mt}{{\mathrm{m_t}}}
\newcommand{\ssqtw}{{\sin^{2}\!\theta_{\mathrm{W}}}}
\newcommand{\csqtw}{{\cos^{2}\!\theta_{\mathrm{W}}}}
\newcommand{\stw}{{\sin\theta_{\mathrm{W}}}}
\newcommand{\ctw}{{\cos\theta_{\mathrm{W}}}}
\newcommand{\ssqtwef}{{\sin}^{2}\theta_{\mathrm{W}}^{\mathrm{eff}}}
\newcommand{\ssqtfef}{{{\sin}^{2}\theta^{\mathrm{eff}}_{f}}}
\newcommand{\ssqtlef}{{{\sin}^{2}\theta^{\mathrm{eff}}_{l}}}
\newcommand{\csqtwef}{{{\cos}^{2}\theta_{\mathrm{W}}^{\mathrm{eff}}}}
\newcommand{\stwef}{\sin\theta_{\mathrm{W}}^{\mathrm{eff}}}
\newcommand{\ctwef}{\cos\theta_{\mathrm{W}}^{\mathrm{eff}}}
\newcommand{\gv}{{g_{\mbox{\tiny V}}}}
\newcommand{\ga}{{g_{\mbox{\tiny A}}}}
\newcommand{\gvel}{{g_{\mbox{\tiny V}}^{e}}}
\newcommand{\gael}{{g_{\mbox{\tiny A}}^{e}}}
\newcommand{\gvmu}{{g_{\mbox{\tiny V}}^{\mu}}}
\newcommand{\gamu}{{g_{\mbox{\tiny A}}^{\mu}}}
\newcommand{\gvf}{{g_{\mbox{\tiny V}}^{f}}}
\newcommand{\gaf}{{g_{\mbox{\tiny A}}^{f}}}
\newcommand{\gvl}{{g_{\mbox{\tiny V}}^{l}}}
\newcommand{\gal}{{g_{\mbox{\tiny A}}^{l}}}
\newcommand{\ghvf}{{\hat{g}_{\mbox{\tiny V}}^{f}}}
\newcommand{\ghaf}{{\hat{g}_{\mbox{\tiny A}}^{f}}}
\newcommand{\gvh}{{\hat{g}_{\mbox{\tiny V}}}}
\newcommand{\gah}{{\hat{g}_{\mbox{\tiny A}}}}
\newcommand{\thw}{{\theta_{\mbox{\mathrm{W}}}}}
\newcommand{\GF}{{G_{\mbox{\tiny F}}}}
\newcommand{\Vub}{{\mathrm{V}_{ub}}}
\newcommand{\Vcb}{{\mathrm{V}_{cb}=}}

\newcommand{\mgen}{{\mathrm{M_{gen}}}}
\newcommand{\ordalph}{{\mathcal{O}(\alpha)}}
\newcommand{\ordalsq}{{\mathcal{O}(\alpha^{2})}}
\newcommand{\ordalcb}{{\mathcal{O}(\alpha^{3})}}
\newcommand{\sigtot}{{\sigma_{\mbox{\tiny TOT}}}}
\newcommand{\sigf}{{\sigma_{\mbox{\tiny F}}}}
\newcommand{\sigb}{{\sigma_{\mbox{\tiny B}}}}
\newcommand{\dsigf}{{\delta\sigma_{\mbox{\tiny F}}}}
\newcommand{\dsigb}{{\delta\sigma_{\mbox{\tiny B}}}}
\newcommand{\dsfbi}{{\delta\sigma_{\mbox{\tiny fb}}^{\mbox{\tiny int}}}}
\newcommand{\AFB}{{A_{\mbox{\tiny FB}}}}
\newcommand{\Afbmm}{{A_{\mbox{\tiny FB}}^{\mbox{\tiny $\mu\mu$}}}}
\newcommand{\Afbtt}{{A_{\mbox{\tiny FB}}^{\mbox{\tiny $\tau\tau$}}}}
\newcommand{\APOL}{{A_{\mbox{\tiny POL}}}}
\newcommand{\pwff}{{\Gamma_{ff}}}
\newcommand{\pwee}{{\Gamma_{ee}}}
\newcommand{\pwmm}{{\Gamma_{\mu\mu}}}
\newcommand{\Af}{{\mathcal{A}_{f}}}
\newcommand{\Ael}{{\mathcal{A}_{e}}}
\newcommand{\Amu}{{\mathcal{A}_{\mu}}}
\newcommand{\dafbint}{{\delta A_{\mbox{\tiny FB}}^{\mbox{\tiny int}}}}
\newcommand{\vcs}{{\left| V_{cs} \right|} }

\newcommand{\Ephot}{{E_{\gamma}}}
\newcommand{\Ebeam}{{E_{\mbox{\tiny BEAM}}}}
\newcommand{\sqs}{{\protect\sqrt{s}}}
\newcommand{\sprime}{{\protect\sqrt{s^{\prime}}}}
\newcommand{\pT}{{\mathrm{p_T}}}
\newcommand{\mmu}{{m_{\mu}}}
\newcommand{\mb}{{m_{b}}}
\newcommand{\thacop}{{\theta_{\mbox{\tiny acop}}}}
\newcommand{\thacol}{{\theta_{\mbox{\tiny acol}}}}
\newcommand{\prad}{{p_{\mbox{\tiny rad}}}}
\newcommand{\lambdabar}{{\lambda \! \! \! \! {\raisebox{+.5ex}{$-$}}}}

\newcommand{\rprg}{{r_{\mbox{\tiny p}}}}
\newcommand{\zprg}{{z_{\mbox{\tiny p}}}}
\newcommand{\thprg}{{\theta_{\mbox{\tiny p}}}}
\newcommand{\phprg}{{\phi_{\mbox{\tiny p}}}}

\newcommand{\mrad}{{\mathrm{mrad}}}
\newcommand{\rad}{{\mathrm{rad}}}
\newcommand{\dgr}{{^\circ}}
\newcommand{\TeV}{{\mathrm{TeV}}}
\newcommand{\GeV}{{\mathrm{GeV}}}
\newcommand{\GeVm}{{\mathrm{GeV/}{\it c}{^2}}}
\newcommand{\GeVp}{{\mathrm{GeV/}{\it c}}}
\newcommand{\MeV}{{\mathrm{MeV}}}
\newcommand{\MeVp}{{\mathrm{MeV/}{\it c}}}
\newcommand{\MeVm}{{\mathrm{MeV/}{\it c}{^2}}}
\newcommand{\KeV}{{\mathrm{KeV}}}
\newcommand{\eV}{{\mathrm{eV}}}
\newcommand{\um}{{\mathrm{m}}}
\newcommand{\umm}{{\mathrm{mm}}}
\newcommand{\uum}{{\mu{\mathrm m}}}
\newcommand{\ucm}{{\mathrm{cm}}}
\newcommand{\ufm}{{\mathrm{fm}}}
\newcommand{\umicrom}{{\mathrm{\mu m}}}
\newcommand{\us}{{\mathrm{s}}}
\newcommand{\ums}{{\mathrm{ms}}}
\newcommand{\uus}{{\mu{\mathrm{s}}}}
\newcommand{\uns}{{\mathrm{ns}}}
\newcommand{\ups}{{\mathrm{ps}}}
\newcommand{\uub}{{\mu{\mathrm{b}}}}
\newcommand{\unb}{{\mathrm{nb}}}
\newcommand{\upb}{{\mathrm{pb}}}
\newcommand{\ipb}{{\mathrm{pb^{-1}}}}
\newcommand{\ifb}{{\mathrm{fb^{-1}}}}
\newcommand{\inb}{{\mathrm{nb^{-1}}}}

\newcommand{\eg}{\mbox{\itshape e.g.}}
\newcommand{\ie}{\mbox{\itshape i.e.}}
\newcommand{\etal}{{\slshape et al\/}\ }
\newcommand{\etc}{\mbox{\itshape etc}}
\newcommand{\cf}{\mbox{\itshape cf.}}
\newcommand{\ffp}{\mbox{\itshape ff}}
\newcommand{\ALEPH}{\mbox{\scshape Aleph}}
\newcommand{\DELPHI}{\mbox{\scshape Delphi}}
\newcommand{\OPAL}{\mbox{\scshape Opal}}
\newcommand{\LTHREE}{\mbox{\scshape L3}}
\newcommand{\CERN}{\mbox{\scshape Cern}}
\newcommand{\LEP}{\mbox{\scshape Lep}}
\newcommand{\LEPONE}{\mbox{\scshape Lep1}}
\newcommand{\LEPTWO}{\mbox{\scshape Lep2}}
\newcommand{\CDF}{\mbox{\scshape Cdf}}
\newcommand{\DO}{\mbox{\scshape D0}}
\newcommand{\SLD}{\mbox{\scshape Sld}}
\newcommand{\CLEO}{\mbox{\scshape Cleo}}
\newcommand{\UAONE}{\mbox{\scshape Ua1}}
\newcommand{\UATWO}{\mbox{\scshape Ua2}}
\newcommand{\TEVATRON}{\mbox{\scshape Tevatron}}
\newcommand{\LHC}{\mbox{\scshape LHC}}
\newcommand{\KORALZ}{\mbox{\ttfamily KORALZ}}
\newcommand{\KORALW}{\mbox{\ttfamily KORALW}}
\newcommand{\ZFITTER}{\mbox{\ttfamily ZFITTER}}
\newcommand{\GENTLE}{\mbox{\ttfamily GENTLE}}
\newcommand{\DELANA}{\mbox{\ttfamily DELANA}}
\newcommand{\DELSIM}{\mbox{\ttfamily DELSIM}}
\newcommand{\DYMU}{\mbox{\ttfamily DYMU3}}
\newcommand{\TANAGRA}{\mbox{\ttfamily TANAGRA}}
\newcommand{\ZEBRA}{\mbox{\ttfamily ZEBRA}}
\newcommand{\PAW}{\mbox{\ttfamily PAW}}
\newcommand{\WWANA}{\mbox{\ttfamily WWANA}}
\newcommand{\FASTSIM}{\mbox{\ttfamily FASTSIM}}
\newcommand{\PYTHIA}{\mbox{\ttfamily PYTHIA}}
\newcommand{\JETSET}{\mbox{\ttfamily JETSET}}
\newcommand{\TWOGAM}{\mbox{\ttfamily TWOGAM}}
\newcommand{\LUBOEI}{\mbox{\ttfamily LUBOEI}}
\newcommand{\SKI}{\mbox{\ttfamily SK-I}}
\newcommand{\SKII}{\mbox{\ttfamily SK-II}}
\newcommand{\ARIADNE}{\mbox{\ttfamily ARIADNE}}
\newcommand{\ARII}{\mbox{\ttfamily AR-II}}
\newcommand{\VNI}{\mbox{\ttfamily VNI}}
\newcommand{\HERWIG}{\mbox{\ttfamily HERWIG}}
\newcommand{\EXCALIBUR}{\mbox{\ttfamily EXCALIBUR}}
\newcommand{\WPHACT}{\mbox{\ttfamily WPHACT}}
\newcommand{\RACOONWW}{\mbox{\ttfamily RACOONWW}}
\newcommand{\YFSWW}{\mbox{\ttfamily YFSWW}}
\newcommand{\QEDPS}{\mbox{\ttfamily QEDPS}}
\newcommand{\CCTHREE}{\mbox{\ttfamily CC03}}
\newcommand{\NCEIGHT}{\mbox{\ttfamily NC08}}
\newcommand{\LUCLUS}{\mbox{\ttfamily LUCLUS}}
\newcommand{\DJOIN}{\mbox{$\mathrm{d_{join}}$}}
\newcommand{\JADE}{\mbox{\ttfamily JADE}}
\newcommand{\DURHAM}{\mbox{\ttfamily DURHAM}}
\newcommand{\CAMBRIDGE}{\mbox{\ttfamily CAMBRIDGE}}
\newcommand{\CAMJET}{\mbox{\ttfamily CAMJET}}
\newcommand{\DICLUS}{\mbox{\ttfamily DICLUS}}
\newcommand{\CONE}{\mbox{\ttfamily CONE}}
\newcommand{\PUFITC}{\mbox{\ttfamily PUFITC+}}
\newcommand{\PHDST}{\mbox{\ttfamily PHDST}}
\newcommand{\SKELANA}{\mbox{\ttfamily SKELANA}}
\newcommand{\DAFNE}{\mbox{\ttfamily DAFNE}}
\newcommand{\CERNLIB}{\mbox{\ttfamily CERNLIB}}
\newcommand{\MINUIT}{\mbox{\ttfamily MINUIT}}
\newcommand{\REMCLU}{\mbox{\ttfamily REMCLU}}
\newcommand{\DST}{\mbox{\ttfamily DST}}
\newcommand{\XSDST}{\mbox{\ttfamily XShortDST}}
\newcommand{\FDST}{\mbox{\ttfamily FullDST}}

\newcommand{\spot}{\mbox{$\bullet \;$}}

\newcommand{\eps}{{\epsilon}}
\newcommand{\erreps}{{\sigma_{\epsilon}}}
\newcommand{\errepsp}{{\sigma_{\epsilon}^{+}}}
\newcommand{\errepsm}{{\sigma_{\epsilon}^{-}}}
\newcommand{\Lmb}{{\Lambda}}
\newcommand{\lmb}{{\lambda}}
\newcommand{\Lmbsq}{{\Lambda^{2}}}
\newcommand{\Lmbisq}{{1/\Lambda^{2}}}
\newcommand{\Lmbpm}{{\Lambda^{\pm}}}
\newcommand{\Lmbp}{{\Lambda^{+}}}
\newcommand{\Lmbm}{{\Lambda^{-}}}
\newcommand{\gc}{{g_{c}}}
\newcommand{\etaij}{{\eta_{ij}}}
\newcommand{\IJ}{{\mathrm{IJ}}}
\newcommand{\IJpm}[1]{{\mathrm{IJ}^{(#1)}}}

\newcommand{\gev}{{\mathrm{GeV}}}
\newcommand{\wpm}{{\mathrm{W}^\pm}}
\newcommand{\wm}{{\mathrm{W}^{-}}}
\newcommand{\w}{{\mathrm{W}}}
\newcommand{\z}{{\mathrm{Z}}}
\newcommand{\ww}{{\mathrm{ W^{+} W^{-} }}}
\newcommand{\tev}{{\mathrm{TeV}}}
\newcommand{\lep}{\mbox{\scshape Lep}}
\newcommand{\lepone}{\mbox{\scshape Lep1}}
\newcommand{\leptwo}{\mbox{\scshape Lep2}}
\newcommand{\delphi}{\mbox{\scshape Delphi}}
\newcommand{\excalibur}{\mbox{\ttfamily EXCALIBUR}}
\newcommand{\wwana}{\mbox{\ttfamily WWANA}}
\newcommand{\fastsim}{\mbox{\ttfamily FASTSIM}}
\newcommand{\phdst}{\mbox{\ttfamily PHDST}}
\newcommand{\skelana}{\mbox{\ttfamily SKELANA}}
\newcommand{\dafne}{\mbox{\ttfamily DAFNE}}
\newcommand{\delana}{\mbox{\ttfamily DELANA}}
\newcommand{\zebra}{\mbox{\ttfamily ZEBRA}}
\newcommand{\delsim}{\mbox{\ttfamily DELSIM}}
\newcommand{\pythia}{\mbox{\ttfamily PYTHIA}}
\newcommand{\jetset}{\mbox{\ttfamily JETSET}}
\newcommand{\ariadne}{\mbox{\ttfamily ARIADNE}}
\newcommand{\cdf}{\mbox{\scshape CDF}}
\newcommand{\gentle}{\mbox{\ttfamily GENTLE}}

\def\Journal#1#2#3#4{{#1}{\bf #2}(#4) #3}
\def\PLB{{ Phys. Lett.}  \bf B}
\def\EUR{{ Eur. Phys. J.} \bf C}
\def\PRL{{ Phys. Rev. Lett.} }
\def\NIMA{{Nucl. Instr. and Meth.} \bf A}
\def\PRD{{ Phys. Rev.} \bf D}
\def\ZPC{{ Zeit. Phys.} \bf C}
\def\CPC{{Comput. Phys. Commun.} }
\def\NUCB{{Nucl. Phys.} \bf B}

\newcommand{\chisqfc}{{\chi^2_{4C}}}
\newcommand{\pkin}{{\bar p_j}}
\newcommand{\WWqqqq}{{\WW~\ra~\qqqq}}
\newcommand{\WWffff}{{\WW~\ra~\ffff}}
\newcommand{\Zqqgam}{{\Z~\ra~\qqgam}}
\newcommand{\ZZqqqq}{{\ZZ~\ra~\qqqq}}


\section{Introduction}
The measurement of the $\W$ boson mass can be used, in combination with other electroweak data, to  test the validity of the Standard Model and obtain estimates of its fundamental parameters. In particular the measurement is sensitive, through loop corrections, to the masses of the top quark and the Higgs boson. 

The $\W$ boson mass and width results presented in this paper are obtained from data recorded by the \DELPHI\ experiment during the 1996-2000 operation of the \LEP\ Collider, known as the \LEP2\ period. This corresponds to a total of 660~$\ipb$ collected over a range of centre-of-mass energies: $\sqs~=~161-209~\GeV$. 

Initially, data were recorded close to the $\WW$ pair production threshold. At this energy the $\WW$ cross-section is sensitive to the $\W$ boson mass, $\mw$. Subsequently, \LEP\ operated at higher centre-of-mass energies, where the $\ee~\ra~\WW$ cross-section has little sensitivity to $\mw$. For these data, which constitute the bulk of the \DELPHI\ data sample, $\mw$ and the $\W$ boson width, $\gw$, are measured through the direct reconstruction of the $\W$ boson's invariant mass from the observed jets and leptons. The analysis is performed on the final states in which both $\W$ bosons in the event decay hadronically ($\WW~\ra~\qqqq$ or  fully-hadronic) and in which one $\W$ boson decays hadronically while the other decays leptonically  ($\WW~\ra~\lnqq$ or  semi-leptonic).

The $\mw$ analyses of the relatively small quantity of data ($\sim20~\ipb$) collected during 1996 at centre-of-mass energies of 161 and 172~$\GeV$ were published in \cite{delpaper161,delpaper172}. These data are not reanalysed in this paper but  are discussed in sections~\ref{sec:xsec} and \ref{sec:172} and included in the final $\mw$ combination.

The data recorded during 1997 and 1998 at $\sqs~=~183$ and $189~\GeV$ have also been the subject of previous \DELPHI\ publications \cite{delpaper183,delpaper189}. These data have been reprocessed and are reanalysed in this paper; the results given here supersede those in the previous publications. Results on the data collected during the final two years of \LEP\ operation are published here for the first time. The data quality, simulation samples and analysis techniques have all been improved with respect to those used in previous \DELPHI\ publications.  The $\W$ mass and width have also been determined by the other \LEP\ collaborations \cite{lepmw} and at hadron colliders \cite{hadmw}.

The results on the $\W$ mass, $\mw$, and width, $\gw$, presented below correspond to a definition based on a Breit-Wigner denominator with an s-dependent width, $|(s-\mw^2) + i s \gw/\mw|$.

After these introductory remarks, the paper starts in section~\ref{sec:lep} by describing the \LEP\ accelerator and the determination of its collison energy. A brief description of the \DELPHI\ detector is provided as section~\ref{sec:det}. This is followed by section~\ref{sec:datasim} which presents the properties of the data sample and of the Monte Carlo simulation samples used in the analysis. 

The analysis method is presented in section~\ref{sec:anal}, first for $\WW~\ra~\lnqq$ events, then for $\WW~\ra~\qqqq$ events. The text describes how the events are selected and the mass and width estimated from $\mw$- and $\gw$-dependent likelihood functions. The potential sources of systematic uncertainty are considered in section~\ref{sec:syst}. These include: inaccuracies in the modelling of the detector; uncertainties on the background; uncertainties on the effects of radiative corrections; understanding of the hadronisation of the $\W$ boson jets; possible cross-talk between two hadronically decaying $\W$ bosons, the effects of which the $\qqqq$ $\mw$ analysis has been specifically designed to minimise; and uncertainty on the \LEP\ centre-of-mass energy determination. The paper concludes in section~\ref{sec:res} with a presentation of the results and their combination.

\section{LEP Characteristics}
\label{sec:lep}

\subsection{Accelerator Operation}

The \LEP2\ programme began in 1996 when the collision energy of the beams
was first ramped to the $\WW$ production threshold of $161~\GeV$
and approximately $10~\rm{pb^{-1}}$ of integrated luminosity
was collected by each experiment.
Later in that year \LEP\ was run at $172~\GeV$ and a dataset
of similar size was accumulated.
In each of the four subsequent years of operation the 
collision energy was raised to successively higher values,
and the accelerator performance improved such that almost half the integrated luminosity was 
delivered at nominal collision energies of $200~\GeV$ and above.
The main motivation for this programme was to improve the sensitivity
of the search for the Higgs boson and other new particles.
The step-by-step nature of the energy increase was 
dictated by the evolving capabilities of the
radio frequency (RF) accelerating system.

%
%

During normal operation the machine would be filled
with 4 electron and 4 positron bunches
at E$_{\rm{beam}} \approx 22~\GeV$, and the beams then
ramped to physics energy, at which point
they would be steered into collision and experimental
data taking begun.   The {\it fill} would last 
until the beam currents fell below a useful level,
or an RF cavity trip precipitated loss of beam.
The mean fill lengths ranged from 5 hours in 1996 
to 2 hours in 1999.   After de-Gaussing the magnets
the cycle would be repeated. 

In 2000, the operation was modified in order 
to optimise still further the high energy reach
of \LEP.
Fills were started at a beam energy safely
within the capabilities of the RF system. 
When the beam currents had decayed significantly,  typically after an hour,
the dipoles were ramped and luminosity delivered at
a higher energy.  This procedure was repeated until the energy was at 
the limit of the RF, and data taken until the beam was lost through a 
klystron trip.  These {\it mini-ramps} lasted less than a minute,  and 
varied in step size with a mean value of 600~$\MeV$.  
The luminosity in 2000 therefore was delivered through a near-continuum of
collision energies between 201 and 209~$\GeV$. 

In addition to the high energy running, 
a number of fills each
year were performed at the $\Z$ resonance.
This was to provide calibration data for the experiments.  
Finally, several fills were devoted to energy calibration activities,
most notably resonant depolarisation (RDP), spectrometer and $Q_s$ measurements (see below for further details).

The machine optics which were used for physics operation  
and for RDP measurements evolved throughout the programme
in order to optimise the luminosity at each energy point.
Certain optics enhanced the build-up of polarisation, and thus
were favoured for RDP measurements. The optics influence
E$_{\rm{beam}}$ in several ways, and are accounted for in the energy model, full details of which are available in ~\cite{lepener}.

\subsection{The LEP Energy Model}

 A precise measurement of the \LEP\ beam energy, and thus the centre-of-mass energy, is a crucial ingredient in
the determination of the $\W$ mass as it sets the overall energy scale. 
The absolute energy scale of \LEP\ is set by the technique of
RDP, which is accurate
to better than 1~$\MeV$. This technique allowed very precise measurements
of the mass and width of the $\Z$ boson to be made at \LEP1.
However, this technique is
only possible for beam energies between  about 41 and 61~$\GeV$.
The \LEP2\ energy scale is set mainly by the nuclear magnetic resonance (NMR) model.   
This makes use of 16 NMR probes, positioned in selected dipoles,
which were used to obtain  local measurements of the bending field.
These probes thus sample the total bending field, which is the 
primary component in determining the beam energy. 
Onto this must be added time-dependent corrections
coming from other sources.
These include effects from earth tides, beam orbit corrections, changes
in the RF frequency, and other smaller effects. Details of all these
can be found in~\cite{lepener}. Using this \LEP\ Energy Model, 
the \LEP\ Energy group provided \DELPHI\
with an estimate of the centre-of-mass energy at the start of each fill and thereafter
in intervals of 15 minutes. For the year 2000 the values before and 
after the mini-ramps were also supplied. No data are used which are taken
during the mini-ramps, as the energy is not accurately known during these
periods.

 The main assumption which is made in the \LEP\ Energy Model is that
the beam energy scales linearly with the readings of the NMR probes.
This assumption of linearity has been tested by three different methods:

\begin{itemize}

\item [1)] {\it Flux Loop.} Each dipole magnet of \LEP\ is equipped with a
single-turn flux loop. Measurements are made for a series of dipole
magnet currents, which correspond roughly to the operating beam 
energies of \LEP2.
This allows the change in flux over almost the entire LEP dipole field
to be measured as the machine is ramped in dedicated experiments.
This change in flux can be compared with the local bending field 
measurements of the NMR probes. The Flux Loop is calibrated against the
\LEP\ energy model in the range 41-61~$\GeV$, using the NMR coefficients
determined from RDP. The measurements from the Flux Loop in the high energy
regime (up to 106~$\GeV$ beam energy) are then compared to those from the
\LEP\ Energy Model. The Flux Loop measurements were made in all years of
\LEP2\ running.

\vspace{0.2cm}

\item [2)] {\it Spectrometer Magnet.} In 1999 a special steel Spectrometer 
Magnet, equipped with three beam position monitors to measure the beam
position both on entry and exit from the magnet, was installed in the \LEP\
ring. The magnetic field of this magnet was carefully mapped before and
after installation in the \LEP\ ring. All these measurements
were very compatible. The beam energy is determined by measuring
the bending angle of the beam in passing through the dipole magnet.
The device was calibrated against RDP in the
41-61~$\GeV$ region and the Spectrometer results were compared to the
\LEP\  Energy Model at beam energies of 70 and 92~$\GeV$. 

\vspace{0.2cm}

\item [3)] {\it $Q_{s}$ versus V$_{RF}$.} 
The synchrotron tune  $Q_s$ can be expressed
as a function of the beam energy and the total RF voltage, V$_{RF}$, plus some additional
small corrections. By measuring $Q_s$ as a function of the total RF voltage
the beam energy can be determined. These measurements were performed in 
1998-2000, at beam energies from 80 to 91~$\GeV$. Again the measurements 
were normalised against RDP in the region 41-61~$\GeV$, and compared to 
the \LEP\ Energy Model at \LEP2\ energies.

\end{itemize}

 The three methods are in good agreement, both with each other and the
\LEP\ Energy Model. Based on these comparisons a small energy offset compared
to the  \LEP\ Energy Model was supplied for each of the
10 beam energies used in \LEP2. This offset is always smaller than 2~$\MeV$.
The estimated centre-of-mass energy uncertainties range between 20 and 40~$\MeV$ and are discussed further in section~\ref{sec:systebeam} .

The \LEP\ centre-of-mass energy has also been determined by the \LEP\ collaborations using \LEP2 events containing on-shell Z bosons and photons (radiative return to the $\Z$ events) \cite{radreturn,radreturnDELPHI}. The \DELPHI\ analysis measured the average difference between the centre-of-mass energy from radiative return events in the $\ee\ra\mumu(\gamma)$ and  $\ee\ra\qq(\gamma)$ channels and the energy reported by the \LEP\ Energy working group,
 \begin{eqnarray*}
   \Delta E_{cm} & = & +0.073 \pm 0.094 (Stat.) \pm 0.065 (Syst.)~ \GeV.
 \end{eqnarray*}

Thus the \DELPHI\ result, relying on similar reconstruction procedures to those described in this paper, is in agreement with the values reported by the \LEP\ Energy working group.

\section{Detector Description}
\label{sec:det}

The $\DELPHI$ detector~\cite{delphi} was upgraded for $\LEP2$.
Changes were made to the subdetectors, the trigger system, 
the run control and the algorithms used in the offline reconstruction
of tracks, which improved the performance compared to the earlier $\LEP1$ period.

The major change was the inclusion of the Very Forward Tracker 
(VFT)~\cite{vft}, which extended the coverage of the innermost 
silicon tracker out to $11\dgr < \theta < 169\dgr$\footnote{The \DELPHI\ 
coordinate system is right-handed with the $z$-axis collinear with 
the incoming electron beam, and the $x$ axis pointing to the centre of the \LEP\ 
accelerator. The radius in the $xy$ plane is denoted $R$ and $\theta$ is used to represent the polar angle to the $z$ axis.}. Together with improved tracking 
algorithms and alignment and calibration 
procedures
optimised for $\LEP2$, these changes led to an improved track reconstruction 
efficiency in the forward regions of $\DELPHI$.

Changes were made to the electronics of the trigger and timing system 
which improved the
stability of the running during data taking. The
trigger conditions were optimised for $\LEP2$ running, to give high
efficiency for Standard Model two- and four-fermion processes and also to give 
sensitivity for events which may be signatures of new physics.
In addition, improvements were made to the operation of the detector during
the $\LEP$ cycle, to prepare the detector for data taking at the very start
of stable collisions of the $\ee$ beams, and to respond to adverse background
from $\LEP$ were they to arise. These changes led to an overall improvement
of $\sim10\%$ in the efficiency for collecting the delivered luminosity from 
$\sim85\%$ in 1995, before the start of $\LEP2$,  to $\sim95\%$ at the end in 2000.

During the operation of the \DELPHI\ detector in 2000 one of the 12 sectors of
the central tracking chamber, the TPC, failed. After the
$1^{\mathrm{st}}$ September 2000 it was not possible to detect the tracks left 
by charged particles inside the broken sector. The data affected correspond
to $\sim 1/4$ of the total dataset of the year 2000. Nevertheless, the redundancy
of the tracking system of \DELPHI\ meant that tracks passing through the sector
could still be reconstructed from signals in any of the other tracking 
detectors. A modified track reconstruction algorithm was used in
this sector, which included space points reconstructed in the Barrel RICH
detector. As a result, the track reconstruction efficiency was only slightly reduced in 
the region covered by the broken sector, but the
track parameter resolutions were degraded compared with the data taken prior to the failure of this sector.

\section{Data and Simulation Samples}
\label{sec:datasim}

\subsection{Data}
\label{sec:lumi}

The $\W$ mass and width are measured in this paper with the data samples collected during the 1996-2000 operation of the $\LEP$ Collider. A summary of the available data samples is reported in table~\ref{tab:enlum}, where
the luminosity-weighted centre-of-mass energies and the 
amount of data collected at each energy are shown.
The luminosity is determined from Bhabha scattering measurements
making use of the very forward electromagnetic calorimetry~\cite{STIC}.
The total integrated luminosity for the \LEP2\ period corresponds 
to approximately 660~pb$^{-1}$.
The integrated luminosities used for the different selections correspond to those data
for which all elements of the detector essential to each specific analysis
were fully functional. The additional requirements on, for example, the status of the calorimetry and the muon chambers mean  that the integrated luminosity of the semi-leptonic analysis is slightly less that that of the hadronic dataset.

All the data taken from the year 1997 onwards have been reprocessed with an
improved reconstruction code, and the analyses on these
data are updated with respect to the previously published
ones  and supersede them. 
The data taken in 1996 have not been reanalysed; the results from this year are taken from the previous publications with minor revisions as reported in section \ref{sec:res}.

In addition to these data taken above the $\WW$-pair production threshold, data were also recorded during this period at the $\Z$ peak.
These samples, containing a total of over 0.5 million collected $\Z$ decays, were taken each year typically at the start and end of the data taking periods. These $\Z$ peak samples were used extensively in the alignment and calibration of the detector and are used in many of the systematic uncertainty studies reported in section~\ref{sec:syst}.

\begin{table}[htb]
\begin{center}
\begin{tabular}{|c|c|c|c|}
\hline
Year & ${\cal L}$-weighted $\sqs$ ($\GeV$) & Hadronic Int. ${\cal L}$
(pb$^{-1}$)&  Leptonic Int. ${\cal L}$ (pb$^{-1}$) \\
\hline
 1996 &  161.31 &  10.1 &  10.1 \\
      &  172.14 &  10.1 &  10.1  \\
\hline
 1997 &  182.65 &   52.5 &  51.8  \\
\hline
 1998 &  188.63 &  154.4 & 152.5 \\
\hline
 1999 &  191.58 &   25.2 &  24.4  \\
      &  195.51 &   76.1 &  74.6 \\
      &  199.51 &   82.8 &  81.6 \\
      &  201.64 &   40.3 &  40.2 \\
\hline
 2000 &  205.86 &   218.4 &  215.9  \\
\hline
\end{tabular}
\vspace{0.2cm}
\caption{\label{tab:enlum} Luminosity-weighted centre-of-mass energies 
and integrated luminosities in the \LEP2\ data taking period. The hadronic integrated luminosity is used for the fully-hadronic channel, the leptonic one is used for the semi-leptonic
channels.}
\end{center}
\end{table}

\subsection{Simulation}
\label{sec:simul}

The response of the detector to various physical processes was
described using the simulation program {\tt DELSIM} \cite{delphi},
which includes modelling of the resolution, granularity and efficiency
of the detector components. In addition, detector correction factors,
described in section~\ref{sec:syst}, were included to improve the description of
jets, electrons and muons. To allow use of the data taken after the
$1^{\mathrm{st}}$ September in 2000, samples of events were
simulated dropping information from the broken sector of the TPC.
A variety of event generators were used to describe all the physics
processes relevant for the analysis. $\WW$ events and all other
four-fermion processes were simulated with the program described in
\cite{delphi4fgen}, based on the {\tt WPHACT}~2.0 generator \cite{wphact}
interfaced with {\tt PYTHIA}~6.156 \cite{pythia} to describe quark
hadronisation and {\tt TAUOLA}~2.6 \cite{tauola} to model
$\tau$ leptons decays. The most recent $\mathcal{O}(\alpha)$
electroweak radiative corrections in the so-called Double Pole
Approximation (DPA) were included in the generation of the signal via
weights computed by {\tt YFSWW}~3.1.16 \cite{yfsww}, 
and the treatment of initial state radiation (ISR)  of this calculation was adopted. The photon
radiation from final state leptons was computed with {\tt PHOTOS}~2.5
\cite{photos}. For systematic studies the alternative hadronisation
descriptions implemented in {\tt ARIADNE}~4.08 \cite{ariadne} and {\tt HERWIG}~6.2
\cite{herwig} were also used. All the hadronisation models were tuned
on the \DELPHI\ $\Z$ peak data \cite{deltune}.

The background process $e^+e^- \rightarrow q\bar{q}(\gamma)$ was
simulated with {\tt KK}~4.14 \cite{kk} interfaced with {\tt PYTHIA}~6.156 for the
hadronisation description. The two-photon events giving rise to
those $e^+e^- q\bar{q}$ final states not described in the four-fermion
generation above were produced with {\tt PYTHIA}~6.143 as
discussed in \cite{delphi4fgen}. The contribution from all other background
 processes was negligible.

The simulated integrated luminosity used for the analysis was about a factor 350
higher than for the real data collected for 4-fermion processes, 
about a factor 60 higher for 2-fermion final states and about 3.5 times greater for $e^+e^-q\bar{q}$ two-photon final states (those not already included in the 4-fermion simulation). 
\section{Analysis Method}
\label{sec:anal}

The measurement of $\mw$ and of $\gw$ are performed on samples of $\WW~\ra~\lnqq$ and $\WW~\ra~\qqqq$ events; these two channels are discussed in turn below. The reconstruction of events where both $\W$s decay leptonically has very limited sensitivity to the $\W$ mass and width, as they contain at least two undetected neutrinos, and hence are not used in this analysis.

The first stage in the analysis is to select events from these decay channels, using either a neural network or a sequential cut-based approach. In some channels, after preliminary cuts, the probability is assessed for each event of how $\WW$-like it is and a corresponding weight is applied in the analysis.

The resolution of the kinematic information extracted from the observed particles in the event can be improved by applying energy and momentum conservation constraints to the event; this is discussed in section~\ref{sec:kfit}. In the fully-hadronic channel the jet directions used as the input to the kinematic fit are also assessed excluding particles from the inter-jet regions. This alternative approach reduces the sensitivity of the $\W$ mass analysis to final state interaction systematics and is discussed in section~\ref{sec:cone}. 

The next stage in the analysis is to produce a likelihood function expressing the relative probability of observing an event as a function of $\mw$ and $\gw$. The likelihood functions used below depend not only on the reconstructed $\W$ mass of the event but make use of other event characteristics to assess the relative weight and resolution of each event. These likelihood functions are then calibrated against simulated events.

The $\W$ mass and width are then extracted by maximising the combined likelihood function of the full observed dataset.

\subsection{Application of Kinematic Constraints to Event Reconstruction}
\label{sec:kfit}
 
The event-by-event uncertainty on the centre-of-mass energy, \ie\ the energy spread, at \LEP\ is typically 0.1\%, while the overall momentum and energy resolution of the observed final state is about 10\%. Hence, the precise knowledge of the kinematics in the initial state can be used to significantly improve the reconstructed kinematic information obtained from the clustered jets and observed leptons in the final state. This is accomplished by means of a $\chi^2$ fit based on the four constraints from the conservation laws of energy and momentum.

The reconstructed jets and leptons of the event may be associated with one of the two hypothesised $\W$ bosons in the event.
A fifth constraint may then be applied to the event by assigning equal masses to these $\W$ boson candidates. As the decay width of the ${\rm{W^{\pm}}}$ bosons is finite, this constraint is non-physical. However, as the event mass resolution and 2~$\GeVm$ $\W$ width are of comparable magnitude in practice this constraint provides a useful approximation. It is of particular use in the semi-leptonic decay channels where, after applying the four-constraints, the event mass resolution is still larger than the $\W$ width and, due to the unseen neutrino, the two fitted masses are strongly anticorrelated. However, in the fully-hadronic decay channel the mass resolution after the four-constraint fit is better and the correlation is less; hence more information is available in the two four-constraint masses than the combined five-constraint event mass.

\vspace{0.2cm}
{\bf Parameterisation of Jets and Leptons}
\vspace{0.2cm}

Each fitted object, jet or lepton, is described by three parameters.
Muons are described by their measured momenta and their polar and azimuthal angles. The uncertainties on these parameters are obtained directly from the track fit. Electrons are characterized by their measured energies and their detected angular position in the electromagnetic calorimeters. The energy uncertainties are obtained
from parameterisations of the responses of the electromagnetic calorimeters, which were tuned to the responses found in Bhabha and Compton scattering events. The angular uncertainties were determined from the detector granularity and were significant only for the forward electromagnetic calorimeter. In semi-leptonic events, the neutrino momentum vector
is considered as unknown, which leads to a reduction by three in the number of effective constraints in the kinematic fit.

\begin{figure}[htp]
\begin{center}
  \includegraphics[height=8cm]{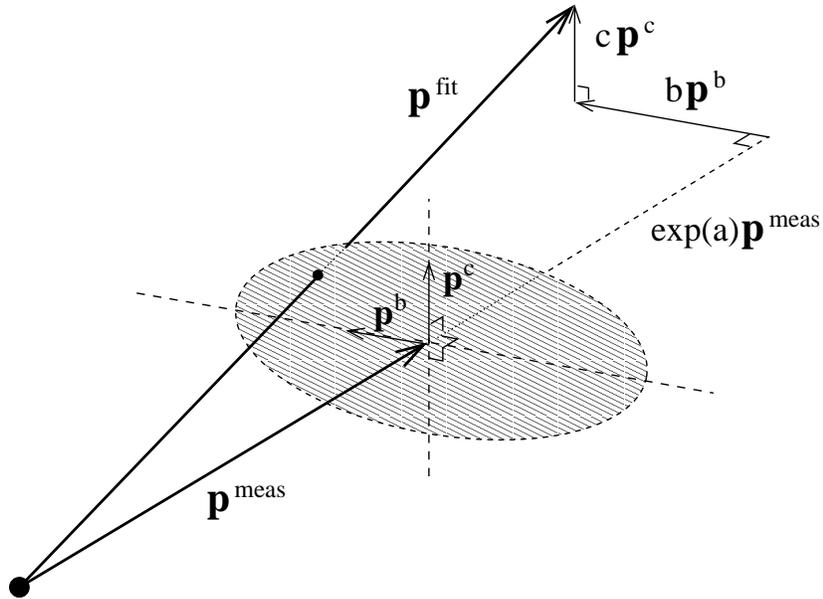}
\caption{Parameterisation used for jets in the constrained fit, as explained in the text and equation~\ref{JetPar}.}
\label{AnaFitPar}
\end{center}
\end{figure}

Each fitted jet momentum ${\vec p_j^{\ f}}$ is projected onto a set of axes with one component
parallel to the measured jet momentum ${\vec p_j^{\ m}}$ and two transverse components, ${\vec p_j^{\ b}}$
and ${\vec p_j^{\ c}}$, each normalized in magnitude to 1~$\GeVp$. In this coordinate system ${\vec p_j^{\ f}}$
can be described by three parameters $a_j$, $b_j$ and $c_j$:

\begin{equation}
 {\vec p}_j^{\ f}= e^{a_j} {\vec p}_j^{\ m} +
 b_j{\vec p_j}^{\ b} + c_j{\vec p_j}^{\ c} ,
\label{JetPar}
\end{equation}

\noindent
where each component is shown in figure~\ref{AnaFitPar}. 
The measured jet energy $E_j^{\ m}$ is rescaled with the same factor $e^{a_j}$ as the jet momentum.
The exponential parameterisation $e^{a_j}$ of the factor in front of ${\vec p}_j^{\ m}$ makes the fit 
more stable and results in uncertainties which have a more Gaussian distribution.
The values of the parameters are determined by performing a
constrained fit, while the transverse directions are given by the
eigenvectors of the momentum tensor described below.

\vspace{0.2cm}
{\bf Form of $\chi^2$}
\vspace{0.2cm}

The algorithm minimizes a $\chi^2$, defined for fully-hadronic events as:
\begin{equation}
 \chi^2 = \sum_{j=1}^{jets} \frac{(a_j-a_0)^2}{\sigma_{a_j}^2} + \frac{{b_j}^2}{\sigma_{b_j}^2} + \frac{{c_j}^2}{\sigma_{c_j}^2},
\label{Chi2}
\end{equation}

\noindent
while forcing the fitted event to obey the constraints. The appropriate terms are included in the $\chi^2$ for events with a leptonic $\W$ decay. The expected energy loss parameter $a_0$ and the 
energy spread parameter $\sigma_{a_j}$, together with the parameters $\sigma_{b_j}$ and $\sigma_{c_j}$, 
are parameterised as functions of the jet polar angles.

\vspace{0.2cm}
{\bf Jet Error Parameterisation}
\vspace{0.2cm}

The jet error parameters, $a_0$, $\sigma_{a_j}$, $\sigma_{b_j}$ and $\sigma_{c_j}$ were obtained from a study of hadronic $\Z$ events. Hadronic $\Z$ events with a two-jet topology were selected from the $\Z$ calibration run data or from the corresponding Monte Carlo simulation. The reconstructed jet energies were compared with the beam energy. In general an energy loss of around 10\% was observed for jets in the barrel region of the detector while this increased to 15\% in the forward regions. A good agreement between the data and simulation was found. The energy loss increases if the event jet topology
becomes less two-jet like, resulting in energy losses of around 15\% for the barrel region and up to 35\% in the forwards regions.


\noindent

The uncertainties on the jet parameters for the first stage of the fit were determined from this study as a function of the polar angle of the jet. However, a dependence of these parameters on the properties of the individual jets has also been observed.

\vspace{0.2cm}
{\bf Jet Breadth}
\vspace{0.2cm}

The dependence of the uncertainties on the individual jet properties is included in a second stage of the fit, where the parameterisation of
the transverse momentum uncertainties depends upon the breadth of the jet. This breadth is 
calculated by projecting the momenta of all particles in the jet on to the plane transverse to the jet
axis. From these projections a two dimensional momentum tensor ${\cal T}_{\beta \gamma}$ is created:

\begin{equation}
{\cal T}_{\beta \gamma} = \sum_{k} p_{\beta}^k p_{\gamma}^k,
\label{momtensor}
\end{equation}

\noindent
where $p_{\beta}^k$ and $p_{\gamma}^k$ are the two components of the projection of the
momentum of particle $k$ in the transverse plane. The normalized eigenvectors of the tensor, ${\vec p}_{j}^{\ b}$ and
${\vec p}_{j}^{\ c}$, reflect the directions where the jet is broadest and slimmest.
The corresponding eigenvalues are $B_b$ and $B_c$. By comparing the resulting jet energies from the first
stage of the fit with the measured ones, an estimate is made of how much energy remained
undetected in the jet, referred to as $E_{j,miss}$. 
The uncertainties on the jet breadths were then parametrised as a function of the eigenvalues, the measured jet energy and the missing energy $E_{j,miss}$.




 

\vspace{0.2cm}
{\bf Use of $\chi^2$}
\vspace{0.2cm}

The $\chi^2$ of the resulting fit is a function of the collection of jet 
parameters $(a_j,b_j,c_j)$  and lepton parameters. The jets and leptons are paired appropriately to each $\W$ boson decay and constraints applied. The total $\chi^2$ is then minimized by an iterative procedure using Lagrange multipliers for the constraints. 

Events for which the $\chi^2$ of the fit is larger than the number of degrees of freedom for the fit, NDF, had their errors scaled by a factor of $\sqrt{\chi^2/NDF}$ in order to take non-Gaussian resolution effects into account.

In the semi-leptonic analysis described in section~\ref{sec:semileptfit} the value of the best fit mass from the $\chi^2$ minimum and the error on this mass is used for each event. In the fully-hadronic analysis described in section~\ref{sec:fullyhadfit} each event uses the $\chi^2$ distribution as a function of the masses of the two $\W$ bosons in the event.

\subsection{Semi-Leptonic Decay Channel}

The $\WW~\ra~\lnqq$ events constitute $44\%$ of all $\WW$ decays. The $\WW$ event candidates are classified according to their leptons and their selection is performed using a neural network. An event $\W$ mass is reconstructed in a kinematic fit, by imposing momentum conservation, the measured centre-of-mass energy and equality of the leptonic and hadronic decay $\W$ masses. An estimate of the mass resolution in each individual event is also obtained from the kinematic fit and an estimate of the event purity is obtained from the neural network output; these quantities are both used in producing the likelihood function from which $\mw$ and $\gw$ are determined.

\subsubsection{Event Selection}

 Events are selected from the recorded data sample requiring that all detectors essential for this measurement were fully efficient: these comprise the central tracking detectors and the electromagnetic calorimeters. The data recorded during the period with a damaged sector of the TPC are also used with matching simulation samples produced. The corresponding integrated luminosities, at each centre-of-mass energy, are given in table~\ref{tab:enlum}.

   Events containing at least three charged particle tracks and with a visible mass greater than 20~$\GeVm$ are considered for analysis. Events containing lepton candidates are then identified in this sample, either by direct lepton identification (electrons and muons), or by clustering the events into a three-jet 
configuration and selecting the jet with the lowest charged multiplicity as the tau candidate. At this stage, events can be considered as candidates in multiple channels.

\vspace{0.2cm}
{\bf{Electron and Muon Identification}}
\vspace{0.2cm}

Charged particles are identified as muons if they are associated
with a hit in the muon chambers, or have an energy deposit 
in the hadron calorimeter that is consistent with a minimum ionising 
particle. Muon identification is performed in the polar angle range between
10$^{\circ}$ and 170$^{\circ}$.
Muons with an unambiguous association \cite{delphi}  with the hits in the muon
chambers, or with a loose association in addition to a good pattern in the hadron calorimeter are classified as good candidates, with the remainder being classified as possible candidates.

Electron identification is performed in the polar angle range between
15$^{\circ}$ and 165$^{\circ}$ by selecting charged particles with a
characteristic energy deposition in the electromagnetic calorimeters.
In the central region of the detector, covered by the HPC electromagnetic
calorimeter, the electron selection followed the criteria described
in \cite{delphi} for candidates below $30~\GeV$. This selection is based
on a neural network using the electron energy to momentum ratio (E/p), the spatial matching between the 
extrapolated track and the shower, the shower shape and the track energy loss per unit path length in the TPC (dE/dx) as the discriminating
variables. Above 30~$\GeV$, a simplified selection is adopted, the main deposit associated with a charged particle track is identified and the surrounding electromagnetic showers are clustered into this electron candidate. Only candidates
 with E/p greater than 0.5 are used.
In the polar angle region corresponding to the forward electromagnetic calorimeter acceptance, below 36$^{\circ}$ and above 144$^{\circ}$, electron candidates are selected from among the calorimetric shower clusters. Only clusters with an energy above 8~$\GeV$ and which could be geometrically associated to extrapolated charged particle tracks are used. The electron candidates are separated into categories of good and possible candidates based on the quality of the track associated with the electron. The association of vertex detector hits to the track is a primary criterion used in assessing the track quality. 

\vspace{0.2cm}
{\bf{Tau reconstruction}}
\vspace{0.2cm}

 As mentioned above, tau candidate events are clustered into a three-jet 
configuration using the \LUCLUS\ \cite{luclus} algorithm. Tracks at
large angle (more than 40$^\circ$ from the nearest jet axis) or which
contribute a large mass to the jet they belong to (${\Delta}M$ bigger than
3.5~$\GeVm$) are removed from the tau candidate. As the tau lepton 
predominantly decays into a final state with one or three charged particles,
with few neutrals, a pseudo-multiplicity defined as the sum of the charged
multiplicity and one quarter of the neutral multiplicity is used and
the jet with the lowest pseudo-multiplicity is chosen as the tau candidate.
Then a further cleaning is applied on this tau candidate : tracks at
more than 20$^\circ$ from the tau axis, or which contribute a large mass
(${\Delta}M$ bigger than 2.5~$\GeVm$) are removed from the tau candidate.
Only tau candidates containing between one and four charged particle tracks
after this cleaning, and with a polar angle between 15$^\circ$ and 165$^\circ$
are kept. Two classes of events are then defined, those with only one charged particle track, and all others.

\vspace{0.2cm}
{\bf{Event Reconstruction and Pre-selection}}
\vspace{0.2cm}

After the lepton identification is performed, the events are reconstructed as the lepton and a two or three jet system. Pre-selection cuts are then applied.

   All tracks not associated to the lepton are clustered using the $\LUCLUS$
algorithm. These jet tracks in semi-leptonic electron and muon decay channel events are clustered with $\DJOIN~=~7.5~\GeVp$,where $\DJOIN$ is a measure of the clusterisation scale used inside $\LUCLUS$. If more than three jets are obtained the tracks are forced into a three-jet configuration. This procedure correctly treats events with hard gluon radiation (the proportion
of three-jet events is about 20\%). In semi-leptonic tau decay events the tracks not associated to the tau candidate are forced into a two-jet configuration.

   A set of pre-selection cuts is then applied. First, a common set of criteria is applied to the system of jets:
\begin{itemize}
  \item visible mass greater than $30~\GeVm$;
  \item at least five charged particle tracks, with at least two with momentum transverse to the beam greater than 1.5 $\GeVp$ and compatible with the primary vertex (impact parameter in $R~<0.15~\ucm$ and in $z~<0.4~\ucm$); 
  \item no electromagnetic cluster with an energy bigger than $50~\GeV$.
\end{itemize}

   Then, for electron and muon semi-leptonic decay channel events, the following additional cuts are used:
\begin{itemize}
  \item energy of the lepton bigger than 20~$\GeV$;
  \item if there is another isolated lepton of the same flavour and opposite
charge, the event acollinearity should be bigger than 25$^\circ$. The acollinearity used here is that between the two `jets' when forcing the event into a two-jet (including the lepton) configuration.
\end{itemize}
 Further cuts are made for electron decay channel events:
 \begin{itemize}
    \item missing transverse momentum should be greater than 8~$\GeVp$;
    \item the cut on missing transverse momentum is increased to 12~$\GeVp$
 for electron candidates in the `possible' class;
    \item angle between the lepton and the nearest jet greater than 15$^\circ$. 
 \end{itemize}
  The cuts specific to the muon decay channel events are: 
  \begin{itemize}
    \item angle between the lepton and the nearest jet greater than 15$^\circ$
 in the case of `possible' class muons;
    \item angle between the missing momentum and the beam axis greater than 10$^\circ$ for muon candidates in the `possible' class. 
\end{itemize}
  While for tau decay channel events, the cuts applied are:
\begin{itemize}
  \item visible hadronic mass smaller than 130~$\GeVm$;
  \item energy of the tau greater than 5~$\GeV$;
  \item fraction of energy of the tau associated to charged tracks greater than 5\%;
  \item at least one of the charged particle tracks from the tau must have a vertex detector hit;
  \item angle between the tau and the nearest jet greater than 15$^\circ$;
  \item angle between the tau and the nearest charged particle greater than 10$^\circ$;
  \item missing transverse momentum greater than 8~$\GeVp$;
  \item the cut on missing transverse momentum is increased to 12~$\GeVp$
 in the case of tau candidates with several charged particles.
\end{itemize}

\par The semi-leptonic electron and muon events are then reconstructed using a constrained fit imposing
conservation of four-momentum and equality of the two $\W$ masses in the event.
As the energy of the tau lepton is unknown, due to the emission of at least
one neutrino in its decay, the mass in the $\tnqq$ channel is entirely 
determined by the jet system and no improvement can be made from applying a constrained fit.

\vspace{0.2cm}
{\bf{Selection}}
\vspace{0.2cm}

The event selection is based upon a multi-layer perceptron neural network \cite{neural}. The network has been optimised separately
for the six classes of events (good and possible $\enqq$, good and possible $\mnqq$, and $\tnqq$ candidates containing either only
one or several charged particles).

\begin{figure}[htp]
  \begin{tabular}{cc}
    \epsfig{file=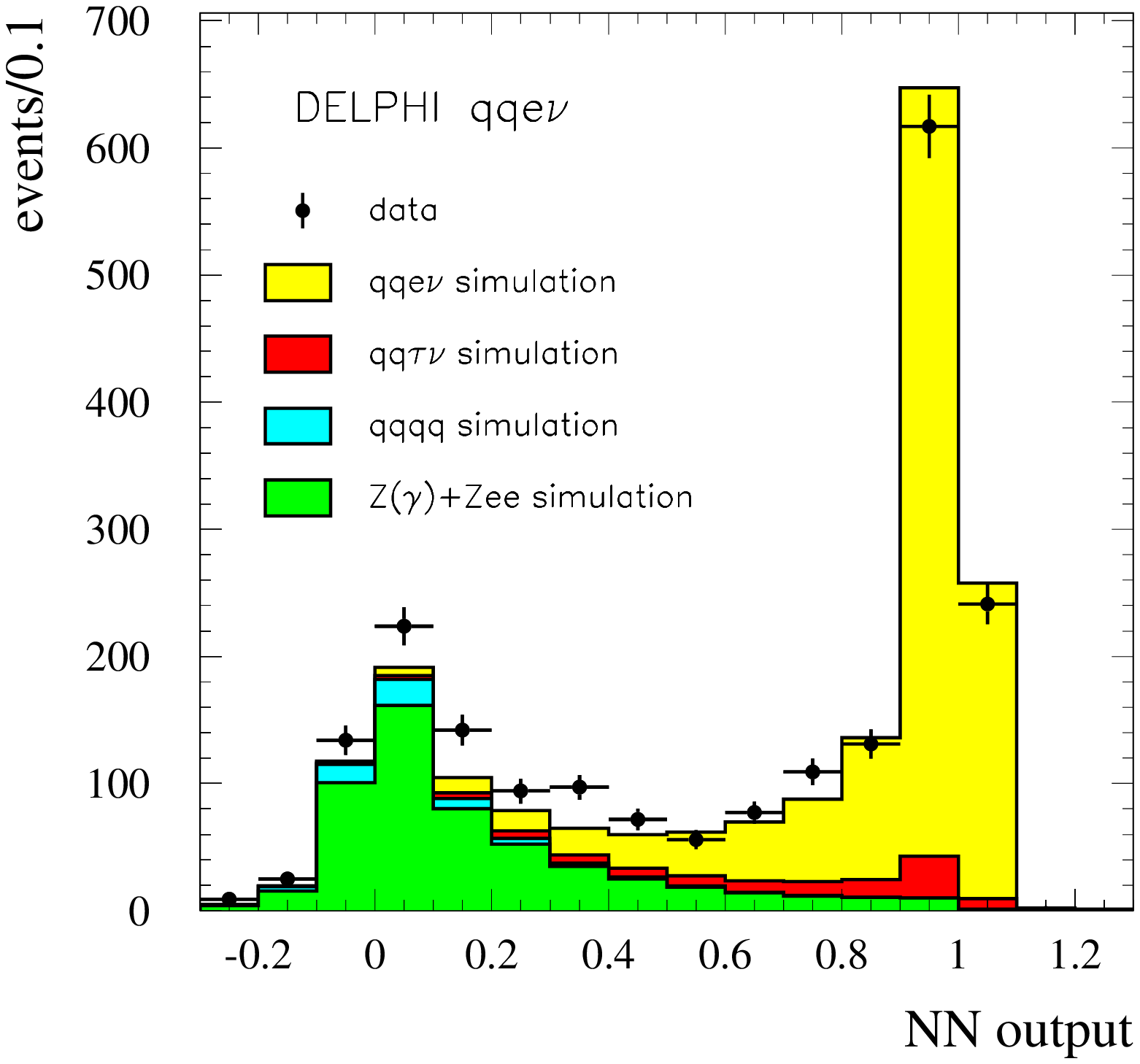,width=0.48\textwidth} &
    \epsfig{file=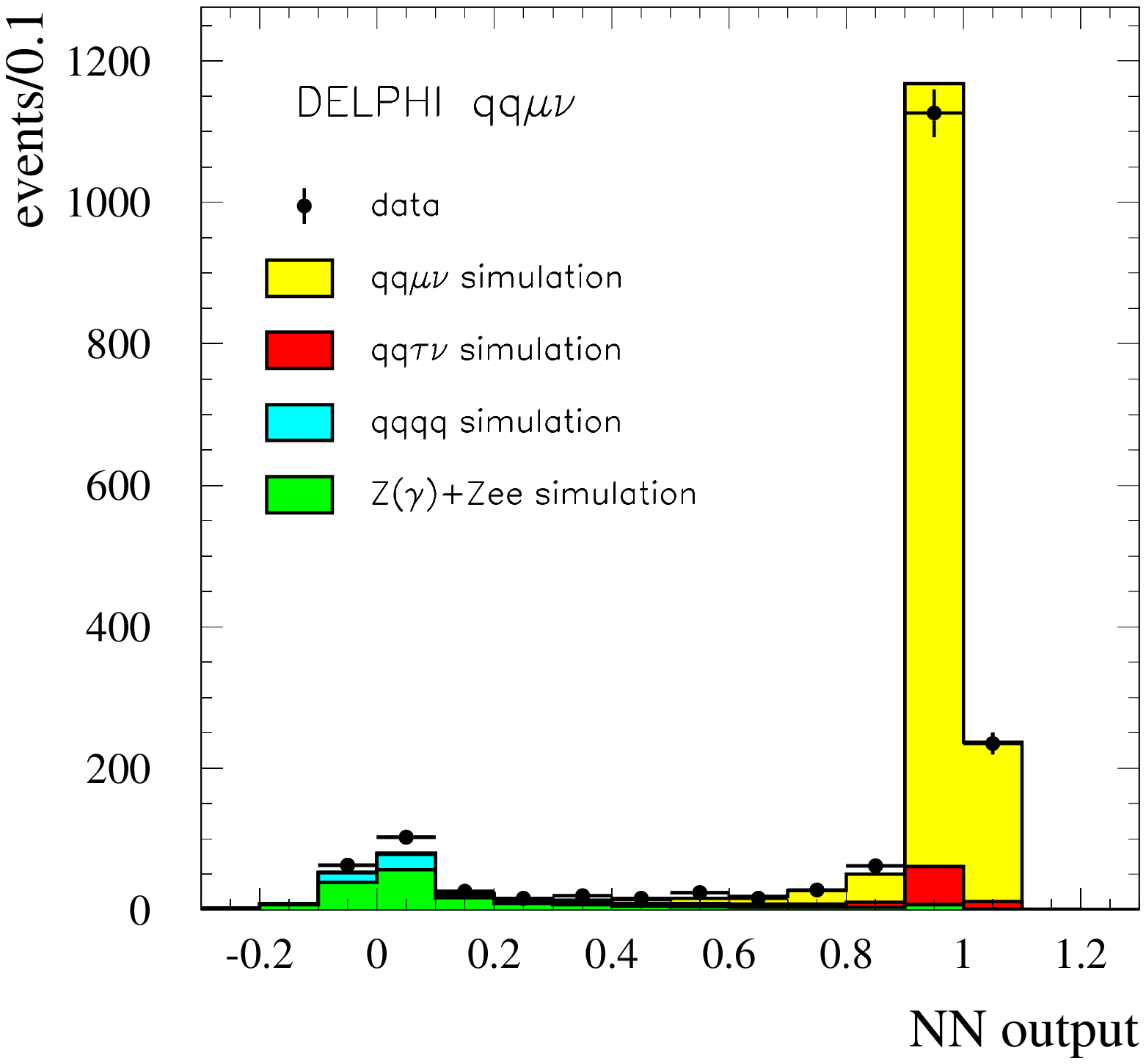,width=0.48\textwidth} \\
   \multicolumn{2}{c}{\epsfig{file=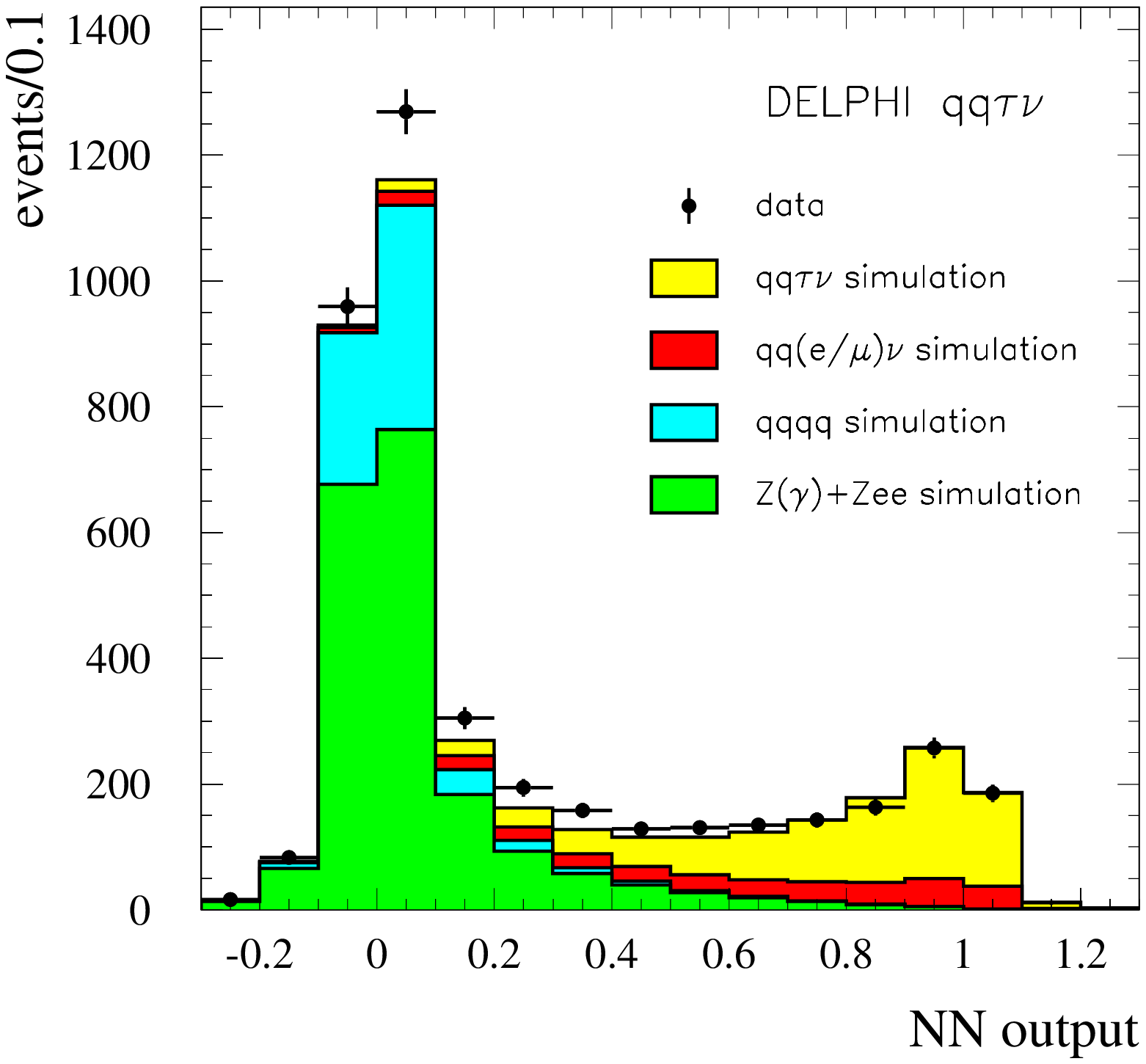,width=0.48\textwidth}} \\
  \end{tabular}
  \caption{The output of the neural network used for the selection of the semi-leptonic
 channels for the data sample recorded at $\sqs~=~183-209~\GeV$.  The data are indicated by the data points with error bars. The histograms show the signal and background simulation contributions normalised to the integrated luminosity of the data sample.
}
 \label{fig:NNout}
\end{figure} 

The choice of the variables used in the neural networks is a compromise between
their independence from the $\W$ mass and their discriminant power. 
The number of input-hidden-output nodes were 12-8-1, 11-7-1 and 17-12-1 for
the e, $\mu$ and $\tau$ channels respectively.
The detailed list of variables is given below.
The network has been tuned on samples of signal and background simulation events,
and examples of the distribution of the neural network output value are shown in figure~\ref{fig:NNout}. The applied selection cut is at 0.40, 0.50 and 0.35 for the e, $\mu$ and $\tau$ channels respectively, independent of the centre-of-mass energy. Any discrepancy in the background rate between data and simulation is accounted for in the systematic uncertainty applied.

The event selection procedure ensures that the events are only selected in one of the channels: events that pass the chosen cut in the muon channel are selected, the remaining events are considered as electron channel candidates and, if they are again rejected, are then analysed under the tau channel hypothesis. This ordering follows the hierarchy of purities in these channels (and is not dependent on the good or possible lepton classes).
After applying the cut on the network output the selection performance is as shown in table~\ref{tab:evtsel}. 
As an example, the global efficiencies for \CCTHREE\ events are 79.8$\%$, 89.8\% and 59.3\% respectively for the $\enqq$, $\mnqq$ and $\tnqq$ events in the data taken at $\sqs~=~189~\GeV$. These numbers are integrated over all event selections as there is a non-negligible cross-contamination of events in the event selections (\eg\ $\enqq$ event selected by the $\tnqq$ selection) which still add useful information in the $\W$ mass and width fits. Here \CCTHREE\ refers to the three charged current processes producing the $\WW$ state for which this analysis is intended:  s-channel photon or Z  production and t-channel $\nel$ exchange.

\begin{table}[ht]
 \begin{center}
  \begin{tabular}{|l|c|c|c|c|c|c|c|}
\hline
                    & \multicolumn{6}{|c|}{Simulation} & \\
1998, 189 $\GeV$    &  (Primary-$l$)$\nqq$  & (Other-$l$)$\nqq$  & $\qqqq$ & $\qq(\gamma)$ & Other 4f & Total &  Data\\
\hline \hline
$\enqq$             &    $257.5$        & $10.5$         & $ 0.7$      &  $9.3$        & $6.5$   & $284.5$ & 256 \\
$\mnqq$             &    $321.2$        & $10.2$         & $ 0.4$      &  $1.1$        & $2.2$   & $335.1$ & 320 \\
$\tnqq$             &    $198.2$        & $56.6$         & $ 3.5$      &  $18.6$       & $10.9$  & $287.9$ & 294 \\
$\qqqq$             &    --             & $34.0$         & $ 1029.9 $  &  $341.6$      & $50.8$  & $1456.3$ & 1506 \\
\hline
2000, 206 $\GeV$    &  &  & &  &  &  &  \\
\hline
$\enqq$             &    $373.9$        & $16.9$        & $1.0$    &  $13.6$     & $11.4$  & $416.8$  & 395 \\
$\mnqq$             &    $457.0$        & $14.8$        & $0.6$    &  $1.7$      & $4.1$   & $478.2$  & 467 \\
$\tnqq$             &    $290.2$        & $87.6$        & $5.7$    &  $22.3$     & $21.4$  & $427.2$  & 426 \\
$\qqqq$             &    --             & $40.6$        & $1514.5$ &  $460.9$    & $107.8$ & $2123.8$ &  2134 \\
\hline
1997-2000  &  &  & &  &  &  & \\ 
183-206 $\GeV$    &  &  & &  &  &  &  \\
\hline
$\enqq$             &    $1091.5$        & $47.7$        & $2.9$    & $39.9$     & $30.7$  & $1212.7$  & 1182 \\
$\mnqq$             &    $1356.7$        & $43.3$        & $1.7$    & $15.2$     & $11.0$  & $1417.8$  & 1402 \\
$\tnqq$             &    $849.3$         & $248.6$       & $16.0$   & $72.2$     & $55.6$  & $1241.6$  & 1270 \\
$\qqqq$             &    --              &  $131.6$      & $4421.0$ & $1399.5$   & $269.8$ & $6222.0$  & 6446 \\
\hline
  \end{tabular}
\caption{ Number of selected events in the decay channel event selections from the 1998 and 2000 data samples and the combined 1997-2000 data sample, and the corresponding number of expected events from the simulation. The table is split into rows giving the results of each of the event selection routines. The primary-$l$ and other-$l$ $\nqq$ columns relate to the nature of the semi-leptonic event selections $\eg$ for the  $\enqq$ selection the results are for the $\enqq$ and $(\mnqq+\tnqq)$ channels respectively. }
  \label{tab:evtsel}
 \end{center}
\end{table}

  For each of the six classes of events, the fraction of semi-leptonic $\WW$
events in the sample has been extracted from simulation as a function of the
neural network output: this is referred to below as the event purity $P_e$.
This feature is particularly useful for the tau selection, where the proportion
of background events is highest.

\subsubsection{Variables used in the Selection Neural Networks}
\label{sec:nnet}
  {\bf{Common Variables for all Leptonic Channels}}
  \begin{itemize}
    \item Polar angle of the leptonic $\W$ (after applying the constrained fit);
    \item angle of the charged lepton with respect to the direction of the leptonic $\W$ (in
 the $\W$ rest frame, and after the constrained fit);
    \item polar angle of the lepton;
    \item polar angle of the missing momentum vector;
    \item angle between the lepton and the nearest jet;
    \item angle between the lepton and the nearest charged hadron track (of
 energy greater than 1~$\GeV$);
    \item missing transverse momentum;
    \item the invariant mass of the measured system of particles $\sprime$ \cite{SPRIM} - this is
 measured using planar kinematics, by forcing the event into 2 jets (using all particles in the event including the lepton) and
assuming a photon is emitted down the beam pipe;
    \item aplanarity (cosine of the angle between the lepton
 and the normal to the plane formed by the jets\footnote{for three-jets events
 in the electron and muon channels, the jets-plane is the plane formed
 by the most energetic jet and the sum of the two others.});
    \item acollinearity (complement of the angle between the two ``jets'' when
 forcing the event into a two-jet configuration);
    \item the minimum $\DJOIN$ distance in the $\LUCLUS$ jet clusterisation algorithm between two jets in the final configuration, where the whole event (hadronic and leptonic system) is forced into three jets. This is known as $\mathrm{d_{j3all}}$.
  \end{itemize}

{\bf{Additional Variable for the Electron Channel Only}}
  \begin{itemize}
    \item Angle between the missing momentum and the nearest jet.
  \end{itemize}

{\bf{Additional Variables for the Tau Channel Only}}
  \begin{itemize}
    \item Angle between the missing momentum and the nearest jet;
    \item fraction of the tau energy coming from charged particle tracks;
    \item missing energy;
    \item reconstructed tau energy;
    \item reconstructed tau mass;
    \item $\mathrm{d_{j4all}}$, as $\mathrm{d_{j3all}}$ (see above) but with the final event configuration forced into four jets.
  \end{itemize}

\subsubsection{Likelihood Function}
\label{sec:semileptfit} 
A likelihood function, ${\cal L}_e (\mw,\gw)$, is evaluated for each selected event with a reconstructed mass in a defined range. The range was 67-91~$\GeVm$ for the data collected in 1997, 67-93~$\GeVm$ for 1998, 67-95~$\GeVm$ for 1999, and 67-97~$\GeVm$ for 2000. The increase in range with rising centre-of-mass energy is to account for the increasing ISR tail. The likelihood function is defined as follows:

\[
{\cal L}_e (\mw,\gw) = P_e \cdot S''(m^{fit},\sigma^{fit},\mw,\gw) + (1-P_e)\cdot B(m^{fit}), \]
 
\noindent where $P_e$ is the event purity, discussed above, $S''$ is the signal function that describes the reconstructed mass distribution
of the semi-leptonic $\W$ decays, and $B$ is used to describe  background processes. The reconstructed event mass $m^{fit}$ and its estimated
error $\sigma^{fit}$ are both obtained from the constrained fit.  The distribution of background events is extracted from simulation as a
function of $m^{fit}$.

The signal function $S''$ is defined in terms of $S$ and $S'$ as discussed below.
The function $S$ relies on the convolution of three components, using $x$ and $m$ as the dummy integration variables:
\begin{eqnarray} 
\label{eqn:sfunc}
\lefteqn{S(m^{fit},\sigma^{fit}|\mw,\gw) =} \\
\nonumber & \int_{0}^{\Ebeam} dm~ G [m^{fit}-m, \sigma^{fit}] \int_{0}^{1} dx~ PS(m(1-x)) ~ BW[m(1-x)|\mw,\gw] ~ R_{ISR}(x).
\end{eqnarray}

\noindent $BW$ is a relativistic Breit-Wigner distribution representing the $\W$ mass distribution,

\begin{equation}
BW(m|\mw,\gw) = \frac{1}{\pi} \frac{\gw}{\mw} \frac{m^2}{(m^2-\mw^2)^2 + \left( m^2 \frac{\gw}{\mw} \right)^2},
\label{eqn:bw}
\end{equation}

\noindent and $PS$ is a phase-space correction factor

\[ PS(m) = \sqrt{1-\frac{4m^2}{s}}. \]

The convolution with the Gaussian function $G$ describes the detector resolution.
The width of the Gaussian depends upon the reconstructed mass error obtained in the
constrained fit for that event.

The ISR spectrum is parameterised as

\[ R_{ISR}(x_{\gamma}) = \beta x_{\gamma}^{(\beta-1)}, \]

\noindent where $x_{\gamma}$ is the ratio of the photon energy to the centre-of-mass
energy and $\beta$ is calculated from the electromagnetic coupling
constant ($\alpha$), the centre-of-mass energy squared ($s$) and the electron mass ($m_{e}$):

\[ \beta = \frac{2\alpha}{\pi}[{\rm ln}(s/m_{e}^2)-1]. \]

Due to the constrained fit, a $\W$ produced at mass $m$ will be reconstructed to a
good approximation as $m/(1-x_{\gamma})$ in the presence of an undetected ISR photon,
giving a tail at high mass in the measured spectrum.
This tail is well described by the integration on the photon spectrum in equation \ref{eqn:sfunc}.

The event selection contains a significant fraction of $\tnqq$ events in
the electron and muon channel samples, and of $\enqq$ events in
the tau sample (see table~\ref{tab:evtsel}). In the tau channel the mass of the
event is determined from the jet system. The behaviour
of true $\tnqq$ and $\enqq$ events in this fit are found to be similar,
and $S''$ = $S$ in this channel.
However, in the electron and muon channel samples the behaviour
of the $\tnqq$ events is somewhat different to that of the $\enqq$, $\mnqq$
events. The $\tnqq$ events have a worse mass resolution and
introduce a small negative offset on the mass. The fraction of tau events, which have been
wrongly classified and are contained in the electron and
muon channel samples, has been parameterised in bins of the lepton energy and
the measured missing mass.  This fraction $P\tau_{e}$
is then taken into account in the likelihood function for the electron and
muon samples, by defining the signal function $S''$ as

 \[ S'' = (1-P\tau_{e}) \cdot S + P\tau_{e} \cdot S' ,\]

\noindent where $S'$ is analogous to $S$, but with the width of the Gaussian
resolution function increased according to a factor determined from simulation studies.
All remaining biases in the analysis due to using this approximate likelihood
description are corrected for in the calibration procedure as
described in section~\ref{sec:massextrac}.

\newpage
\subsection{Fully-Hadronic Decay Channel}

The $\WW~\ra~\qqqq$ events constitute $46\%$ of all $\WW$ decays. The event masses can be reconstructed from the observed set of jets. The kinematics of the jets can be significantly over-constrained in a kinematic fit, improving the event mass resolution, by imposing momentum conservation and the measured centre-of-mass energy. The influence of the many ambiguities in the event reconstruction, which dilute the statistical information, is minimised by optimally weighting the different hypotheses in the likelihood fit of $\mw$ or $\gw$.

The dominant systematic error is due to the possible influence of final state interference effects between particles from the two decaying $\W$s. Reconstructing the jet directions using only the particles from the core of the jet reduces the possible effects of these final state interference effects. This technique and the mass estimator based on all observed particles are both discussed in section~\ref{sec:cone}.

\subsubsection{Event Selection}
\label{sec:qqqqselect}

As in the semi-leptonic analysis, appropriate criteria were imposed on the functionality of the detector when selecting the data sample for analysis. The corresponding integrated luminosities, at each centre-of-mass energy, are given in table~\ref{tab:enlum}.

The event selection can be separated into three stages. First a pre-selection is performed to reduce the data sample to events with a high multiplicity and high visible energy. In the second stage events with a four or five jet topology are retained. The observables on which the selection is made are chosen to be, to a good approximation, independent of the centre-of-mass energy $\sqs$: the same selection criteria are used for all energies for the pre-selection and jet topology selection. The final stage of the event selection is to use the inter-jet angles and jet momenta to estimate the probability that this was a $\WW~\ra~\qqqq$ event.

\begin{figure}[ht]
\begin{tabular}{cc}
\epsfig{file=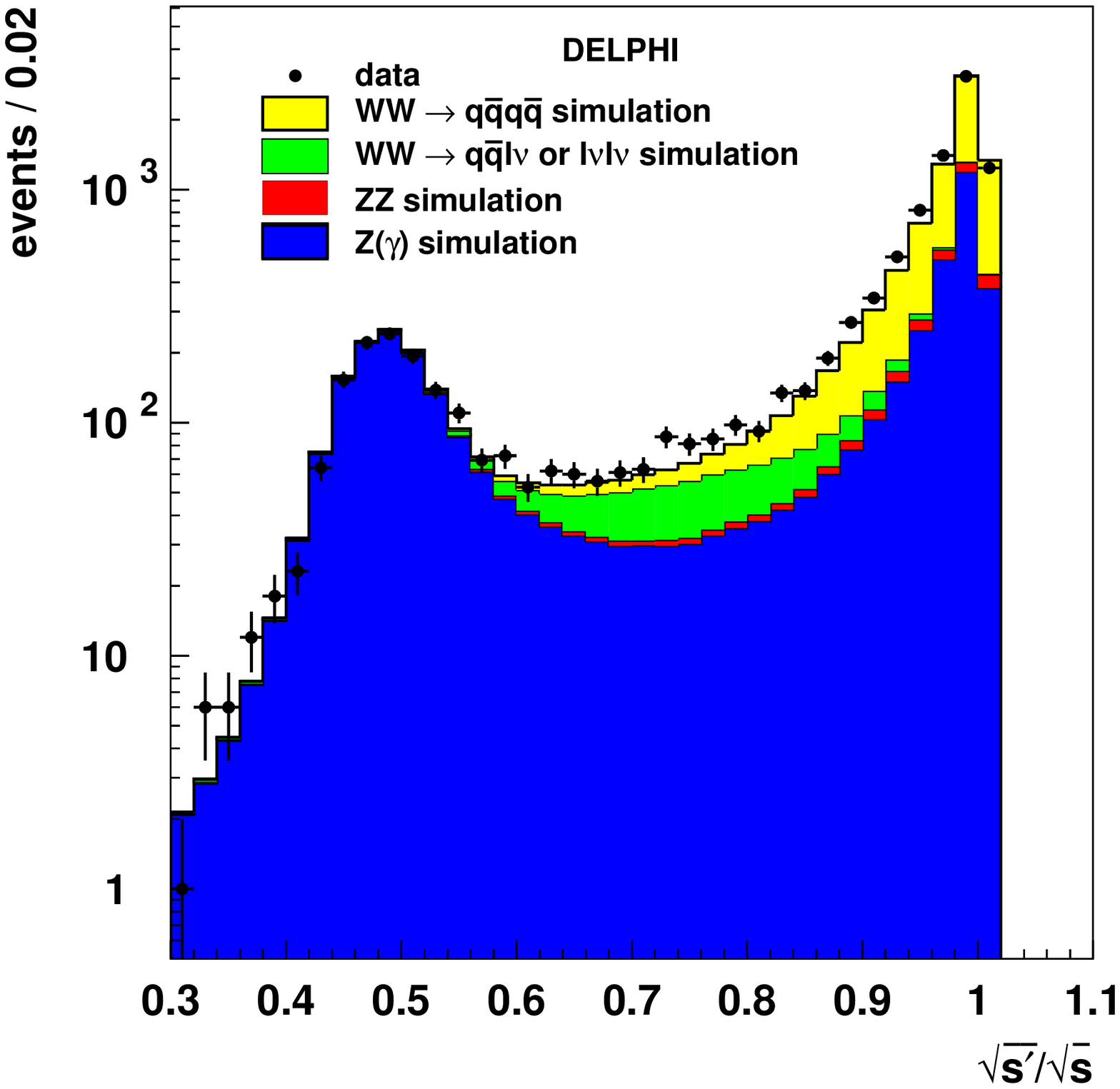,width=0.45\textwidth} &
\epsfig{file=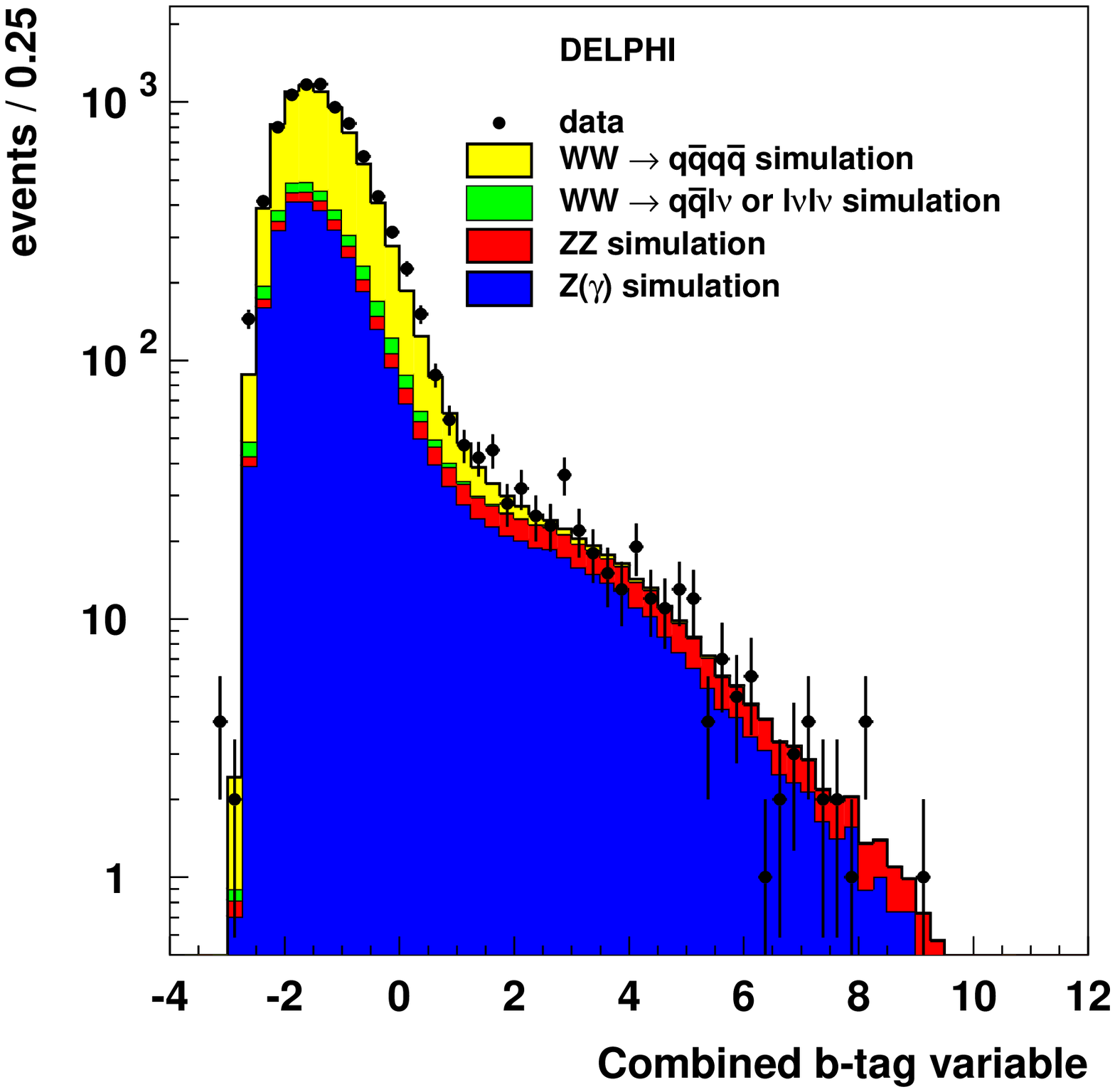,width=0.45\textwidth}
\end{tabular}
\caption{The distribution of two event selection variables for candidate $\qqqq$ events from the full $\LEP2$ data sample and the corresponding simulation samples. The left hand plot shows the scaled effective centre-of-mass energy, the right hand plot the combined b-tag variable. The distributions are shown after the cuts on all other pre-selection variables have been applied.}
\label{fig:sel}
\end{figure}

The pre-selection cuts applied are:
\begin{itemize}
\item the charged particle multiplicity should be larger than 13;
\item the total visible energy of the event must exceed 1.15 $\frac{\sqs}{2}$;
\item the scaled effective centre-of-mass energy  $\frac{\sprime}{\sqs}$ \cite{SPRIM} is required to be equal to or larger than 0.8;
\item rejection of events tagged as likely to be containing b quarks \cite{aabtag}.
\end{itemize}
The last criterion removes $7\%$ of the remaining $\Z \ra \qq (\gamma)$ and $18\%$ of the remaining $\ZZ$ events, while changing the signal selection efficiency by less than $1\%$. The distributions of data and simulation events for the scaled effective centre-of-mass energy and combined b-tag variable are shown in figure~\ref{fig:sel}; the cut on the combined b-tag variable retains all events below 2.

The remaining events are then clustered using the $\DURHAM$~\cite{DURHAM} jet clustering algorithm with a fixed $y_{cut}$ of 0.002. 
The jets obtained are required to have an invariant mass of greater than 1~$\GeVm$ and contain at least three particles. 
If the jets do not meet these criteria or more than five jets are obtained, the clustering is continued to higher values of  $y_{cut}$.
Events which cannot be clustered into either four or five jets that fulfill these criteria are rejected. 
The initial $y_{cut}$ value of this procedure was optimised for maximal sensitivity to $\mw$ and results in a sample of approximately 50\% four and 50\% five jet events. 

\begin{figure}[htp]
\begin{tabular}{cc}
\epsfig{file=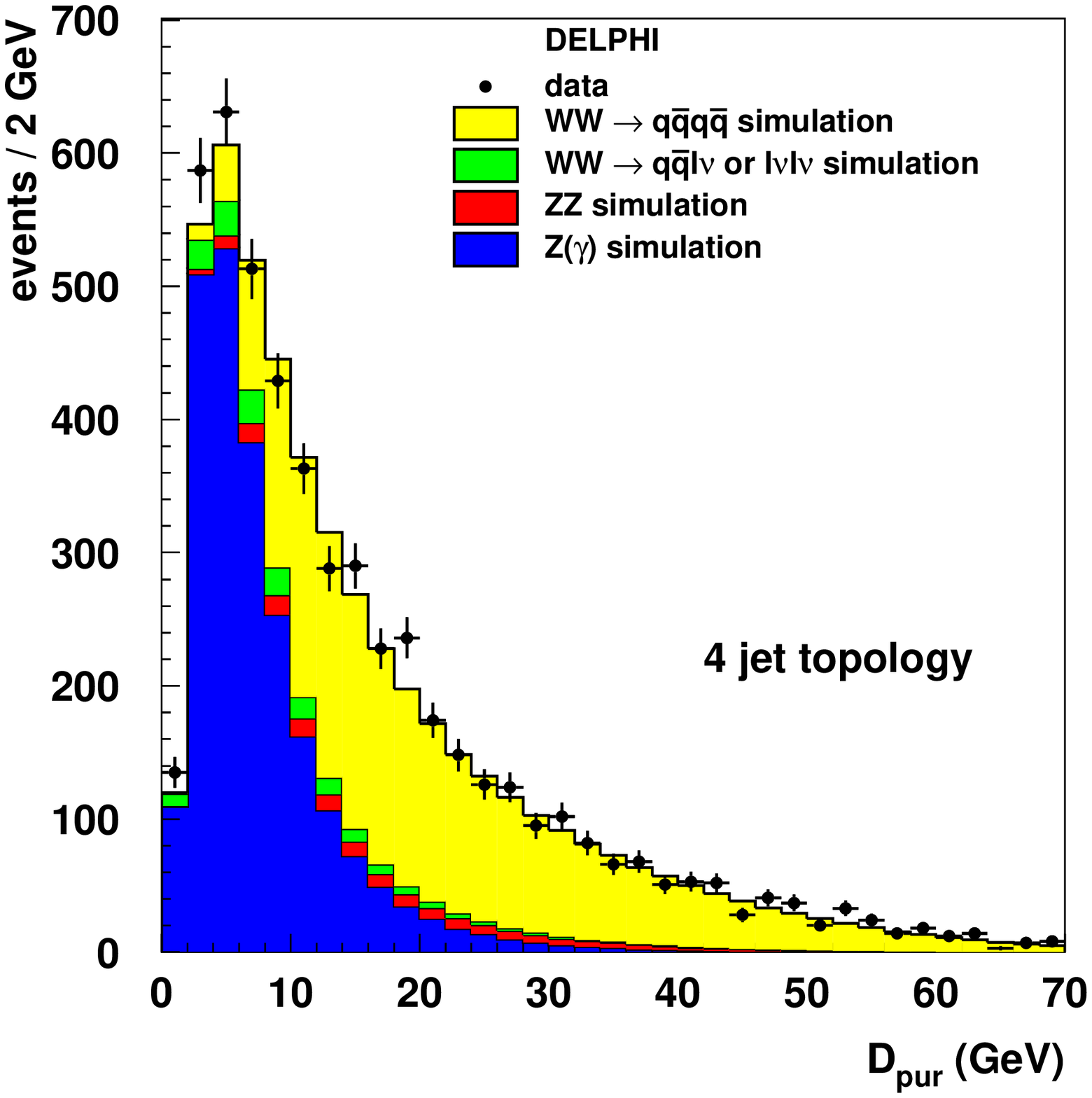,height=90mm,width=74mm} &
\epsfig{file=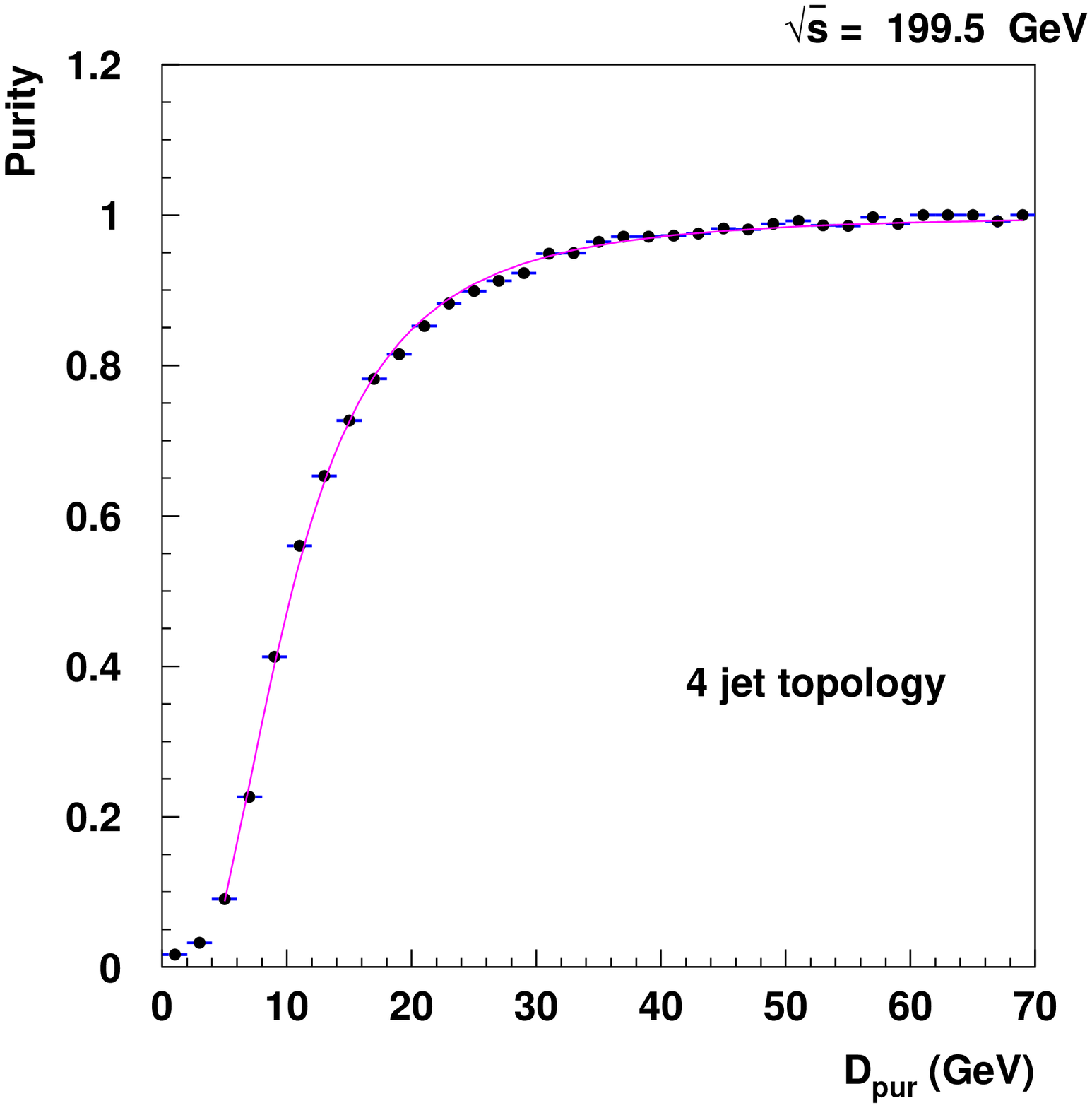,height=90mm,width=74mm}
\\
\epsfig{file=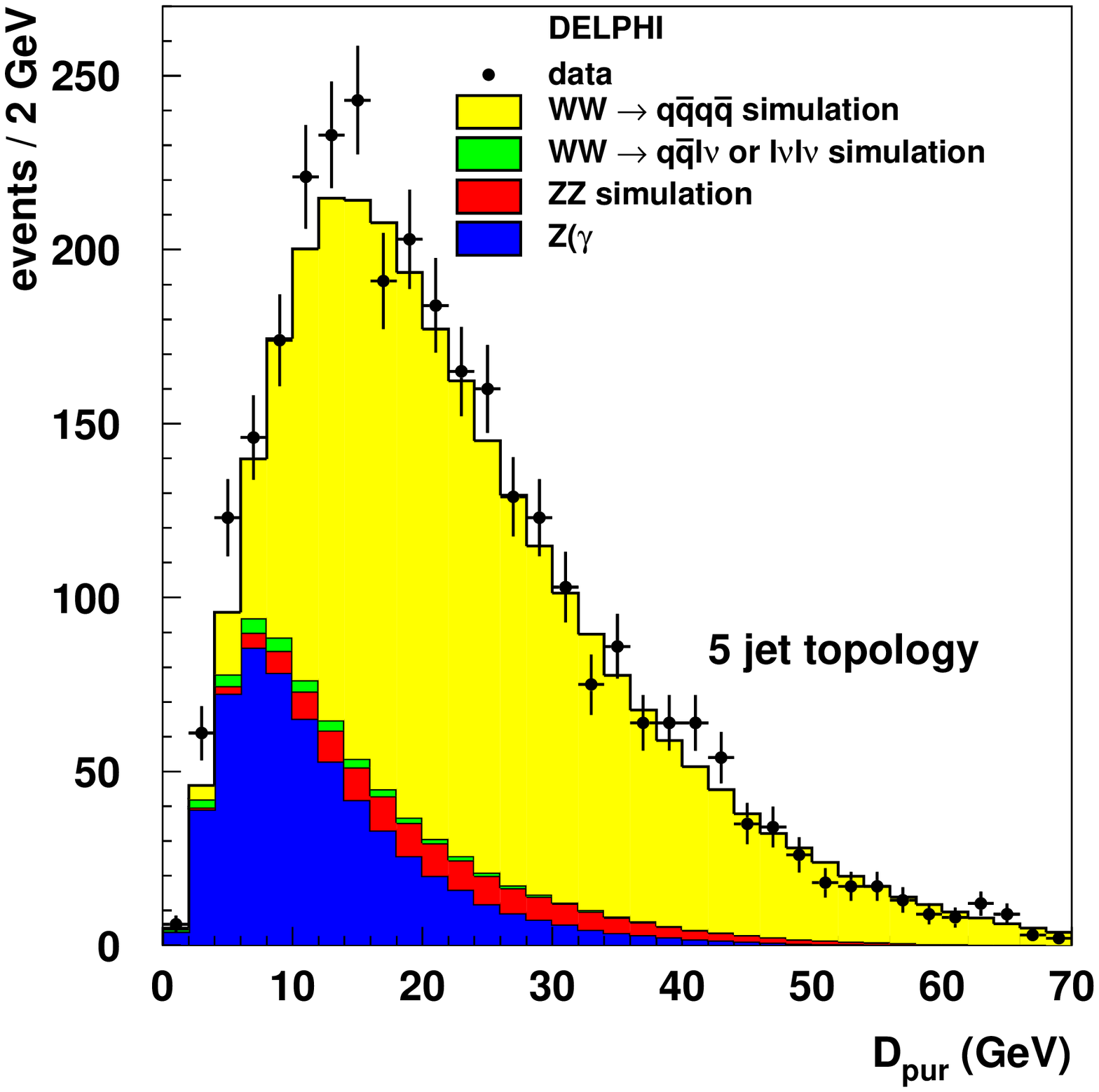,height=90mm,width=74mm} &
\epsfig{file=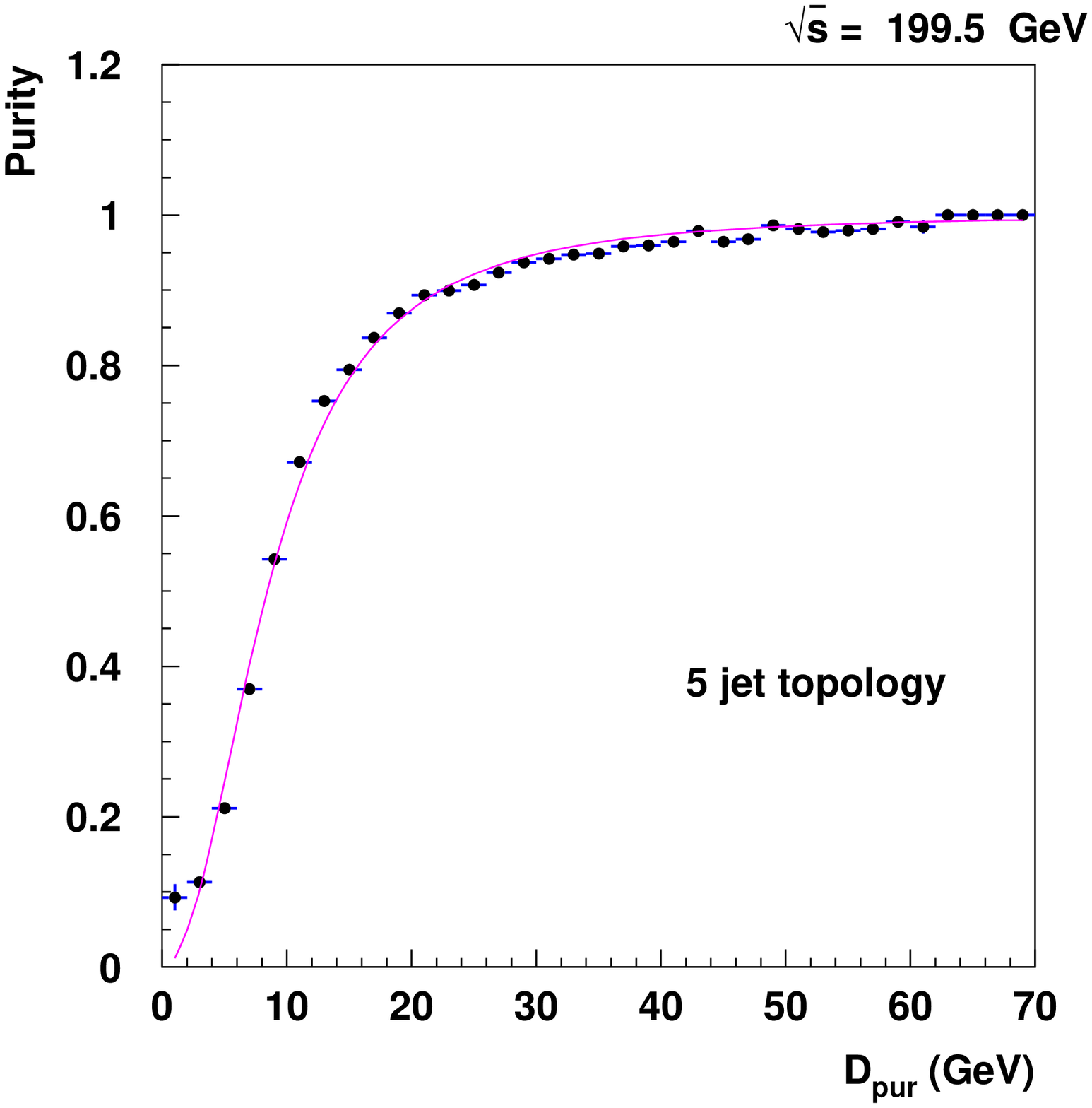,height=90mm,width=74mm}
\end{tabular}
\caption{The left hand plots show the distribution of the $D_{pur}$ variable for four jet (top) and five jet (bottom) events from the full $\LEP2$ data sample and the corresponding simulation samples. The right hand plots show the distribution of the four-fermion event purity with this variable at a centre-of-mass energy of $199.5~\GeV$ extracted from simulation events. The fitted parameterisation of this distribution is given by the line.}
\label{fig:p4f}
\end{figure}

The jets obtained from this procedure are then used in a constrained fit, described in section~\ref{sec:kfit}, where momentum conservation and the measured centre-of-mass energy are enforced.
From the fitted jets a topological observable, $D_{pur}$, was formed to discriminate between signal events and $\Z \ra \qq$ events with hard gluon radiation:

\[ D_{pur} = \theta^{fit} \cdot E^{fit} \cdot
              \sqrt{{\tilde \theta}^{fit} \cdot {\tilde E}^{fit}}
\]

\noindent where $E^{fit}_{j}$ and ${\tilde E}^{fit}_{j}$ are the smallest and second smallest fitted jet energies and $\theta^{fit}_{ij}$ and  ${\tilde \theta}^{fit}_{ij}$ are the smallest and second smallest fitted inter-jet angles.
The expected fraction of $\qqqq$ events ($\WW$ or $\ZZ$) in the selected sample, the event purity $P^{4f}$, is parameterised as a function of this variable. This fraction of $\qqqq$ events, \ie\ doubly-resonant events rather than just $\WW$ events, is used in the theoretical distribution function described below. 
Events with an estimated purity below $25\%$ are rejected.
The distribution of the $D_{pur}$ observable is shown in figure~\ref{fig:p4f} for both the 4 and 5 jet topology events, and the numbers of selected events are given in table~\ref{tab:evtsel}. An excess of data events over the expected number of simulation events was observed.

\subsubsection{Cone Jet Reconstruction}
\label{sec:cone}

The largest contribution to the systematic uncertainty in the fully-hadronic decay channel arises from the hypothesis, used throughout the likelihood construction, that the fragmentation of the partons from both $\W$ bosons happens independently. 
However, Bose-Einstein Correlations (section~\ref{sec:bec}) and colour reconnection (section~\ref{sec:cr}) effects may result in cross-talk between the two $\W$ systems. 
A jet reconstruction technique is presented here which has been designed to have reduced sensitivity to colour reconnection effects. 

Conventionally, as used for the jets in the semi-leptonic analysis, the particles in the event are clustered into jets using a jet clustering algorithm and the energy, magnitude of the momentum and direction of the jet are reconstructed from the clustered particles. 
The jet momentum and energy are then used as the input to the kinematic fit. This technique is referred to in this paper as the standard reconstruction method and provides the optimal statistical sensitivity.

\begin{figure}[tp]
\begin{center}
\vspace{1.cm}
  \includegraphics[height=8.5cm]{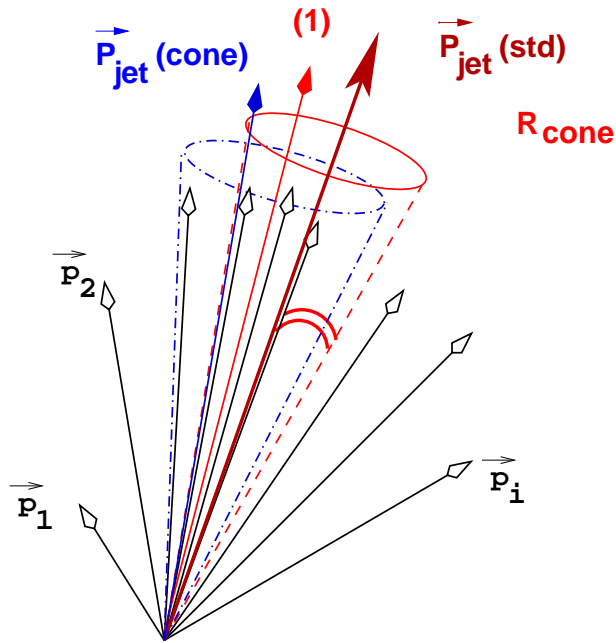}
  \caption{Illustration of the iterative cone jet reconstruction algorithm used for the fully-hadronic $\W$ mass analysis as discussed in the text.}
\label{fig:conemw}
\end{center}
\end{figure}

In the alternative reconstruction algorithm discussed here the effect of particles in the inter-jet regions on the reconstructed jet direction is reduced. This is achieved by using a cone algorithm. The initial jet direction  ${\vec p}^{\, jet}$ is defined by the standard clustering algorithms ($\DURHAM$~\cite{DURHAM}, $\CAMBRIDGE$~\cite{CAMJET} or $\DICLUS$~\cite{DICLUS}) and a cone of opening angle $R_{cone}$ defined around this as in figure~\ref{fig:conemw}.  The jet direction is recalculated (direction (1) on the figure) using those particles which lie inside the cone. This process is iterated
by constructing a cone (of the same opening angle $R_{cone}$) around this new jet direction and the jet direction is recalculated again. The iteration is continued until a stable jet direction ${\vec p}^{\, jet}_{cone}$ is found. Only the jet direction is changed in this procedure, the magnitude of the momentum and the jet energy are rescaled to compensate for the lost energy of particles outside the stable cone. The value of the cone opening angle $R_{cone}$ is set to $0.5$~rad, a value optimised for the measurement of the colour reconnection effect as reported in \cite{delphicr}.

This cone jet reconstruction technique reduces the sensitivity to the colour reconnection effect (see section~\ref{sec:cr}) at the expense of some statistical sensitivity. The expected statistical uncertainty increases by approximately $14\%$. This technique has been applied only to the $\W$ mass and not to the $\W$ width analysis.

This technique of jet reconstruction should not be confused with the alternative jet clustering algorithms ($\DURHAM$, $\CAMBRIDGE$ or $\DICLUS$) used in the analysis (see below). The alternative jet clustering algorithms are used as the starting point for the cone jet reconstruction in order to improve the statistical sensitivity of the analysis rather than to reduce the sensitivity to colour reconnection effects.

\subsubsection{Likelihood Function}
\label{sec:fullyhadfit}

{\bf Event Ideograms}
\vspace{0.2cm}

Each of the selected events is analysed through the use of a likelihood ratio function, which we will label here as the event ideogram. The final ideogram for each event consists of the weighted sum of the ideograms produced using a range of event reconstruction hypotheses $h_i$. These reconstruction hypotheses, including for example the possible different associations of the jets to their parent $\W$ bosons, are discussed below. The details of how these hypotheses are combined is then described below under the heading of `Ideogram Sum'. 
  
The ideogram reflects the relative compatibility of the kinematics of the event with the premise that two heavy objects, with masses $m_x$ and $m_y$, were produced. The ideogram is based on the least-square, $\chisqfc$, of the energy and momentum constrained fit of the observed set of jet kinematics, $\{\pkin\}$, of the reconstructed final state. 

Thus, for each pair of test masses ${\vec m}$ = $(m_x,m_y)$, we can obtain the $\chi^2_{4C}(\{{\bar p_j}\}|{\vec m},h_i)$. As the calculation of the $\chisq$ over the full mass ${\vec m}$ plane is computationally intensive we apply the following approximation in the analysis. 
The $\chi^2$ is only calculated once per hypothesis $h_i$ at the minimum of the $\chi^2_{4C}({\vec m})$ in the full ${\vec m}$-space. 
The probability in all other points ${\vec m} = (m_x,m_y)$
is calculated using a Gaussian approximation for the $\chi^2({\vec m})$ given by:

\[
\chi^2_i(m_x,m_y) \simeq \chi^2_{4C} + ({\rm \bf m} - {\rm \bf m}^{\rm fit})^T {\rm \bf V}^{-1} ({\rm \bf m} - {\rm \bf m}^{\rm fit}),
\]

\noindent
with
\begin{eqnarray*} 
{\bf m}   &=& \left( \begin{array}{c}
                     m_x\\
                     m_y
                     \end{array}\right), \nonumber \\
                                    & & \nonumber \\
{\bf m^{\rm fit}} &=& \left( \begin{array}{c}
                     m_x^{\rm fit}\\
                     m_y^{\rm fit}
                     \end{array}\right) \nonumber. \\
\end{eqnarray*}

The masses $m_x^{fit}$, $m_y^{fit}$, and the covariance matrix $\bf {V}$ are taken from the 4C kinematic fit. When the $\chi^2_{4C}$ is larger than the number
of degrees of freedom (NDF=4), the $\chi^2_i(m_x,m_y)$ is rescaled with a factor NDF/$\chi^2_{4C}$ in order
to compensate for non-Gaussian resolution effects.

This procedure decreases the computing time taken by an order of magnitude compared with the full six constraint fit \cite{delpaper183}, while resulting in only a minimal reduction in the $\W$ mass precision obtained ($2\pm1\%$).

We denote the ideogram of the event under hypothesis $h_i$ as  $P(\{{\bar p_j}\}|{\vec m},h_i)$. Assuming a Gaussian form, this is calculated from the $\chisq$  as follows:

\[
P(\{{\bar p_j}\}|{\vec m},h_i) \ d{\vec m}  \ = \ {\rm exp}\left( -\frac{1}{2} \cdot  \chi^2_{4C}(\{{\bar p_j}\}|{\vec m},h_i) \right) d{\vec m}.
\]

\begin{figure}[ht]
\begin{tabular}{cc}
\epsfig{file=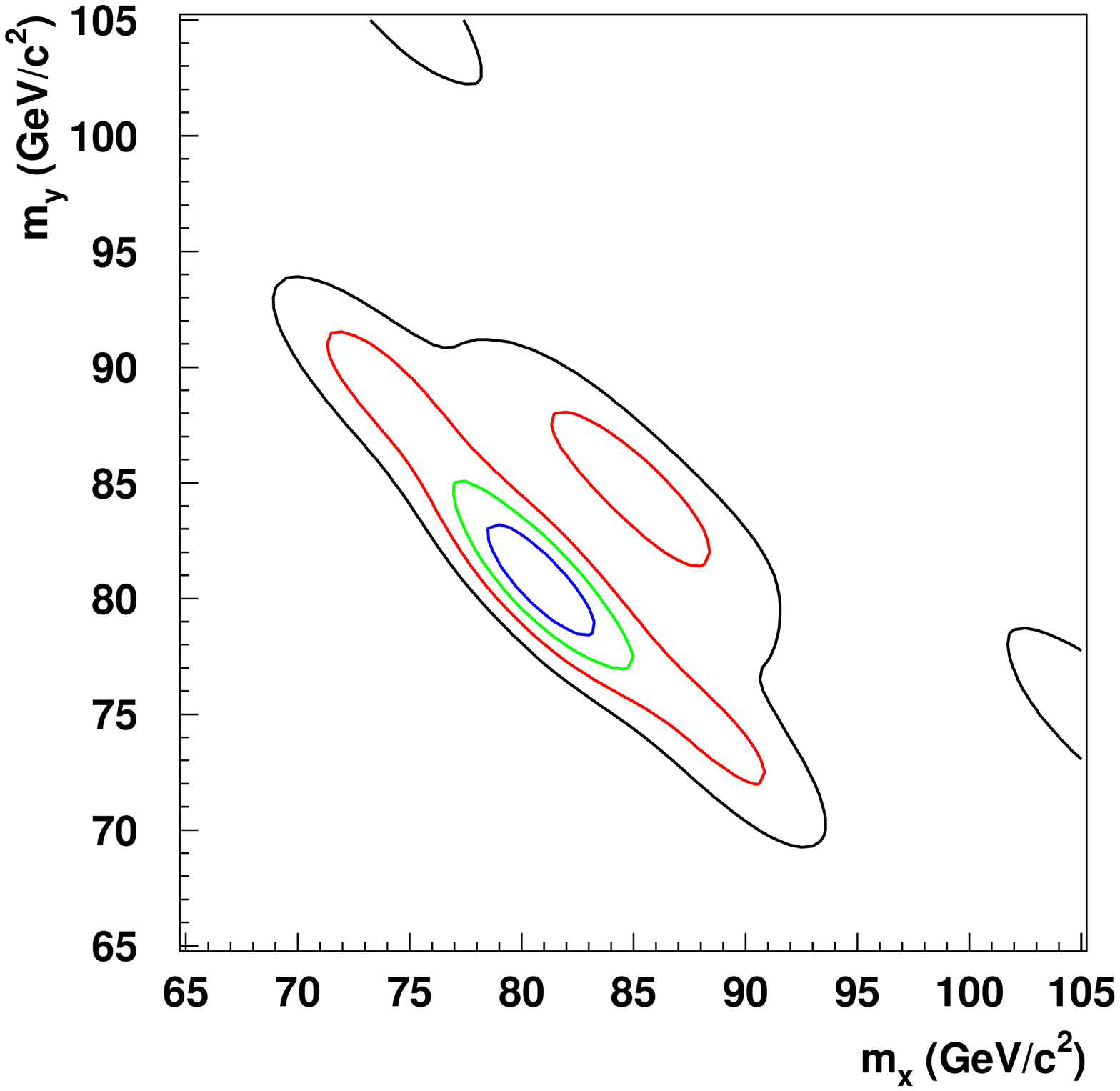,width=0.45\textwidth} &
\epsfig{file=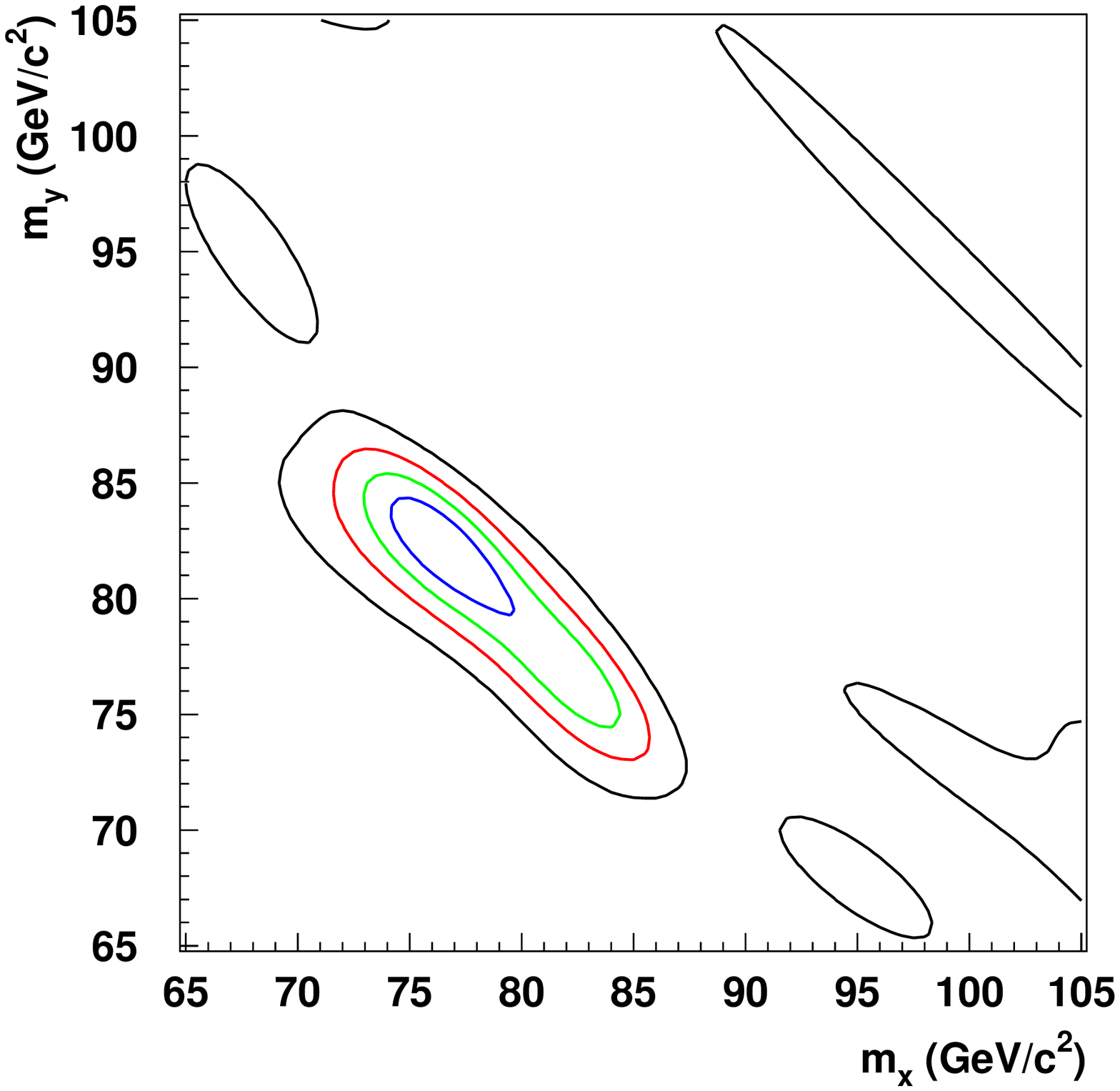,width=0.45\textwidth}
\end{tabular}
\caption{Examples of a reconstruction hypothesis weighted sum of two-dimensional probability ideograms (see text) for a four-jet (left) and five-jet (right) hadronic event. The ideograms include terms from each potential jet-pairing, three jet clustering algorithms and possible ISR emission.  The 1,2,3 and 4-sigma contours are shown.}
\label{fig:ideogram}
\end{figure}

Example ideograms are shown in figure~\ref{fig:ideogram}. These ideograms show the weighted sum of the reconstruction hypothesis ideogram terms for an individual event. 
The reconstruction hypotheses, which we will discuss in the following sections, include a range of options for the jet clustering algorithms that assign particles to jets, the possible associations of jets to $\W$ bosons, and a treatment for events that may have significant initial state radiation.

\newpage
{\bf Jet Pairings} 
\vspace{0.2cm}

As discussed in section~\ref{sec:qqqqselect}, the reconstructed particles in the event were clustered into four or five jets. 
These jets can then be associated to their parent $\W$ bosons. 
For events clustered into four (five) jets there are three (ten) combinatorial possibilities for the jet pairing into $\W$ bosons. The relative probability of each of these jet pairings to be the correct jet association is estimated.

This jet to $\W$ boson association weight, $w_k$, is estimated as a function of the reconstructed polar angle of the $\W$ boson and the estimated charge difference between the two reconstructed $\W$ bosons in the event. 
For the five jet events the transverse momentum of the gluon jet is also used.

The production angle $\theta_{W}$ of the $\W^+$ ($\W^-$) boson is correlated with the flight direction of the incoming $e^+$ ($e^-$) beam. For each jet pairing the $\W$ boson polar angle was calculated and its probability $P_{\theta}(\theta_W)$ assessed from a centre-of-mass dependent parameterisation of correctly paired simulation events.

The jet charge $Q_{jet}^i$ for jet $i$ in the clustered event can be measured as:

\[
Q_{jet}^i = \frac{\sum_{n=1}^{n_{jet}} |{\vec p_n}|^{0.5} \cdot q_n}{\sum_{n=1}^{n_{jet}} |{\vec p_n}|^{0.5}}
\]
where $n_{jet}$ are all charged particles in jet $i$, while $q_n$ and ${\vec p_n}$ are their charge and momentum.
For each association $k$ of the jets to their parent $\W$ bosons the charge difference $\Delta Q_k = Q_k^{\W_1} - Q_k^{\W_2}$ is obtained. Again, the probability of this being the correct jet assignment is assessed using a Monte Carlo simulation-derived parameterisation.
The relative weight for each jet pairing $k$ can be expressed as:
\[
w^{W}_{k} = P_{W^+}(\Delta~Q_{k}) \cdot P_{\theta}(\theta^{k}_{W_1}) + ( 1 - P_{W^+}(\Delta~Q_{k}) ) \cdot P_{\theta}(\pi - \theta^{k}_{W_1}).
\]

In five jet events, a two jet and a three jet system are considered. The three jet system is considered as comprising a $\qq$ pair and a gluon jet. The probability of emission of a gluon from a $\qq$ pair is approximately inversely proportional to the transverse momentum  of the gluon with respect to the original quarks. Hence, the most probable gluon jet in the three jet system is the jet with the smallest transverse momentum ($k_T$) with respect to the two other jets in the candidate $\W$ boson rest frame. Each of the ten possible jet associations, in this five jet event, is then given a relative weight from its most probable gluon jet of $w_{k}^{gluon} = 1/k_T$. 

The combined relative jet pairing weight of each combination is given by multiplying the jet pairing weights $w^{W}_{k}$ and, for five jet events, also multiplying by the $w_{k}^{gluon}$ weight.  The relative weights are then normalised so that the sum of the weights for all the jet-paring combinations of the event is 1, giving combination weights $w_k$. The use of all the jet pairings, rather than simply picking the best one, improves the statistical precision of this analysis by 4\%.

\vspace{0.2cm}
{\bf Jet Clustering Algorithms}
\vspace{0.2cm}

Several standard jet clustering algorithms are used in this analysis. Whilst the overall performances of the algorithms are similar, the reconstruction of an individual event can differ significantly. In this analysis, the event ideograms were reconstructed with three clustering algorithms $\DURHAM$, $\CAMBRIDGE$ and $\DICLUS$. The ideograms resulting from each clustering algorithm are summed with fixed optimised relative weights, $w_c$, determined from simulation events. The sum of the three jet clustering weights for one event is normalised to 1.

The use of a range of jet clustering algorithms, rather than taking only one, improves the statistical precision of this analysis by 5\%.

\vspace{0.2cm}
{\bf Initial State Radiation Hypotheses}
\vspace{0.2cm}

A kinematic fit (see section~\ref{sec:kfit}) is performed with modified constraints and an extra free parameter $p_z^{\rm fit}$ to account for the possible emission of an ISR photon of momentum $p_z$ inside the beam pipe. The modified constraints are:

\[ \sum_{i=1}^{n_{\rm objects}} (E, p_x, p_y, p_z)_i =
(\sqs-|p^{\rm fit}_z|,0,0,p^{\rm fit}_z). \]

The probability that the missing momentum in the $z$ direction is indeed due to an unseen ISR photon was extracted from the simulation as a function of $ |p^{\rm fit}_z| / \sigma_{p_z}$, where $\sigma_{p_z}$  is the estimated error on the fitted $z$ momentum component; only events with this ratio greater than 1.5 are treated with the mechanism described below.

\begin{figure}[ht]
\begin{center}
\epsfig{file=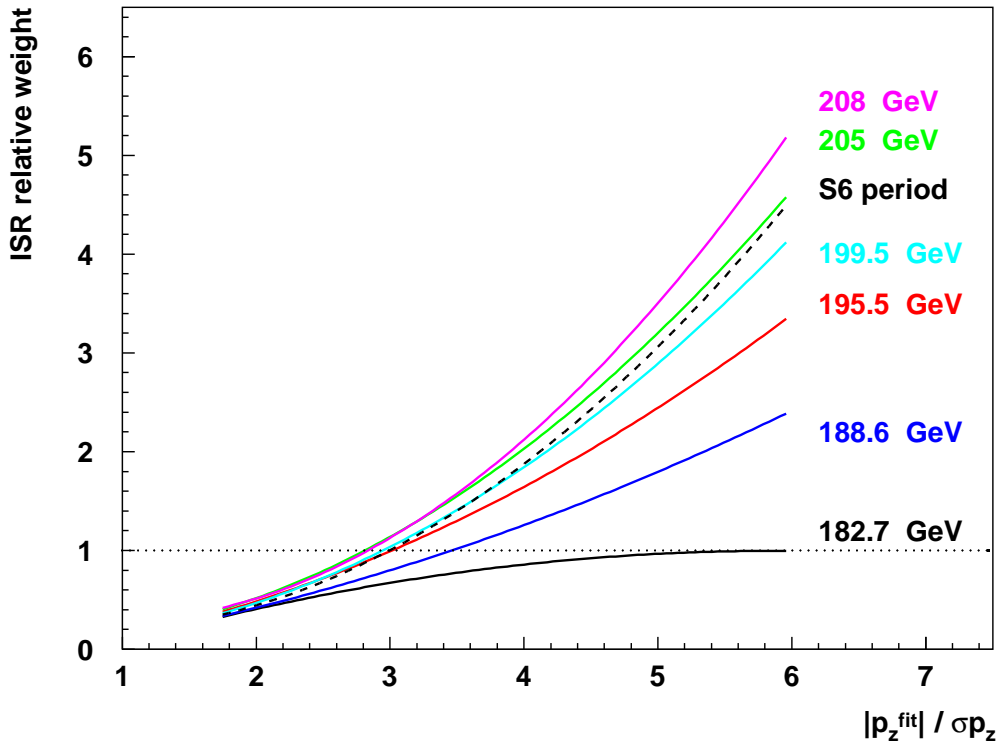,width=0.9\textwidth}
\end{center}
\caption{Parameterised weight given to the ISR solution of the kinematic fit, relative to the unity weight of the no ISR solution, as a function of the $ |p^{\rm fit}_z| / \sigma_{p_z}$ value of the event for different centre-of-mass energies. The period with a damaged TPC sector (S6) is indicated with a dashed line.}
\label{fig:isr}
\end{figure}

Additional ideograms are then calculated for these events, with a relative weight factor derived from the ISR hypothesis probability. The ideogram obtained without the ISR hypothesis is given a relative weight 1, while the other ideograms obtained from this procedure are given relative weight factors according to the distribution shown in figure~\ref{fig:isr}. The weights are then normalised such that the sum of the ISR and no ISR hypotheses for an event sum to 1, giving ISR weights $w_{isr}$.

This treatment is applied to $15\%$ of the events and results in an improvement of the expected $\W$ mass error for these events of $15\%$.  

\vspace{0.2cm}
{\bf Ideogram Sum}
\vspace{0.2cm}

An ideogram is produced for each event under each of the possible reconstruction hypotheses. For four jet events there are three jet association hypotheses to be performed with three clustering algorithms and maximally two ISR hypotheses, giving a total of eighteen ideograms. For five jet events there are sixty possible ideograms. The final ideogram for each event is produced as a weighted sum of these:

\[
P(\{{\bar p_j}\}|{\vec m},\{h_i\}) \ = \ \sum_{k=1}^{\rm 3 \ or \ 10} \ \sum_{isr=1}^{2} \ \sum_{c=1}^{3} w_k \cdot w_{isr} \cdot w_c \cdot P(\{{\bar p_j}\}|{\vec m},h_{k,isr,c}),
\]

\noindent where the sum over $k$ takes into account the three or ten possible jet pairings in the event, the sum over $isr$ the two different initial state radiation hypotheses used in the kinematic fit and the sum over $c$ the three jet clustering algorithms. The sum of all weights for each event is fixed to unity, so that while possible reconstruction hypotheses within an individual event have different weights the overall weight for each event is the same.

\vspace{0.2cm}
{\bf Likelihood}
\vspace{0.2cm}

To obtain information about $\mw$ and $\gw$ a theoretical probability distribution function, $P({\vec m}|\mw,\gw)$, is required predicting the population density in the ${\vec m}$-plane of the event ideogram. 
The ideogram in ${\vec m}$-space can then be transformed into a likelihood, ${\cal L}_e(\mw,\gw)$, in the ($\mw$,$\gw$)-space by convoluting it with this expected distribution $P({\vec m}|\mw,\gw)$:

\begin{equation}
{\cal L}_e(\mw,\gw) = \int_{m_{\rm min}}^{m_{\rm max}} \int_{m_{\rm min}}^{m_{\rm max}} P(\{{\bar p_j}\}|{\vec m},\{h_i\}) \cdot P({\vec m}|\mw,\gw) \ d{\vec m},
\label{eqn:likeqqqq}
\end{equation}

\noindent
where the two-dimensional integral is over the relevant kinematic region in the ${\vec m}$-space.
This region is taken to be $m_{\rm min}$ = 60~$\GeVm$ and $m_{\rm max}$ = 110~$\GeVm$, and
the combined ideogram is normalized to unity in the same region:

\[
 \int_{m_{\rm min}}^{m_{\rm max}} \int_{m_{\rm min}}^{m_{\rm max}} P(\{{\bar p_j}\}|{\vec m},\{h_i\}) \ d{\vec m} \ = \ 1.
\]

\vspace{0.2cm}
{\bf Theoretical Distribution Function}
\vspace{0.2cm}

The theoretical probability distribution function, $P({\vec m}|\mw,\gw)$, predicts the population density in the ${\vec m}$-plane of the event ideogram for a given $\mw$ and $\gw$. To provide an accurate description of the data the form assumed for $P({\vec m}|\mw,\gw)$ must take into account not only the expected distribution for the $\WWqqqq$ signal events but also that of the background events in the selected sample. The two principal components of the background, $\Zqqgam$ and  $\ZZqqqq$, are considered. 

The background process $\Zqqgam$ does not have a doubly resonant structure and a uniform population
of these events is expected in the ${\vec m}$-space independent of the values of the parameters ($\mw$,$\gw$). Therefore, the probability density function from this background source is assumed to be a constant denoted $B$. The probability ($P^{4f}$) that a given event is a  $\qqqq$ event was calculated from the event topology as described in section~\ref{sec:qqqqselect}.

The $\WWqqqq$ and $\ZZqqqq$ events both have a doubly resonant Breit-Wigner structure in the $\vec m$-plane,  modulated by a phase-space correction
factor $PS({\vec m}|\sqs)$ due to the nearby kinematic limit $m_{W^+} + m_{W^-} \leq \sqs$. The probability density function component used to model four-fermion events is given by:
\begin{eqnarray*}
\lefteqn{ S({\vec m}|\mw,\gw) = PS({\vec m}|\sqs) \cdot }\nonumber \\
 & \cdot \left[ \frac{{\tilde{\sigma}}_s^{\rm WW}}{{\tilde{\sigma}}_s^{\rm WW} + {\tilde{\sigma}}_s^{\rm ZZ}} \cdot BW_{\rm WW}({\vec m}|\mw,\gw) + \frac{{\tilde{\sigma}}_s^{\rm ZZ}}{{\tilde{\sigma}}_s^{\rm WW} + {\tilde{\sigma}}_s^{\rm ZZ}} \cdot BW_{\rm ZZ}({\vec m}|\mz,\gz) \right],
\end{eqnarray*}

\noindent
where ${\tilde{\sigma}}_s^{\rm WW}$ and ${\tilde{\sigma}}_s^{\rm ZZ}$ reflect the accepted cross-sections, calculated from simulation, of respectively the $\WW$ and the $\ZZ$ final states. These cross-sections are centre-of-mass energy dependent but are independent of the reconstructed event topology. 

The two-dimensional Breit-Wigner distribution is approximated as the product of two one-dimensional Breit-Wigners:

\[
BW_{\rm WW}({\vec m}|\mw,\gw) = BW_{\rm W}(m_{W^+}|\mw,\gw) \cdot BW_{\rm W}(m_{W^-}|\mw,\gw),
\]

\noindent
with $BW_{\rm W}$ given by the expression in equation~\ref{eqn:bw} of section~\ref{sec:semileptfit}.
An expression of the same form is assumed for the $\ZZ$ component.

A dependence on the centre-of-mass energy is also introduced into $S({\vec m}|\mw,\gw)$ through the phase space correction factor $PS({\vec m}|\sqs)$:

\[
PS({\vec m}|\sqs) = \frac{1}{s} \sqrt{(s-m_{W^+}^2-m_{W^-}^2)^2 - 4 m_{W^+}^2 m_{W^-}^2}.
\]

The combined density function is then constructed from the signal and background terms:
\[
P({\vec m}|\mw,\gw,\sqs) = P^{4f} \cdot S({\vec m}|\mw,\gw,\sqs) + (1-P^{4f}) \cdot B.
\]

Utilising this probability density function, and the event ideogram, equation~\ref{eqn:likeqqqq} may be used to calculate the event likelihood function. The extraction of the parameters of interest, $\mw$ and $\gw$, from the event likelihood functions are discussed below.

\subsection{Mass and Width Extraction}
\label{sec:massextrac}

The mass and width of the $\W$ boson are extracted from maximum likelihood fits to data samples. This section describes this procedure, the calibration applied and the cross-checks of this method that have been performed. 

\begin{figure}[htp]
 \begin{tabular}{cc}
  \epsfig{file=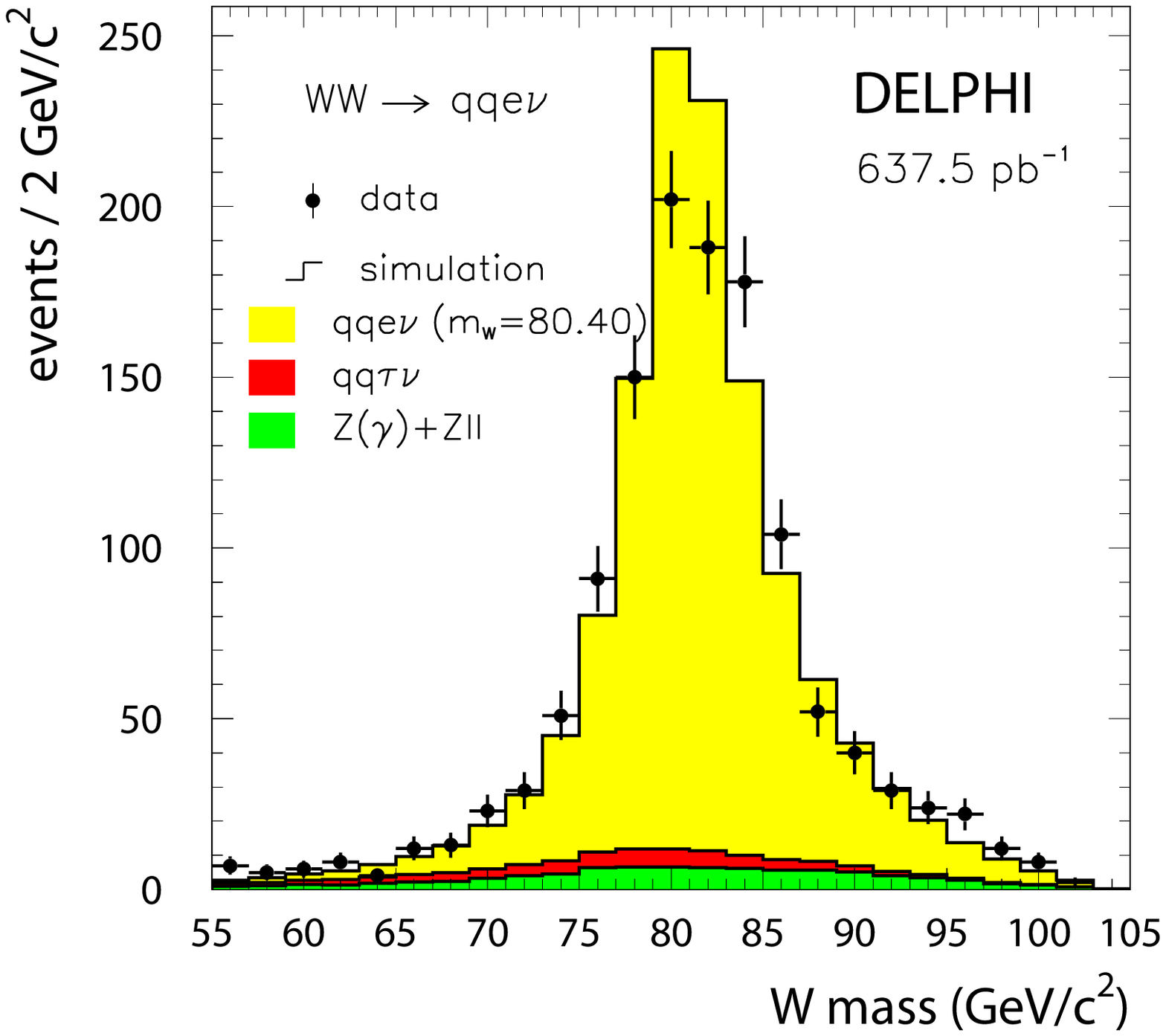,width=0.48\textwidth} & 
  \epsfig{file=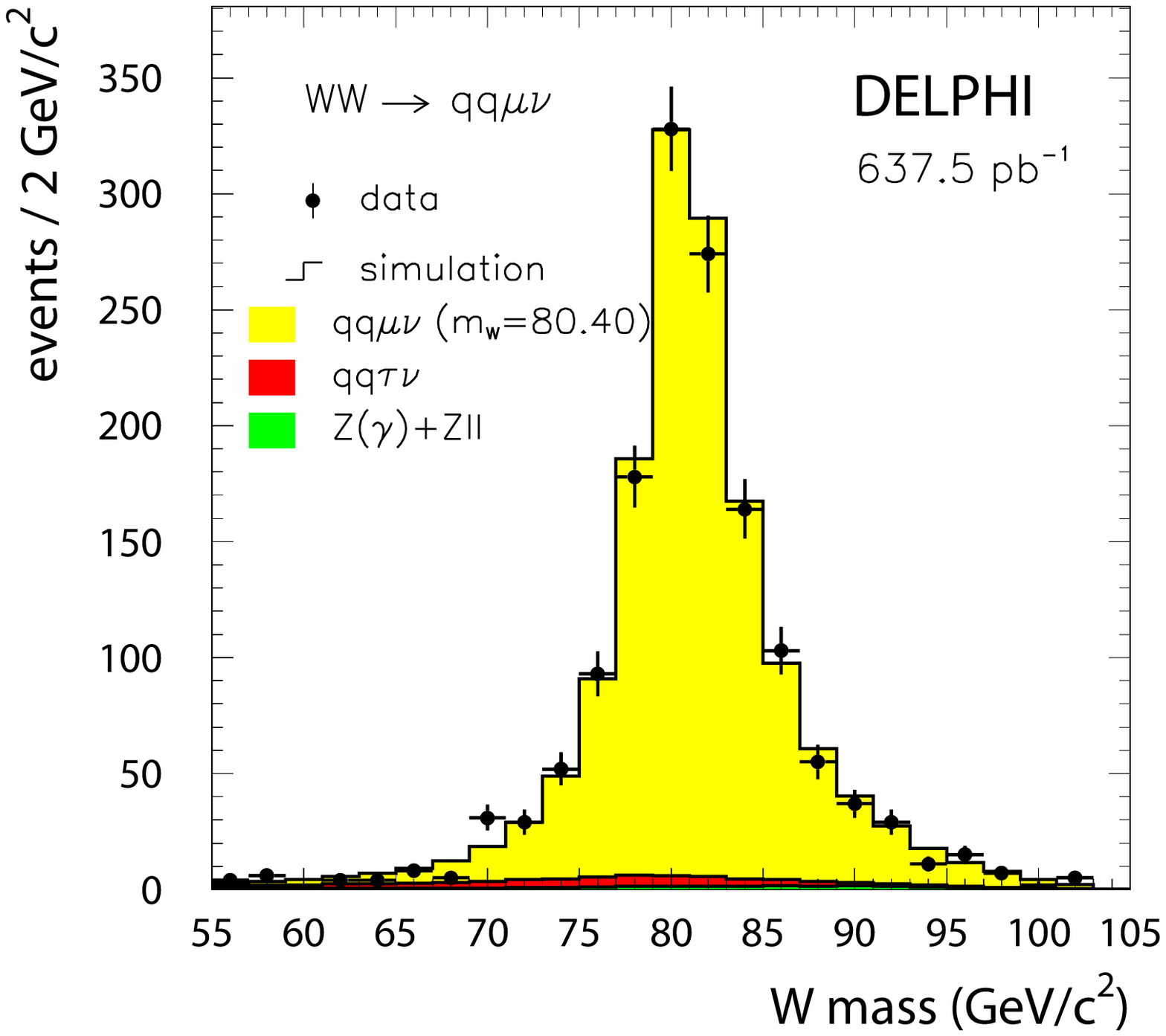,width=0.48\textwidth} 
\\
(a) & (b) \\
\multicolumn{2}{c}{\epsfig{file=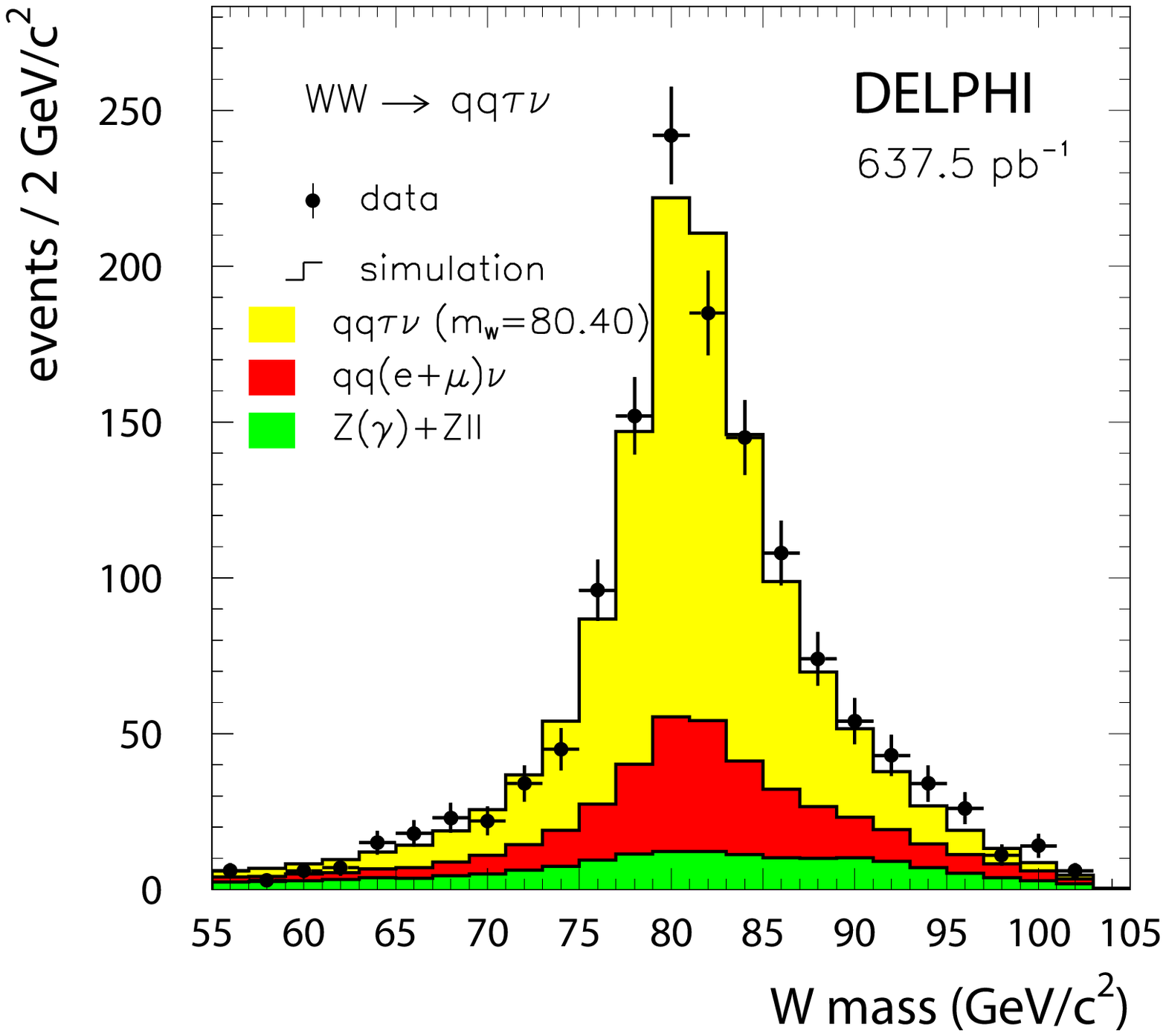,width=0.48\textwidth}} \\
\multicolumn{2}{c}{(c)} \\
  \epsfig{file=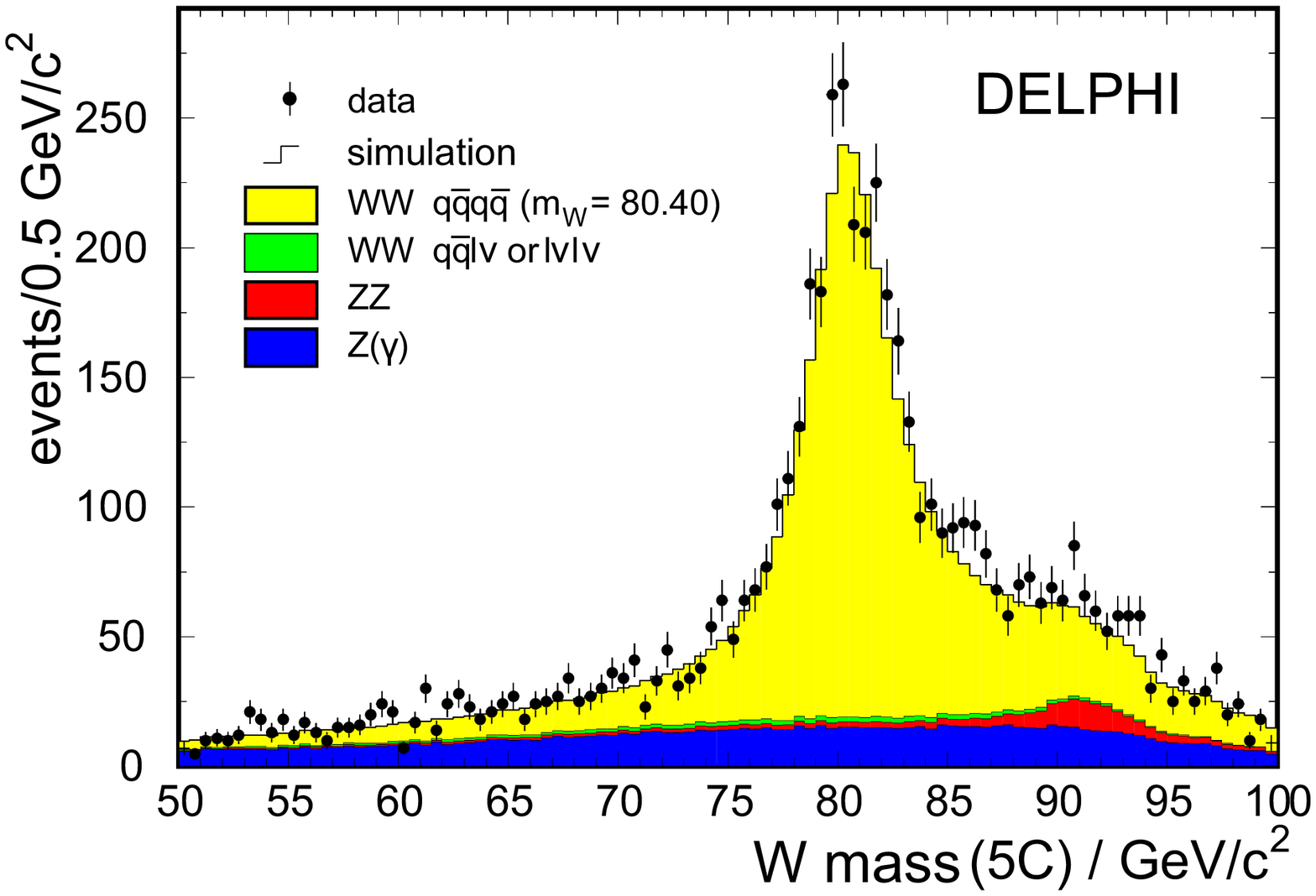,width=0.48\textwidth} &
  \epsfig{file=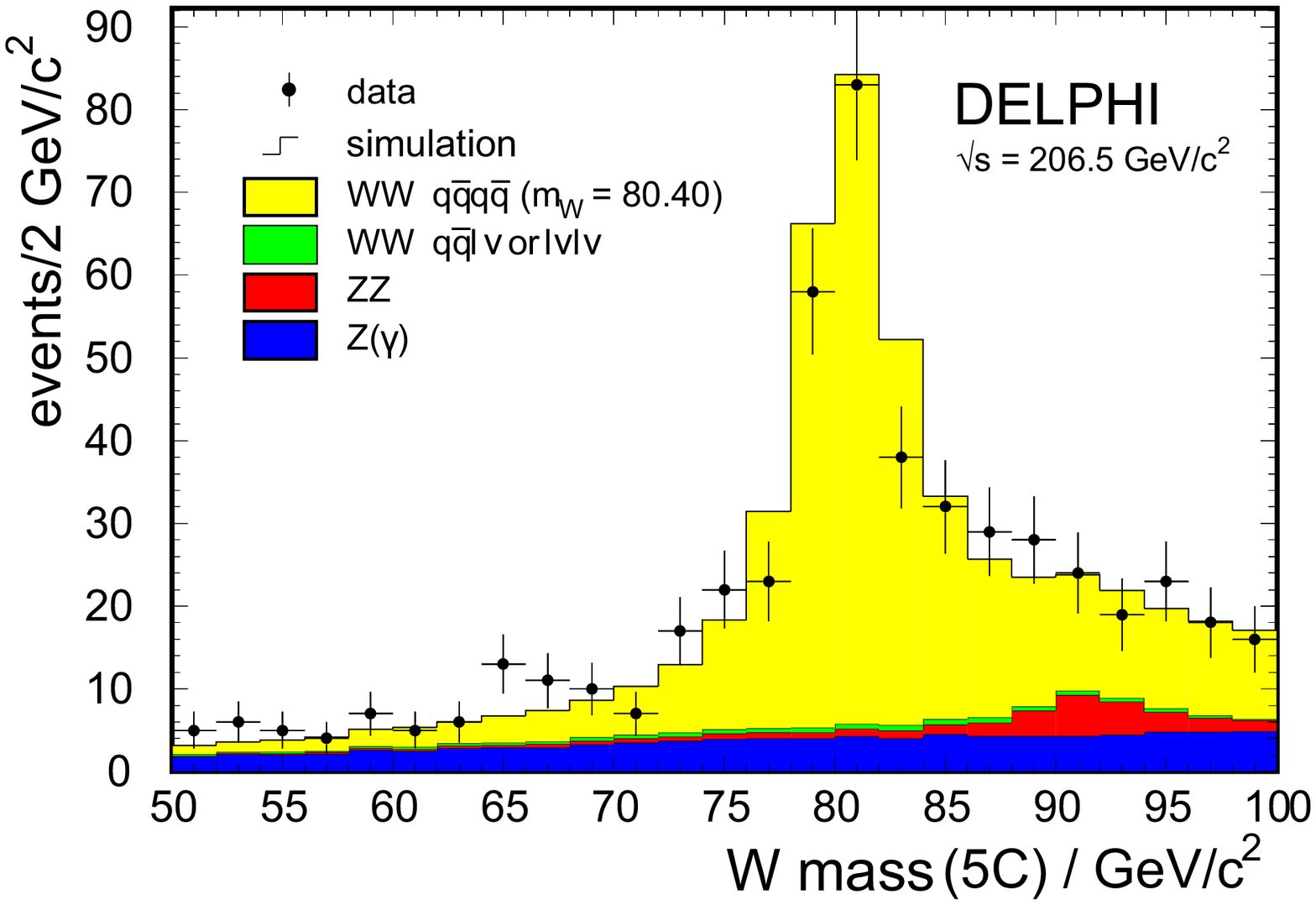,width=0.48\textwidth}
\\
(d) & (e) \\
 \end{tabular}
  \caption{The distribution of the reconstructed $\W$ masses from a kinematic fit with five constraints imposed in the (a) $\enqq$, (b) $\mnqq$, (c) $\tnqq$ and (d) and (e) $\qqqq$ analysis channels at all energies. (d) shows the data sample taken at all energies until September 2000, the data taken after that with a damaged TPC sector is shown in (e). In (d) and (e) only the jet pairing with the highest probability is included in the figures. The simulation samples have been normalised to the same integrated luminosity as the data.} 
  \label{fig:mass}
 \end{figure}

The distribution of the reconstructed invariant masses of the selected events after applying a kinematic fit, imposing four-momentum conservation and the equality of the two di-jet masses, are shown in figure~\ref{fig:mass}. This figure is provided for illustrative purposes only, the mass and width fitting procedure is described below.

The combined likelihood of the data can be obtained from the product of the event likelihoods described above. In practice this is achieved by performing the sum of the logarithms of the individual event likelihoods. The fitted data samples are divided by data taking year and applied event selection. For the mass fit the data from the fully-hadronic event selection and the electron, muon and tau semi-leptonic selections are all fitted separately. In the determination of the $\W$ width, where the relative precision is much worse, the data are divided only into fully-hadronic and semi-leptonic selection samples. The procedure for combining the results from  each of these fits is discussed in section~\ref{sec:comb}.

The $\W$ mass and width are extracted from maximum likelihood fits. The $\W$ mass fit is performed assuming the Standard Model value for the $\W$ width (2.11~$\GeVm$).  The $\W$ width was obtained assuming a mass of $80.4~\GeVm$. The correlation between $\mw$ and $\gw$ was found to have a negligible impact on the extracted mass and width value: the current uncertainty of $44~\MeVm$ on $\gw$  \cite{pdg} gives rise to a $0.6~\MeVm$ uncertainty in the extracted $\mw$.   

The terms used in the likelihood and described above are functions which approximate a description of the underlying physics and detector response. Hence, this approach necessitates a calibration of the analysis procedure. The calibration is performed using signal and background simulation events for which the true mass and width values are known. Rather than regenerating the events at a range of mass and width values, the calibration of the analysis uses reweighted events. The reweighting was performed using the extracted matrix element of the  \WPHACT\ and \YFSWW\ generators. The reweighting procedure is cross-checked using independent simulation events generated at three $\W$ mass and width values. In the fully-hadronic channel where both the standard method and the cone-jet reconstruction technique are applied to the W Mass measurement, both analyses are calibrated separately: the illustrative values reported in this section are for the standard analysis.


 
\begin{figure}[ht]
 \begin{center}
  \epsfig{file=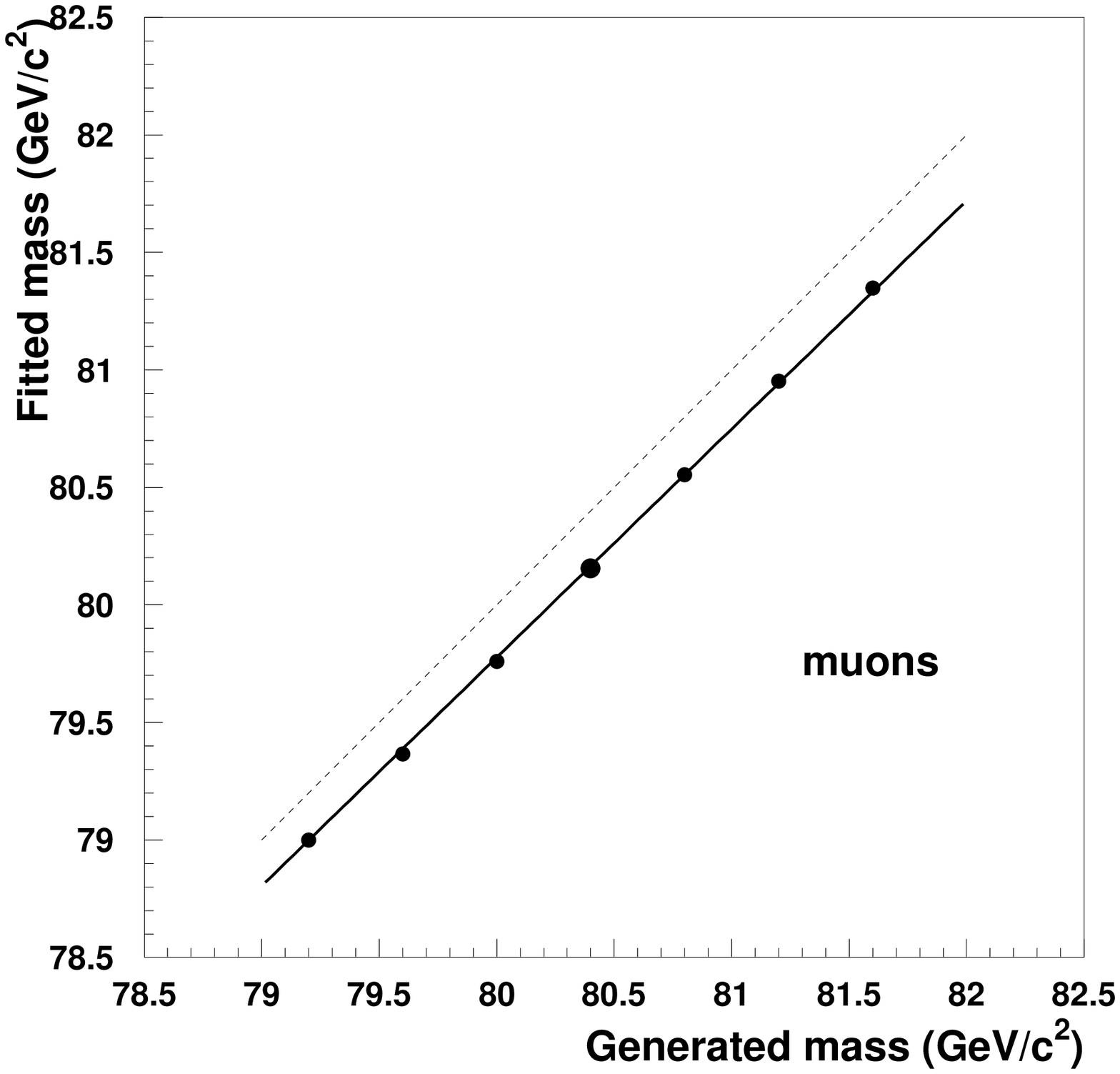,height=85mm,width=85mm}
  \caption{$\W$ mass calibration curve in the $\mnqq$ channel at $\sqs = 189~\GeV$. The dashed line indicates the result that would be obtained without any analysis bias.}
\label{fig:qqmvBias}
\end{center}
\end{figure}

A high statistics simulation sample is used to calibrate the analysis, comprised of an appropriate mixture of signal and background events.
The result of the likelihood fit as a function of the simulated $\W$ mass is
shown in figure~\ref{fig:qqmvBias} for the $\mnqq$ channel analysis at
$\sqs = 189~\GeV$. The analysis has a linear behaviour in the
mass window of interest, and the calibration curves are defined
by two parameters :
\begin{itemize}
  \item the slope of the generated mass against fitted mass line;
  \item the offset defined at a fixed reference point. This point is chosen 
to be the value used in our simulation; 80.4~$\GeVm$ for the mass and 2.11~$\GeVm$ for the width.
\end{itemize}

  The slopes at different energies are found to be compatible, and their
mean values are respectively $0.984 \pm 0.013$, $0.993 \pm 0.006$ and
$0.963 \pm 0.013$ in the $\enqq$, $\mnqq$ and $\tnqq$ analyses. In the $\qqqq$ analysis the slope was compatible with unity to within $2\%$ at all centre-of-mass energies and no slope calibration was applied.

The highly linear behavior, with a value of the slope close to unity is an {\it a posteriori} justification of the fitting functions used in the likelihood fit and described in section~\ref{sec:fullyhadfit}. The remaining effects not taken into account by these fitting functions give rise to the offset.  As an example, the calibration offsets
at $\sqs = 189~\GeV$ are respectively $-0.108 \pm 0.012$,
$-0.215 \pm 0.010$, $-0.252 \pm 0.015$ and $-0.222 \pm 0.006~\GeVm$
in the $\enqq$, $\mnqq$, $\tnqq$ and $\qqqq$ analyses for the mass. The offsets vary slightly with the centre-of-mass energy.

The same procedure is also applied for the $\W$ width analyses. In the $\lnqq$ channel a slope of $0.894 \pm 0.008$ is obtained independent of the centre-of-mass energy and the offset at $\sqs = 189~\GeV$ was $+0.065 \pm 0.015~\GeVm$ . However, in the $\qqqq$ analysis the slope is found to be dependent on the  centre-of-mass energy, the slopes at  $\sqs = 189~\GeV$ and 205~$\GeV$ are approximately $1.1$ and $1.2$  respectively and furthermore the relation between the reconstructed and generated $\gw$ is not perfectly linear. Hence the offset is parameterised as a function of the generated $\W$ width and the centre-of-mass energy. The calibration offset at  $\sqs = 189~\GeV$ is $183 \pm 13~\MeVm$ at the reference width.

The analyses are corrected with these calibration results, and the statistical error on the offset is included in the systematic error (see below). 

\begin{figure}[htp]
 \begin{center}
  \epsfig{file=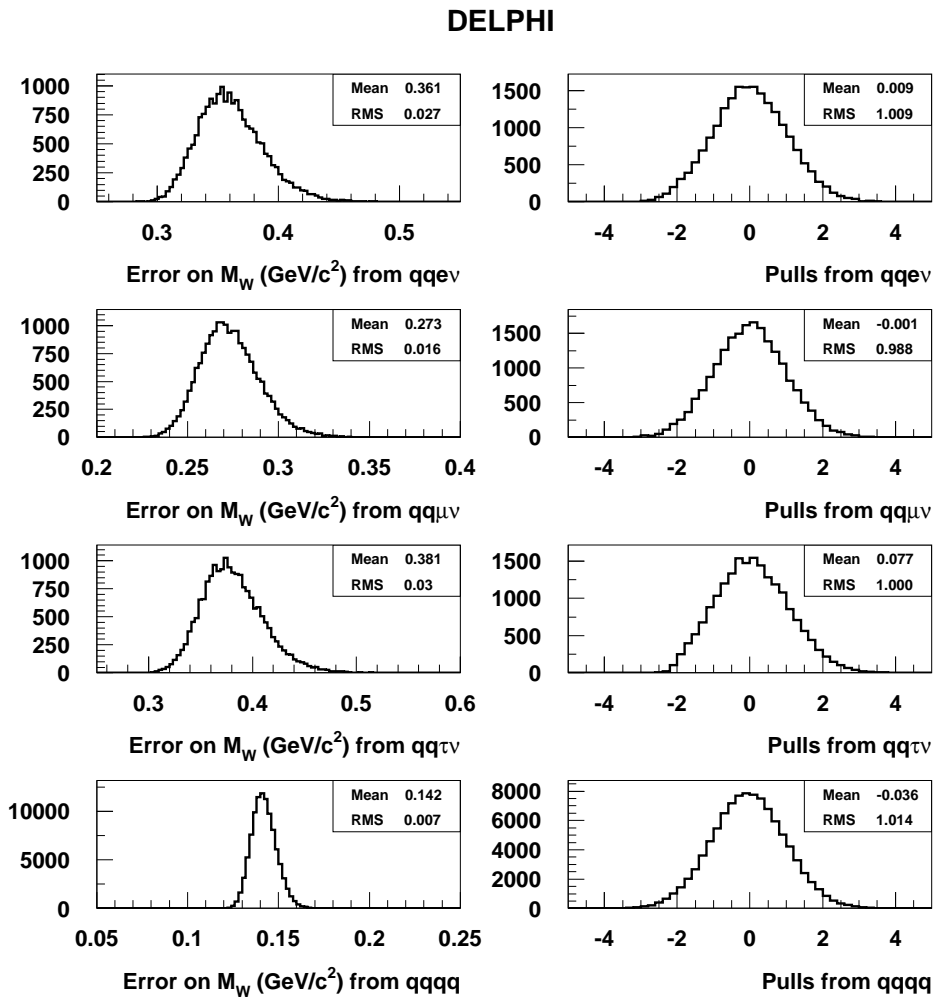,width=0.9\textwidth}
  \caption{The errors (left) and pulls (right) of the $\W$ mass fits for each semi-leptonic analysis channel and the fully-hadronic channel. These plots 
were obtained using simulated event samples with the same statistics as the data sample collected at $200~\GeV$. The errors obtained on the fits to the data samples were $365~\MeVm$ for the $\enqq$ analysis, $282~\MeVm$ for $\mnqq$, $438~\MeVm$ for $\tnqq$ and $149~\MeVm$ for the standard $\qqqq$ analysis.}
 \label{fig:pullslept}
\end{center}
\end{figure}

After applying the calibration procedure, the consistency of the analyses is checked. Sets of simulation events, with a sample size the same as the data, containing the expected mixture of signal and background events were used to test the analyses. Figure~\ref{fig:pullslept} shows error and pull plots from analysing 20000 or more such samples, where the pull is defined as

\[ \rm{pull} = \frac{(\mw_{\rm fit} - \mw_{\rm gen})}{\sigma_{\rm fit}}, \]
here the subscript `fit' and `gen' distinguish the result from the calibrated analysis fit and the generated parameter in the simulation respectively. The $\sigma_{\rm fit}$ is the error estimated by the analysis. This error has been scaled in the analysis to obtain a Gaussian width of one for the pull distributions, as shown in the plots. These plots were produced at all centre-of-mass energies for both parameters. 
 The error distributions in figure~\ref{fig:pullslept} also demonstrate that this quantity is in good agreement with the value obtained from the data.

\newpage
\section{Systematic Uncertainties}
\label{sec:syst}

The sources of systematic error that have been considered for the $\W$ mass and width determinations are described in the subsections below. The results of these studies at example centre-of-mass energies are summarised in tables~\ref{tab:systmw189},~\ref{tab:systmw205} and \ref{tab:systgw}. In the fully-hadronic channel the standard method  and the cone jet reconstruction technique have been applied as described in section~\ref{sec:cone}. The systematic uncertainties are in agreement between these two techniques except for the error sources from final state interactions (FSI), where separate values for the two techniques are given.

\subsection{Calibration }

The analysis calibration procedure is described above in section~\ref{sec:massextrac}. The accuracy with which the offset of the analyses can be determined is limited by the size of the generated simulation samples. Sufficient events were generated to limit this error to  $5\%$ or less of the statistical error on the mass or width determination in any given channel.

\subsection{Detector Effects - Muons}

Contributions to the systematic error on the $\W$ mass and width due to the reconstruction of muons are considered in this section. These were evaluated using the $\Z \rightarrow \mumu$ events collected at the $\Z$ peak during the \LEP2\ period. The systematic uncertainties determined by these studies for the $\W$ mass analysis are presented in table~\ref{tab:systleptonmw}.

\vspace{0.2cm}
{\bf{Inverse Momentum Scale}}
\vspace{0.2cm}

The primary sources of systematic error on the muon momentum scale are the detector alignment or possible reconstruction distortions (particularly in the TPC). As a result of these effects, we may also anticipate an opposite bias on the measured track curvature for positive and negative muons. 

Corrections to the inverse momentum scale, $1/p$, are calculated from the selected $\mumu$ samples. The mean inverse momentum, $<1/p>$, is calculated separately for positive and negative muons in different bins of the polar angle, and a correction for the positive muons is defined as
\begin{equation}
   \frac{1}{2} (<\frac{1}{p-}> - <\frac{1}{p+}>),
\end{equation}
with the opposite sign correction applied to negative muons. These corrections are typically of the order $1$ to $2 \times 10^{-4}~\GeV^{-1}{\it c}$, except in the polar angle regions at the junction between the barrel and endcaps where the correction can reach $10^{-3}~\GeV^{-1}{\it c}$ in the worst case.
In the simulation this correction is, as expected, compatible with zero.
After applying the corrections $<1/p>_{data}$ and $<1/p>_{simulation}$ are 
found to be in agreement within $0.2\%$, and this value is used to calculate the systematic on the muon inverse momentum scale. The systematic uncertainty on the positive and negative muon inverse momentum scale difference is estimated by varying the correction by $\pm50\%$ of its value. 

\vspace{0.2cm}
{\bf{Inverse Momentum Resolution}}
\vspace{0.2cm}

The momentum resolution (typically $0.001~\GeV^{-1}{\it c}$ in
$1/p$) was found to be commonly around $10\%$ better in simulation events than in the data. This discrepancy, determined for all years of \LEP2\ and polar angle regions, is corrected by smearing the simulation with a Gaussian. An additional smearing of $\pm0.0003~\GeV^{-1}{\it c}$ in $1/p$ is used to estimate the systematic error resulting from this correction. This systematic does not affect the $\mw$ determination but is a small component of the $\gw$ measurement uncertainty for events containing muons. 

\subsection{Detector Effects - Electrons}

Contributions to the systematic error on the $\W$ mass and width due to the reconstruction of electrons are considered in this section. These were evaluated using the Bhabha and Compton events collected at the $\Z$ peak and high energies during the \LEP2\ period. The systematic uncertainties determined by these studies for the $\W$ mass analysis are presented in table~\ref{tab:systleptonmw}.

\vspace{0.2cm}
{\bf{Energy Scale}}
\vspace{0.2cm}

\begin{figure}[ht]
  \begin{tabular}{cc}
    \epsfig{file=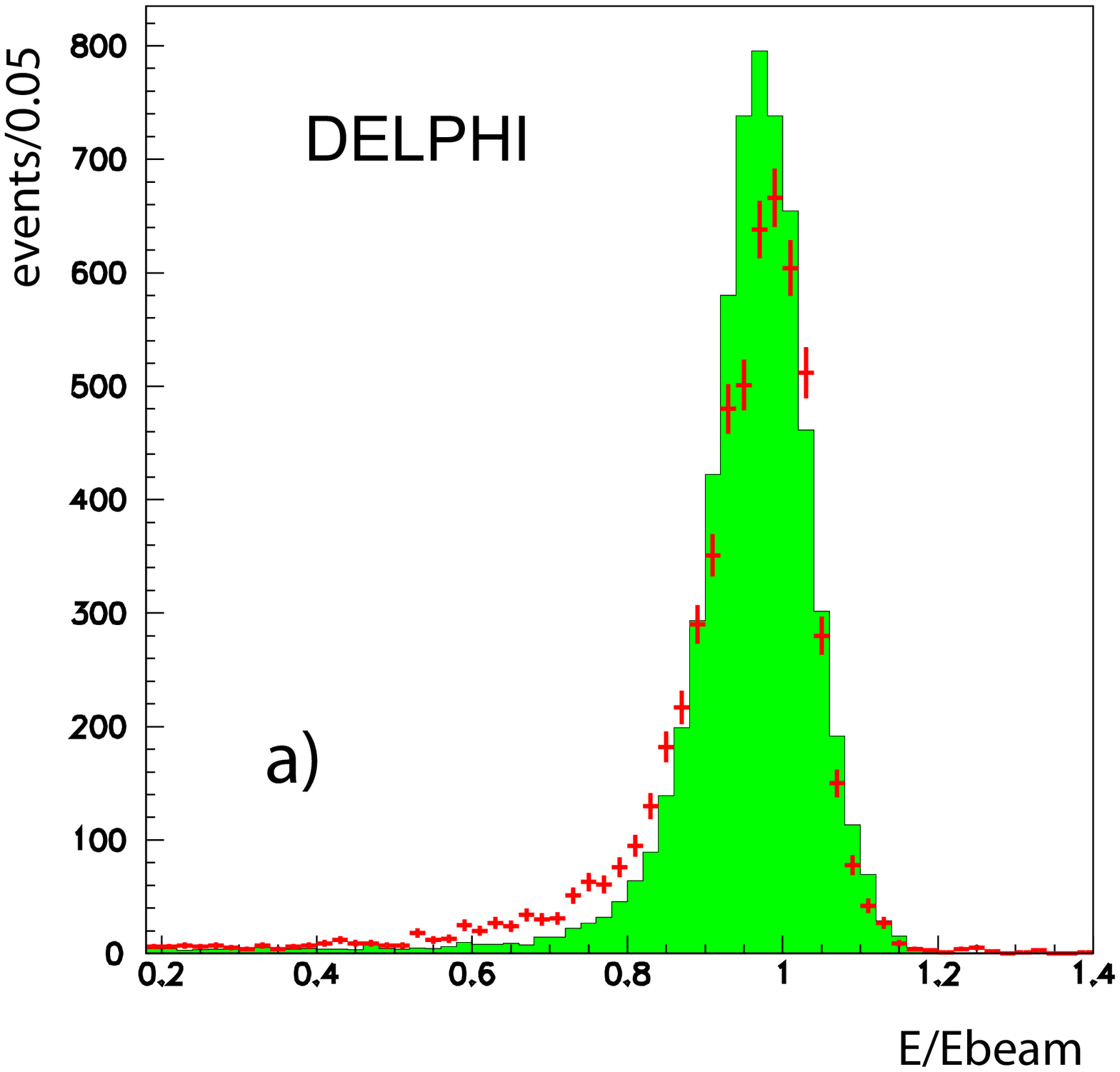,width=75mm} &
    \epsfig{file=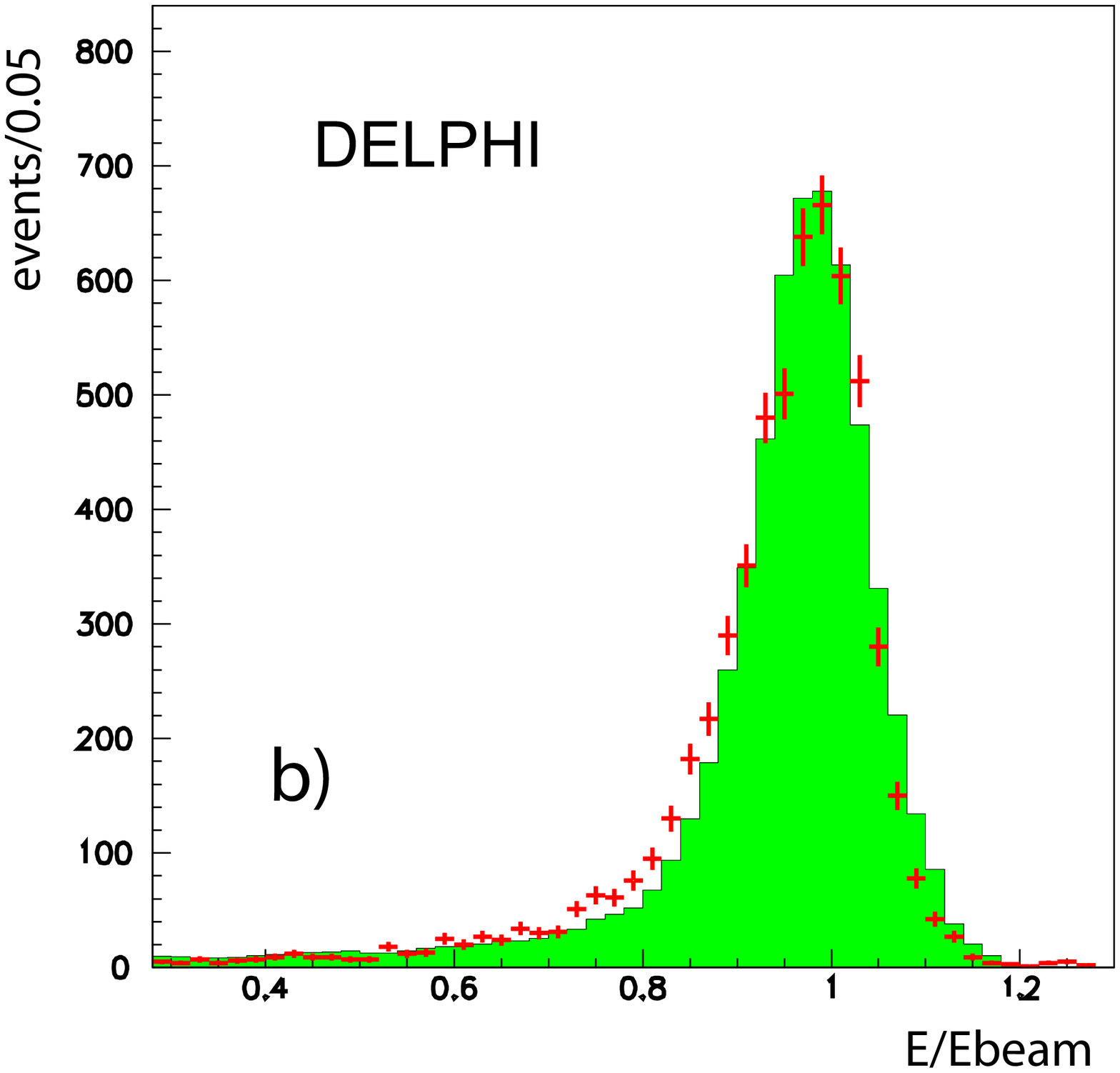,width=75mm}    
  \end{tabular}
  \caption{The ratio E/${\rm E}_{\rm beam}$ for electrons in the endcaps from Bhabha events
    recorded at the $\Z$ peak in 1998. The shaded histogram is the simulation and
    the points are the data. Plot (a) shows the raw distribution, while plot (b) gives this after the bremsstrahlung correction discussed in the text. The resolution correction (see text) has also been applied. }
  \label{fig:BremElec}
\end{figure}

The reconstructed energy of electrons was compared between data collected at the $\Z$ peak and fully simulated samples of Bhabha events. In the barrel region of the detector the data and simulation are in good agreement. However, in the forward directions a slight difference is observed between the data and simulation (see figure~\ref{fig:BremElec}) and attributed to an under-estimation of the quantity of material in the simulation before the electromagnetic calorimeter in the \DELPHI\ endcaps. A correction is applied to the simulation by introducing the effect of extra bremsstrahlung emission corresponding to an additional 3\% of a radiation length.
Following \cite{brem}, the probability $w$ that an electron of initial energy $E_0$ has an observed energy between E and E+dE after traversing a thickness of $t$ radiation lengths is 
\begin{equation}
    w(E_{0},E,t) dE = \frac{dE}{E_0} \frac{[ln(E_{0}/E)]^{(t/ln2)-1}}{{\Gamma}(t/ln2)}.
\end{equation}
For each event, the corrected energy $E$ is chosen randomly according to the
distribution $w$. The optimal value of the parameter $t$ was adjusted from the data and simulation comparison. 

After the endcap correction was applied, good agreement between data and simulation was obtained throughout the detector. The residual systematic error on this absolute energy scale is estimated to be $\pm 0.3\%$ of the measured energy and is estimated from the selection cut stability and statistical precision of the data and simulation comparison.

\vspace{0.2cm}
{\bf {Energy Resolution}}
\vspace{0.2cm}

The resolution on the reconstructed electron energies was also compared between the data and simulation Bhabha samples. The agreement is improved by applying a Gaussian smearing to the simulation with a width varying between 1 and 2\% of the measured electron energy in the barrel, and 2 to 4\% in the endcaps, depending on the year of data taking. The systematic error on this smearing Gaussian width is estimated to be $\pm 1\%$ of the measured energy. This systematic does not affect the $\mw$ determination but is a small component of the $\gw$ measurement uncertainty for events containing electrons.

\vspace{0.2cm}
{\bf {Energy Linearity}}
\vspace{0.2cm}

\begin{figure}[ht]
  \begin{tabular}{cc}
    \epsfig{file=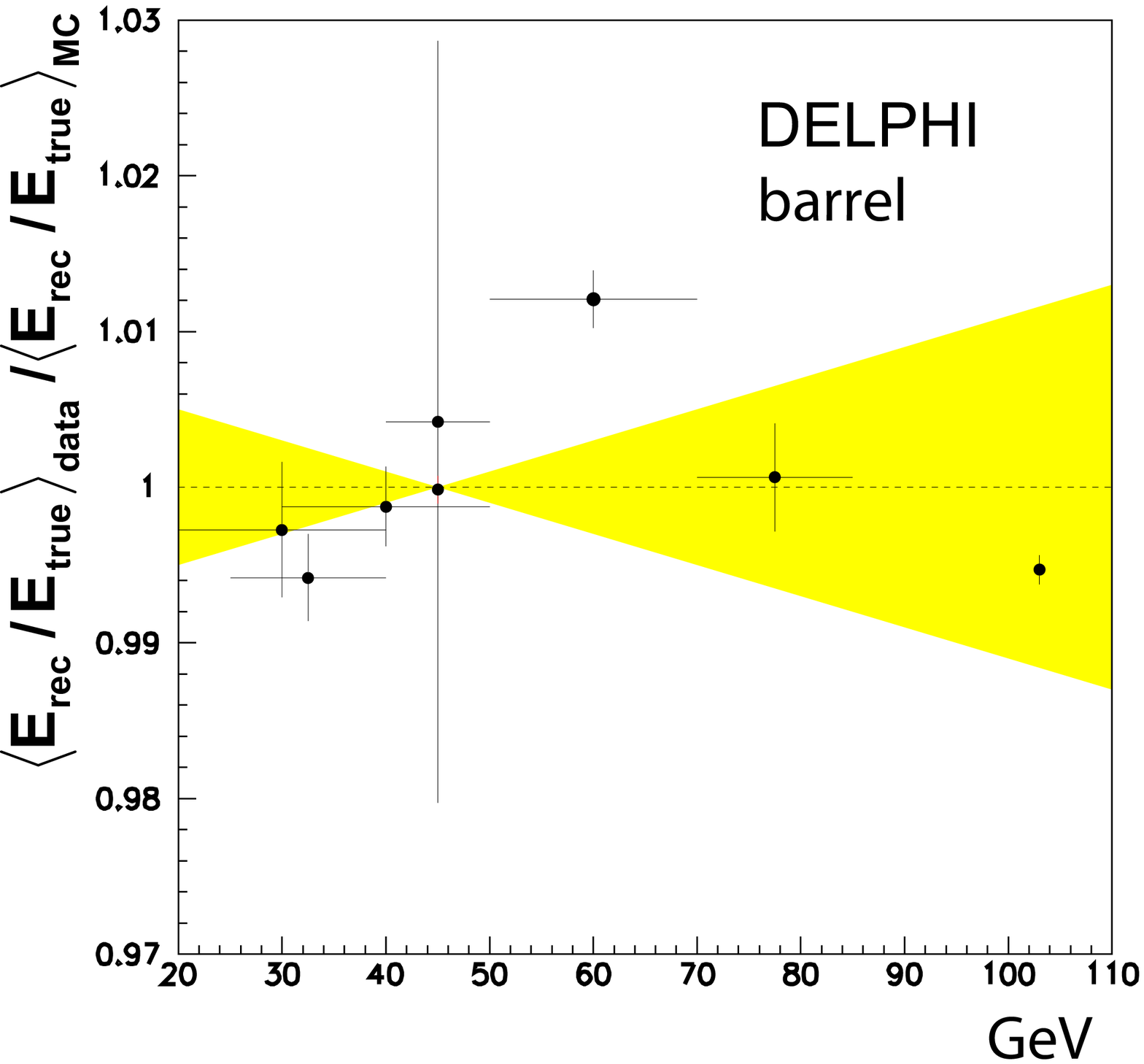,width=75mm} &
    \epsfig{file=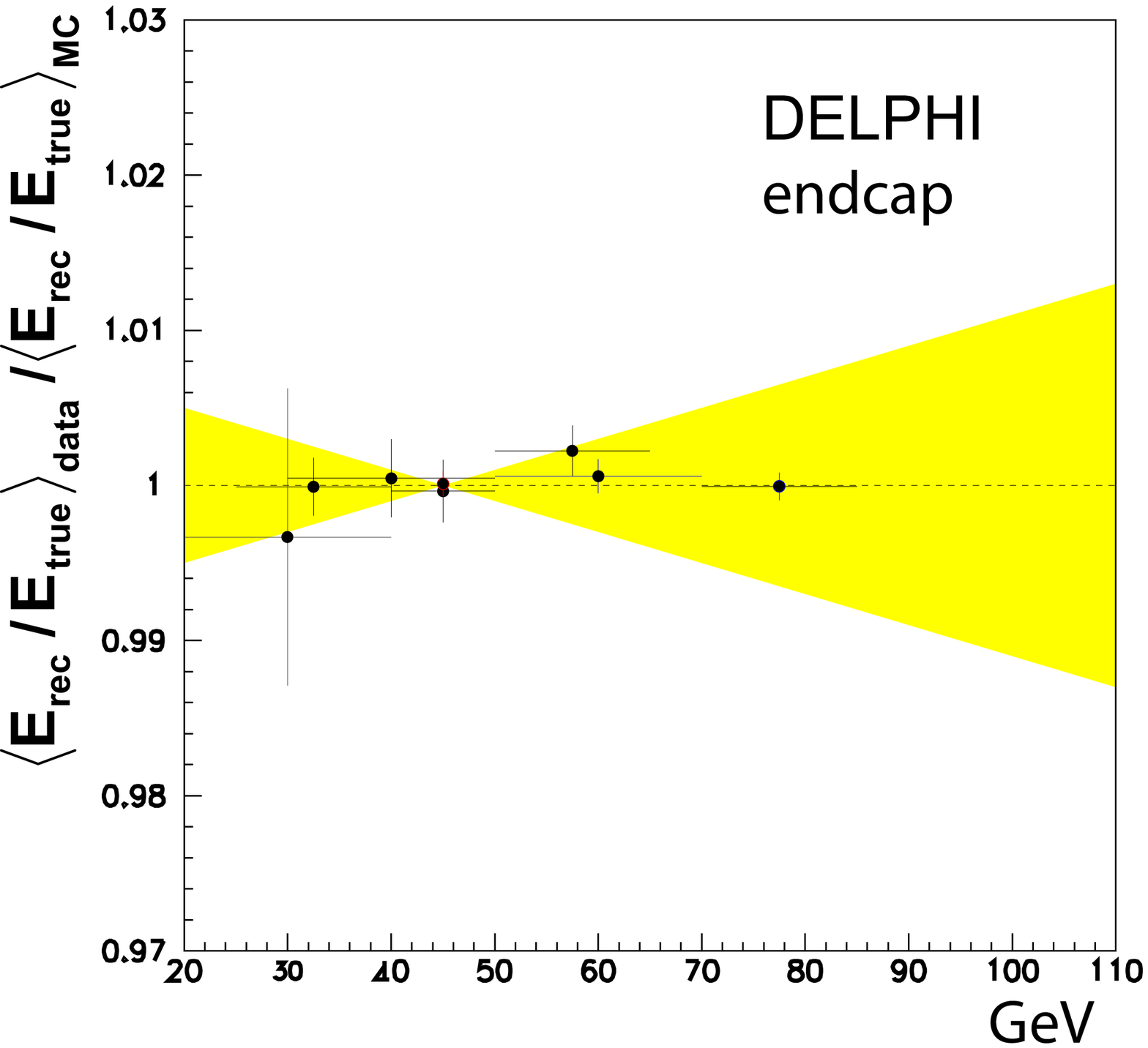,width=75mm}
  \end{tabular}
  \caption{The double ratio of reconstructed and true average energy values in data and simulation,
$\langle~E_{rec}/E_{true}~\rangle_{data}/\langle~E_{rec}/E_{true}~\rangle_{MC}$, for data taken in 2000. The shaded area represents the quoted systematics due to a possible dependence of the 
    energy calibration with the electron energy. The left hand plot is for electrons observed in the barrel electromagnetic calorimeter and the right hand plot for electrons in the endcap. Note that, by construction, the Bhabha point at 45~$\GeV$ is at one.}
\label{fig:CalibElec}
\end{figure}
 
The reconstructed electron energy was also studied as a function of the true energy. The $\Z$ peak and high energy running provided high statistic Bhabha samples with which to study electrons of 45~$\GeV$ and above 100~$\GeV$ energy. For these samples the ``true'' electron energy is taken from the beam energy.  
The reconstructed electron energy was also checked using low energy electrons from
Compton events at the $\Z$ peak, and high energy electrons from radiative Bhabha
scattering at high centre-of-mass energy. In these cases the true energy of the
lepton is deduced from 3-body kinematics using only the angular information and assuming that the unseen particle was along the beam axis. Figure~\ref{fig:CalibElec} shows the compatibility of the reconstructed electron energy in data and simulation, only statistical errors are shown. One of the three points measured for radiative Bhabhas in the Barrel shows a discrepancy but this effect is not confirmed by the better measured high energy (non-radiative) Bhabha point, whereas physical calibration problems such as threshold effects or leakage in the calorimeter would be expected to increase in size with energy. Hence, no additional corrections are applied. A systematic error is estimated assuming a deviation of the energy calibration slope  $\rm{E_{data}} / \rm{E_{simulation}}$ versus $\rm{E_{simulation}}$  of 1\% over the range 25 to 70~$\GeV$. These values approximately correspond to the relevant energy range for the observed electrons in the analysis.

\subsection{Detector Effects - Taus}

The $\tnqq$ channel differs from the other $\WW$ semi-leptonic decay channels as these events contain two (or three for leptonic tau decays) neutrinos in the final state. Thus, the mass of the event can be determined only from the decay products of the other $\W$. As a result the lepton systematics described in the preceding sections are not relevant to the $\tnqq$ channel. The only relevant systematic involving the tau decay products arises from uncertainties in the assignment of the reconstructed tracks between the tau product and the hadronically decaying $\W$. This effect is small compared with the overall uncertainty on the jet energy and direction, the systematic on which is considered in the sections below.

 \begin{table}[ht]
 \begin{center}
  \begin{tabular}{|l|c|c|}     \hline
\multicolumn{3}{|c|}{$\mw$  Lepton Correction Systematic Errors ($\MeVm$)} \\ \hline
Sources of Systematic Error & $\enqq$ 189 $\GeV$ & $\enqq$ 205 $\GeV$ \\ \hline \hline
Electron Energy Scale             & 18 & 22  \\
Electron Energy Resolution        & -  &  -  \\
Electron Energy Linearity         & 16 & 11 \\ \hline \hline
                            & $\mnqq$ 189 $\GeV$ & $\mnqq$ 205 $\GeV$ \\ \hline 
Muon 1/p Scale                    & 16   &  21 \\
$\mu^{+}$ $\mu^{-}$ 1/p Difference    & 1   &   4 \\
Muon 1/p Resolution               & -   &  - \\
\hline
\end{tabular}
  \caption{Contributions to the systematic error on the $\W$ mass measurement at 189 and 205~$\GeV$ related to the lepton reconstruction. The uncertainties on each of these numbers is typically 3 $\MeVm$.}
  \label{tab:systleptonmw}
 \end{center}
\end{table}

\subsection{Jet Description}
\label{sec:jetdesc}

Jets are composite objects, and the detector and analysis response to
them can be dependent on their internal structure. Therefore it is not
straightforward to separate in a clean way uncertainties arising from
the modelling of the detector in the simulation from those due to
the theoretical description of the jet structure.

Moreover this description is not based on exact calculations, whose
uncertainty can be in principle reasonably well estimated, but on
phenomenological models tuned to best reproduce the data at the $\Z$
peak: the Lund model as implemented in {\tt PYTHIA} is the standard
choice for this analysis. In this situation the comparison of
different models may be a useful tool to understand which parts of the
fragmentation description the measurement is sensitive to, but only a
direct comparison of the chosen model with well understood data
samples, in particular $\Z$ hadronic decays, can give the ultimate
estimate of the uncertainty from the observed data-simulation disagreements.

The jet studies performed are described in the text below and the
corresponding jet correction systematic errors are provided in table~\ref{tab:systjetmw}. The most relevant jet characteristics were
calibrated on real data control samples, and uncertainties on these
calibrations are propagated through the analysis.

\vspace{0.2cm} {\bf{Energy Scale}} \vspace{0.2cm}

The absolute jet energy scale was studied in on-peak $\Z \rightarrow
q\bar{q}$ decays, by comparing the reconstructed energies, $E_{rec}$, in
data and simulation in selected two jets events. The $b-$tagging
technique is used to remove $b$ quark jets which are essentially not
present in $\WW$ decays. The true jet energy in these events is
assumed to be the beam energy $E_{beam}$, under the assumption that the
bias introduced by QED ISR is described with negligible error in the
simulation (the {\tt KK2f} generator was used for these events).  The
double ratio of average values
$\langle~E_{rec}/E_{beam}~\rangle_{data}/\langle~E_{rec}/E_{beam}~\rangle_{MC}$
was evaluated as a function of the jet polar angle and applied as a
scale factor correction to the four-momentum components of the jet in
simulated events. The correction value depends on the year as well as
the angular region, with the deviation from unity ranging typically
from a few per mille up to 3-4\% in the most forward region.

The systematic uncertainty on this correction is determined by the
limited on-peak $\Z$ statistics, and it is estimated to be $\pm
0.3\%$.

\vspace{0.2cm} {\bf{Energy Resolution}} \vspace{0.2cm}

The same event sample used to study the jet absolute energy scale was
also used to calibrate the jet energy resolution in the simulation.  A
Gaussian smearing was determined from the data and is applied to the 
simulated jet energy with a magnitude dependent on the ratio of the reconstructed and true jet energies. This procedure takes into account the asymmetric shape of the
jet energy observable. When applying the correction to the simulated
$\WW$ events an estimate of the true jet energy is required. When the
event is reconstructed with two jets from each hadronically decaying
$\W$, the generated quark energies are used. However, when gluon
radiation has given rise to an additional jet the true jet energy
estimate is determined by applying the same clustering algorithm as
used in the analysis to the simulated partons prior to the detector
simulation.  In both cases the association of the true and
reconstructed jets is performed according to geometric criteria.

The average resolution correction ranges from 4.5\% of the jet energy
in the barrel to 6.6\% in the endcaps. The correction is also
dependent on the year. The systematic uncertainty on the correction is
estimated to be $\pm 2\%$ of the jet energy.

\vspace{0.2cm} {\bf{Energy Linearity}} \vspace{0.2cm}

The dependence of the energy calibration as a function of the jet
energy was checked using low energy jets from $q\bar{q} + \rm{gluon}$
events at the $\Z$ peak and high energy jets from $\ee \rightarrow
q\bar{q}$ decays at high energy.

In the first case, the true jet energy is determined using three-body
massless kinematics. The jet energy range used in this study is
restricted to the region where the data and simulation true energy
distributions do not show sizeable discrepancies. This energy selection
avoids introducing an unnecessary sensitivity in this analysis to the
modelling of hard gluon radiation in the simulation.

In the second high-energy jet case the effective hadronic mass
$\sprime$ is required to be such that $\sqrt{s'/s} > 0.95$. The true
jet energy is then again determined using three-body massless
kinematics but now the third object is an hypothetical ISR photon
emitted along the beam pipe. The difference between the estimated jet
energy and the nominal beam energy is constrained to be smaller than
10~$\GeV$.

A jet energy linearity slope in $\rm{E_{data}} / \rm{E_{simulation}}$ versus $\rm{E_{simulation}}$ is then determined. The study was performed separately in the barrel and endcap regions of the detector and for each data taking year. The results from the different data taking years are compatible within statistical errors. The study showed agreement in the slope at typically the $0.5\%$ level over the range 25 to 75~$\GeV$, and this deviation value is used to determine the systematic uncertainty.

\vspace{0.2cm} {\bf{Angular Bias}} \vspace{0.2cm}

As reported in \cite{radreturnDELPHI}, the reprocessing of data and simulation 
used for this analysis has a noticeable excess of tracks at low polar angles (forward tracks) in data as compared to the simulation. The most likely cause of
this effect is an underestimation in the simulation of the track
reconstruction efficiency for low-momentum particles at low polar
angle.
                                                                               
This effect introduces a small bias in the distribution of the jets' reconstructed polar angle in the simulation compared with data. In order to evaluate the effect of such a bias, a systematic shift of the jets' polar angle is applied to the simulation. The shift as a function of the polar angle itself has been
determined using on-resonance $\Z$ hadronic decays, and is found to
have the form $0.008 \cos{\theta_j}^{5.3}$ where $0 < \theta_j <
\pi/2$ is the polar angle of the jet. The corresponding W mass and width shifts have been evaluated and symmetric systematic errors of these values applied. The W mass uncertainty is reported in table~\ref{tab:systjetmw}.

\vspace{0.2cm} {\bf{Angular Resolution}} \vspace{0.2cm}

A study of the acollinearity of jets in on-peak $\Z \rightarrow
q\bar{q}$ events was performed and appropriate smearings to the simulation of the jet angular direction, dependent on the polar angle of the jet, were estimated. The smearings on the polar angle are typically 5~mrads. A systematic error is estimated by applying an extra 5~mrad angular smearing.

\vspace{0.2cm} {\bf{Jet Mass}} \vspace{0.2cm}

The jet mass is known not to be exactly described in the simulation;
both inaccuracies in the fragmentation description (related to the jet
breadth due to soft and hard gluon radiation) and imperfections in the
modelling of the detector response (reconstruction efficiencies and
noise) are responsible for these discrepancies. However, only those
data-simulation differences in the jet mass which are not compensated
by differences in the inter-jet angle are relevant for the systematic
uncertainty, since these cause systematic biases in the reconstructed
$\W$ mass. 

For this reason the fragmentation-induced differences are only
marginally relevant for the mass measurement. Furthermore, the calibration
procedure adopted, in particular for the energy and angular smearing, 
corrects for most of the effects given by the differences in jet breadth.
The jet breadth is relevant as broader jets are worse reconstructed: they are 
detected with larger uncertainties on the jet direction; are likely to lose more energy due to the imperfect hermeticity of the detector; and cause more confusion in the jet clustering.

The jet correction procedure described above, as well as the
constrained kinematic fit, modifies all the four-momentum components of
the jet but leaves unchanged the jet boost, i.e. the $E/m$ ratio. It is
therefore useful to study this observable, instead of the simple jet
mass.

Detector noise is a source of data-simulation discrepancy which
clearly biases the reconstructed boson mass, since it changes the mass
and boost of the jets while leaving, on average, the inter-jet angle
unchanged.  Significant data-simulation differences in low energy
neutral clusters, both in the electromagnetic and hadronic
calorimeters, are attributed primarily to an imperfect noise
description, while the discrepancies in the charged particles of jets
are considered to be almost entirely due to the modelling of the fragmentation.

The average effect of removing low energy neutrals below 2~$\GeV$ on
the jet $m/E$ was evaluated as a function of the polar angle and of
the $m/E$ of the jet itself, since the impact of the noise depends on
the breadth of the jet. The expected effect on the neutrals from
fragmentation was subtracted. The fragmentation effect was obtained
from charged particles, suitably scaled for the relative neutral and
charged particle multiplicity.

This $m/E$ effect was then propagated in the full analysis chain to
extract the relative systematic uncertainty on the full mass and width
measurements.

 \begin{table}[ht]
 \begin{center}
  \begin{tabular}{|l|ccc|c|}     \hline
\multicolumn{5}{|c|}{$\mw$  Jet Correction Systematic Errors ($\MeVm$)} \\ \hline
Sources of Systematic Error & \multicolumn{4}{|c|}{189 $\GeV$}  \\ \hline \hline
                         & $\enqq$ & $\mnqq$ & $\tnqq$ & $\qqqq$ \\
Energy Scale             &  8      & 6       &  11     & 8  \\
Energy Resolution        &  3      & 3       &   5     & 9 \\
Energy Linearity         & 12      & 9       &  12     & 16\\
Angular Bias             &  3      & 5       &   5     & 2\\
Angular Resolution       &  -      & -       &   -     & 8\\
Jet Mass                 &  9      & 8       &   8     & 10\\ \hline \hline

                         & \multicolumn{4}{|c|}{205 $\GeV$}  \\ 
                         & $\enqq$ & $\mnqq$ & $\tnqq$ & $\qqqq$ \\ \hline 
Energy Scale             &  11     & 9       &  16     &  8  \\
Energy Resolution        &   8     & 5       &   8     & 10 \\
Energy Linearity         &  15     & 11      &  20     & 8\\
Angular Bias             &   9     &  8      &   7     & 19\\
Angular Resolution       &  -      & -       &  -      & 1\\
Jet Mass                 &  13     & 12      &  17     & 13\\

\hline
\end{tabular}
  \caption{Contributions to the systematic error on the $\W$ mass measurement at 189 and 205~$\GeV$ related to jet reconstruction. The uncertainties on each of these numbers is typically 6 $\MeVm$.}
  \label{tab:systjetmw}
 \end{center}
\end{table}

\vspace{0.2cm} {\bf{Fragmentation Model}} \vspace{0.2cm}

The effect of using different hadronisation models on the analysis was
studied by replacing the standard choice, {\tt PYTHIA}, with both
the {\tt ARIADNE} and {\tt HERWIG} models, each tuned by \DELPHI\ to best match
experimental data. The mass and width shifts were evaluated at 189~$\GeV$ and 207~$\GeV$ centre-of-mass 
energies and are reported in tables~\ref{tab:fragmodelsmw} and~\ref{tab:fragmodelsgw}. Detailed studies
performed at the $\Z$ peak showed that for several observables all the
models showed disagreements with the data and that these disagreements 
were all in the same direction: the jet mass variable, discussed in the 
previous paragraph, is a clear example. Hence the results of the 
hadronisation model comparison were used only to investigate
the sensitivity of the analysis to specific features of the models,
and not used directly as an evaluation of the systematic uncertainty due to the
choice of model.

The biggest difference was found to be between {\tt PYTHIA} and {\tt
  HERWIG}, and was shown to be largely due to the different production
  rates of heavy particles, mainly kaons, protons and neutrons. At
  parton level these differences modify not only the jet masses but also
  change the jet-jet angles accordingly, leaving the
  bosons invariant masses unchanged.   However, the reconstruction and analysis
  procedure breaks this compensation since in the fully-hadronic event
  reconstruction all charged particle tracks are assigned the pion mass,
  and all neutrals are assumed to be massless (photon-like). 
 In the semi-leptonic analysis, the nominal masses are used in the 
jet reconstruction for those particles
with a positive identification, i.e. for charged kaons and protons identified
by the RICH and for $K^{0}_{S}$ and Lambdas reconstructed as secondary vertexes
from their decay products \cite{delphi}.

The {\tt HERWIG} version used, although tuned to best reproduce the
$\Z$ peak \DELPHI\ data, is known to describe the particle
production rates poorly. This is especially the case for baryons, therefore
using { \tt HERWIG} accentuates this particle mass assignment effect.
Generally the measured particle rates are closer to those in {\tt PYTHIA}
and {\tt ARIADNE}. Reweighting in the models the production rates of the 
most abundant heavy particles species, kaons and protons, 
reduces the disagreement among the different models,
bringing it to the level of the statistical uncertainty of the fit. Tables~\ref{tab:massrew} and \ref{tab:widthrew} show the residual 
discrepancies obtained between the models after they have been reweighted to the \PYTHIA\ values. The component of the fragmentation systematic error which is not due to the heavy particle multiplicity effect is obtained from these numbers. The largest value - either the central value or its uncertainty - from either model is taken as the systematic error estimate.

The component of the fragmentation error that is due to the heavy particle rate was also evaluated for the $\W$ mass analysis; this small component of the error is neglected for the $\W$ width analysis. The $\W$ mass shift was evaluated between the \DELPHI\ tune of \PYTHIA\ and the same events reweighting to the measured particle rates $\pm 1 \sigma$ of their uncertainty. The average of the modulus of the two shifts is reported in table~\ref{tab:particlerate} and is taken as the estimate of the fragmentation error due to the heavy particle multiplicity. 

The combined fragmentation error was evaluated for the $\W$ mass by adding the particle reweighting effects and the model variation uncertainty in quadrature. This fragmentation error is listed separately from the other jet description uncertainties in the systematic uncertainty summary tables~\ref{tab:systmw189},~\ref{tab:systmw205} and~\ref{tab:systgw}.

\begin{table}[tb]
\vspace{0.5cm}
 \begin{center}
  \begin{tabular}{|l|c|c|c|c|}
\hline
                       & \multicolumn{4}{|c|}{$\Delta \mw $ $\MeVm$} \\ 
                       &  $\enqq$    & $\mnqq$     & $\tnqq$     & $\qqqq$ \\ 
\hline
  HERWIG - PYTHIA      & $-7 \pm  10$ & $-16 \pm 9$ & $-17 \pm 13$ & $-9 \pm 5$  \\
  ARIADNE - PYTHIA     & $-11 \pm 9$  & $-12 \pm 9$ & $-10 \pm 12$ & $-15 \pm 5$ \\
\hline
  \end{tabular}
  \caption{Effect of different fragmentation models on the $\W$ mass determination.}
  \label{tab:fragmodelsmw}
 \end{center}
\end{table}

\begin{table}[tb]
\vspace{0.5cm}
 \begin{center}
  \begin{tabular}{|l|c|c|}
\hline
                     &  \multicolumn{2}{|c|}{$\Delta \gw $ $\MeVm$} \\ 
                           &  $\lnqq$ & $\qqqq$ \\ 
\hline
 HERWIG - PYTHIA      & $+46 \pm 13$ &  $-2 \pm 11$ \\
 ARIADNE - PYTHIA     & $-9 \pm 15$ &   $+1 \pm 11$\\

\hline
  \end{tabular}
  \caption{Effect of different fragmentation models on the $\W$ width determination.}
  \label{tab:fragmodelsgw}
 \end{center}
\end{table}

\begin{table}[tb]
\vspace{0.5cm}
 \begin{center}
  \begin{tabular}{|l|c|c|c|c|}
\hline
                       & \multicolumn{4}{|c|}{$\Delta \mw $ $\MeVm$} \\ 
                             &  $\enqq$       & $\mnqq$      & $\tnqq$     & $\qqqq$ \\ 
\hline

HERWIG Rew. - PYTHIA    & $-2 \pm 10 $ & $-8 \pm 9$  & $-5 \pm 13$ &  $-11 \pm 6$\\
ARIADNE Rew. - PYTHIA   & $-10 \pm 9 $ & $-10 \pm 9$ & $-10 \pm 12$ & $-1 \pm 4$ \\

\hline
  \end{tabular}
  \caption{Effect of different fragmentation models on the $\W$ mass determination, after reweighting the heavy particle species rates in the Monte Carlo simulations to the measured rates.}
  \label{tab:massrew}
 \end{center}
\end{table}

\begin{table}[tb]
\vspace{0.5cm}
 \begin{center}
  \begin{tabular}{|l|c|c|}
\hline
                       &  \multicolumn{2}{|c|}{$\Delta \gw $ $\MeVm$} \\ 
                       &  $\lnqq$ & $\qqqq$ \\ 
\hline
HERWIG Rew. - PYTHIA Rew.   & $ +29 \pm 13$ & $+3 \pm 8$ \\
ARIADNE Rew. - PYTHIA Rew.  & $ -11 \pm 15$ & $-1 \pm 8$ \\
\hline
  \end{tabular}
  \caption{Effect of different fragmentation models on the $\W$ width determination, after reweighting the heavy particle species rates in the Monte Carlo simulations to the measured rates.}
  \label{tab:widthrew}
 \end{center}
\end{table}

\begin{table}[tb]
\vspace{0.5cm}
 \begin{center}
  \begin{tabular}{ |l|c|c|c|c|}
\hline
              & \multicolumn{4}{|c|}{$\Delta \mw$ $\MeVm$} \\ 
Particle Type &  $\enqq$       & $\mnqq$      & $\tnqq$     & $\qqqq$ \\ 
\hline
$\rm{K}^{\pm}$ & $0.1 \pm 0.3$ & $0.9 \pm 0.3$ & $1.5 \pm 0.4$ & $0.2 \pm 0.5$\\
Proton         & $2.0 \pm 0.4$ & $1.5 \pm 0.3$ & $3.2 \pm 0.5$ & $3.5 \pm 0.5$\\
\hline
\end{tabular}
  \caption{Effect on the $\W$ mass of reweighting the heavy particle species rates in the Monte Carlo simulations. The mass shifts were evaluated between 
 the DELPHI tune of PYTHIA and versions reweighted to 1 sigma above 
 and below the measured particle rates. The shift value reported is the average of the modulus of these two shifts. The measured charged multiplicity in a $\Z$ peak event for kaons is $2.242 \pm 0.063$ \cite{pdg}, 
whereas for protons the measured multiplicity is $1.048 \pm 0.045$ \cite{pdg}. 
}
  \label{tab:particlerate}
 \end{center}
\end{table}

\subsection{Mixed Lorentz Boosted $\Z$s} 

An alternative method of evaluating the jet description systematic is to use the technique of mixed Lorentz boosted $\Z$s (MLBZ). This method attempts to emulate $\WW$ events using two on-peak $\Z$ events. The emulated $\WW$ events are constructed both from simulated events and the large statistics sample of $\Z$ peak data events. Standard $\W$ mass and $\W$ width analyses can then be performed on these event samples. Hence, the MLBZ method provides a direct comparison between data and the simulation model of choice. The difference between the measurements made from the data and simulation MLBZs can be interpreted as primarily providing a statistically sensitive cross-check of the fragmentation systematic assigned to the $\W$ mass and width measurements. This method would also identify some sources of detector modelling error.

A $\WWffff$ event is emulated by selecting two $\Z$ events and
rotating and Lorentz boosting them so that their superposition reflects a true $\WW$ event. The mixture of quark species will not be the same as in true $\WW$ events, it will however be the same between the data and simulated $\Z$ samples that are used in the comparison. To emulate a $\qqqq$ event two hadronically decaying $\Z$ events were used. To emulate a $\lnqq$ event one $\Z$ decaying into hadrons and one $\Z$ decaying into charged leptons was used. One hemisphere of the $\Z\ra\lplm$ decay is removed to represent the $\W\ra\len$ decay. The emulation process is performed by manipulating the reconstructed tracks and calorimeter energy clusters.

A realistic distribution of $\WW$ events is obtained by using event templates. The four momenta of the four primary fermions in a {\tt WPHACT} $\WW$ event are used as the event template. The $\Z$ events are chosen such that they have a thrust axis direction close to the polar angle of one of the $\W$ fermions. This ensures that the distribution in the detector of the tracks and energy clusters selected in the $\Z$ event follows that expected in $\WW$ events. Each of the template $\W$s is then boosted to its rest frame. The particles in a final state of a selected $\Z$ event are rotated to match the rest-frame direction of the fermions from the template $\W$. The energy and momentum of the $\Z$ events are then rescaled to match the kinematic properties of the $\W$ boson decay. The two $\Z$ events are then each boosted into the lab frame of the template $\WW$ event and mixed together. The same $\WW$ event templates are used for the construction of both the data and Monte Carlo simulation MLBZ events, thus increasing the correlation between both emulated samples. 

Tests were performed to confirm the reliability of the MLBZ method in assessing systematic errors. MLBZs were produced using $\Z$s with the {\tt PYTHIA}, {\tt HERWIG} and {\tt ARIADNE} models and the observed mass shifts were compared and found to agree with the statistically limited mass shifts observed in $\WW$ simulation events. A significant mass shift (300~$\MeVm$) was introduced by using the cone rejection algorithm (discussed in section~\ref{sec:cone}) for the $\W$ mass measurement in the $\qqqq$ channel. The real and simulated MLBZs and $\WW$ events agreed on the estimated size of the mass shift between the standard and cone estimators at the $15\%$ level.

The MLBZ method was used to create emulated $\WW$ event samples. The $\Z$ events were selected from data recorded during the \LEP2\ calibration runs of the same year or from the corresponding Monte Carlo simulation samples. Values for the $\mw$ and $\gw$ estimators were determined separately for the data and simulation samples. This method has been applied on a cross-check analysis in the semi-leptonic channels and to the standard fully-hadronic analysis. The results from the fully-hadronic analysis are shown in Table~\ref{tab:mlbzresults}. The semi-leptonic cross-check analysis applied the MLBZ procedure to the W mass determination separately in the electron, muon, and tau channels with uncertainties of around $8~\MeVm$ being obtained and the results being compatible with the systematic uncertainties quoted in this paper. The MLBZ method provides a useful cross-check of the size of the systematic uncertainty arising from fragmentation and other jet description errors reported in the previous section. From the values obtained from the MLBZ method we conclude that the systematic uncertainties have not been significantly underestimated. 

\begin{table}[tb]
\vspace{0.5cm}
 \begin{center}
  \begin{tabular}{|l|c|rcr|rcr|}
\hline
          & $\sqs$  &  \multicolumn{3}{|c|}{$\Delta \mw$}  &   \multicolumn{3}{|c|}{$\Delta \gw$} \\
                &  $\GeV$        &  \multicolumn{3}{|c|}{$\MeVm$}     &   \multicolumn{3}{|c|}{$\MeVm$} \\
\hline
\multicolumn{8}{|c|}{MLBZ} \\
\hline
$\qqqq$ Data - PYTHIA        & 206.5 & \ \  -7.9     & $\pm$ &    4.9  \ \  &  \ \   20.1    & $\pm$ &  10.5  \ \      \\

\hline
  \end{tabular}
  \caption{Results obtained with the MLBZ method (see text).}
  \label{tab:mlbzresults}
 \end{center}
\end{table}

\subsection{Electroweak Radiative Corrections}
\label{sec:radcorr}

The measurements of the $\W$ mass and width described in this paper rely upon 
the accuracy of the event description provided by the simulation. Hence, the modelling accuracy of the electroweak radiative corrections implemented in the event generator is a source of systematic uncertainty.

The radiative corrections for 4-fermion events are described
in~\cite{delphi4fgen} and in section~\ref{sec:simul}. For $\WW$ (\CCTHREE)
events, the signal used in this analysis, the corrections are based on {\tt
  YFSWW}~\cite{yfsww} and the effect of the theoretical uncertainties
in it on the $\W$ mass measurement were initially studied
in~\cite{wmasssys} at pure event generator level.

In~\cite{eweak} this study has been performed in the context
of the full \DELPHI\ simulation and analysis procedure; furthermore the
main uncertainties due to non-\CCTHREE\ 4-fermion background events have
been studied. Radiative corrections uncertainties on non 4-fermion
background events are included in the uncertainty estimated on the background.

Several categories of uncertainty sources have been studied, which are considered here in turn.

\vspace{0.2cm} {\bf{$\bfWW$ Production: Initial State Radiation (ISR)}}
\vspace{0.2cm}

ISR plays a key role in the $\W$ mass analysis as it is one of
the main sources of the bias on the fitted result with respect to the
true value. This bias, which is removed by calibrating the fits with the simulation, is due to the energy-momentum conservation constraint used in
the kinematical constrained fits. The ISR is computed in the {\tt YFS}
exponentiation approach, using a leading logarithm (LL)
${\mathcal{O}}(\alpha^3)$ matrix element.

The difference between the best result, obtained from implementing the
${\mathcal{O}}(\alpha^3)$ ISR matrix element, and the
${\mathcal{O}}(\alpha^2)$ one provides an estimate of the effect
of missing the matrix element for higher orders. The missing higher orders 
lead to the use of a wrong description for events with more than three hard 
photons or more than one photon with high $p_t$.  

The difference between the best result and the $\mathcal{O}(\alpha)$ result
includes the previous study, and can be used as an estimate of the upper limit
of the effect of missing the non-leading logarithm (NLL) terms at
${\mathcal{O}}(\alpha^2)$; this effect of missing NLL terms is expected to be smaller than the effect from the LL terms given by this ${\mathcal{O}}(\alpha^3)$ to ${\mathcal{O}}(\alpha)$ difference.

Also taking into account the study performed in~\cite{wmasssys}, the
ISR related uncertainty can be conservatively estimated at 1~$\MeVm$ for
the mass and 2~$\MeVm$ on the width.

\vspace{0.2cm} {\bf{$\bfW$ Decay: Final State Radiation (FSR)}}
\vspace{0.2cm}

The FSR description and uncertainty is tightly linked to the final
state considered. QED FSR from quarks is embedded in the parton shower
describing the first phase of the hadronisation process. It is
therefore essentially impossible to separate it from the rest of the
hadronisation process, and the related uncertainty is considered as
included in the jet and fragmentation related systematics.

FSR from leptons is described by {\tt PHOTOS}. The difference between
the best result, based on the NLL treatment, and the LL one can give
an estimate of the effect of the missing part of the
$\mathcal{O}(\alpha)$ FSR correction. While the result depends on the 
semi-leptonic channel, the difference is always less than $1~\MeVm$.

In~\cite{wmasssys} the effect of the missing higher orders beyond
${\mathcal{O}}(\alpha^{2})$ has been found to be negligible at generator
level.  Simple perturbative QED considerations suggest that the size
of the effect should not exceed the size of the effect from the missing part of the $\mathcal{O}(\alpha)$ FSR correction; therefore conservatively the $1~\MeVm$ can be doubled to take into account both of these components of the uncertainty.

\vspace{0.2cm} {\bf{Non-factorizable QED Interference: NF
    $\mathbf{\mathcal{O}(\alpha)}$ Corrections}} \vspace{0.2cm}

Non-factorizable ${\mathcal{O}}(\alpha)$ QED interference between $\W$s
is effectively implemented through the so-called Khoze-Chapovsky~\cite{KC} (KC)
ansatz.

The effect of using the KC ansatz with respect to the Born calculation, 
where this interference is not described, can be considered as an upper
limit of the missing part of the full ${\mathcal{O}}(\alpha)$
calculation and of the higher order terms. A dedicated study shows
that the effect is less than $2~\MeVm$ for all the measurements.

\vspace{0.2cm} {\bf{Ambiguities in Leading Pole Approximation (LPA)
    definition: Non Leading (NL) $\mathbf{{\mathcal{O}}(\alpha)}$ Corrections}}
\vspace{0.2cm}

Two sources of uncertainties are considered, following the study
in~\cite{wmasssys}. The effect of missing higher orders can be, at least
partly, evaluated by changing the electroweak scheme used in the
$\mathcal{O}(\alpha)$ calculation. This essentially means changing the
definition of the QED fine structure constant used in the
$\mathcal{O}(\alpha)$ matrix element. The effect is very small, at the
limit of the fit sensitivity, both for the mass and the width.

The second, more relevant, source of uncertainty connected to the LPA
is in its possible definitions, i.e. the ambiguity present in the way of
expanding the amplitude around the double resonant $\W$ pole. The
standard {\tt YFSWW} uses the so called $\mbox{LPA}_A$ definition; a
comparison with the $\mbox{LPA}_B$ one can give an estimate of the
effect from the intrinsic ambiguity in the LPA definition. A
dedicated study has been performed evaluating the difference:

\[
\Delta {\mathcal{O}}(\alpha) (\mbox{LPA}_A - \mbox{LPA}_B)  =  
\Delta (\mbox{Best LPA}_A - \mbox{no NL LPA}_A) - 
\Delta (\mbox{Best LPA}_B - \mbox{no NL LPA}_B) 
\]

\noindent in order to evaluate only the effect of the different scheme on the
radiative corrections (and not at Born level).  The size of the effect
is less than 1~$\MeVm$ for the mass and less than 4~$\MeVm$ for the width.

\vspace{0.2cm} {\bf{Radiative Corrections on 4-$\mathbf{f}$ Background
    Diagrams: Single $\W$}} \vspace{0.2cm}

The Double Pole Approximation~(DPA) is known to be valid within a few W widths of the double resonant pole. 
The DPA correction is applied only to the \CCTHREE\
part of the matrix element (and partly to the interference,
see~\cite{delphi4fgen}); non-\CCTHREE\ diagrams contributions are not
directly affected by the DPA uncertainty (except for possible effects
in the interference term which is relevant for the electron channel).

It is clear that this procedure still leaves the problem of the approximated
radiative corrections treatment for the non-\CCTHREE\ part of the matrix
element (and the interference).  The ISR studies previously discussed
can reasonably cover the most relevant part of the electroweak
radiative corrections uncertainties present also for the $\WW$-like
4-$f$ background diagrams, e.g. the non-\CCTHREE\ part. There is, however, a
notable exception: the so called single $\W$ diagrams
for the $q\bar{q'}e\nu$ final state.

The bulk of single $\W$ events are rejected in the $\W$ mass and width
analysis, since the electron in these events is lost in the beam pipe.
But the \CCTHREE\ - single $\W$ interference is sizeable, and it has a strong
impact on the $\W$ mass result in the electron channel.  The situation
is different in the $\W$ width analysis, where in $\enqq$ events
reconstructed by the electron analysis the effects of non-\CCTHREE\ diagrams and the \CCTHREE\ - non-\CCTHREE\ interference are opposite in sign and almost
completely cancel.

The situation is made even more complex by the cross-talk between
channels, e.g. events belonging in reality to one channel but
reconstructed as belonging to another one. This cross-talk is
particularly relevant between semi-leptonic electron and tau decays, 
and this explains why the $\tau$ channel analysis is also sensitive to this 
uncertainty source.

The effect of this uncertainty has been studied in two ways. Firstly, since the
uncertainty on the single $\W$ rate associated to radiative corrections
is known in literature to be about $4\%$, the non-\CCTHREE\ part of the matrix element, assumed to be dominated by the single $\W$ contribution, has been varied by $4\%$ for $q\bar{q'}e\nu$ final states.  Another possible source of
uncertainty related to 4-$f$ background is estimated by partly
applying the DPA correction to the interference term (see the
discussion in~\cite{delphi4fgen}). The effect of this way of computing
the corrections can be considered as another estimate of the
uncertainty related to the 4-$f$ background presence.

The maximal size of these effects is about 6~$\MeVm$ (for the mass in
$qqe\nu$  and the width in $qq\tau\nu$).

\vspace{0.2cm} {\bf{Total Uncertainty}} \vspace{0.2cm}
  
  The results of all the studies presented are combined in a
  single uncertainty for each channel. Tables~\ref{tab:deltamw}
  and~\ref{tab:deltagw} present the estimates for the mass and width from the different sources
  of uncertainties discussed above.

\begin{table}[hbt]
\begin{center}
\begin{tabular}{|l|c|c|c|c|} 
\hline 
\multicolumn{5}{|c|}{$\mw$ Electroweak Correction Systematic Errors ($\MeVm$)} \\
\hline 
Uncertainty Source & $\enqq$ & $\mnqq$ & $\tnqq$ & $\qqqq$ \\
\hline
ISR                      & 1.0 & 1.0 & 1.0 & 1.0 \\
FSR                      & 0.5 & 0.5 & 1.0 & - \\
NF ${\mathcal{O}}(\alpha)$ & 1.0 & 1.0 & 1.0 & 2.0 \\
NL ${\mathcal{O}}(\alpha)$ & 1.0 & 1.0 & 1.0 & 1.0  \\
4-$f$ Background         & 5.5 & 0.5 & 1.0 & 0.5 \\   
\hline
Total & 9 & 4 & 5 & 4.5 \\
\hline
\end{tabular}
\caption{Summary of the systematic uncertainties on the $\W$ mass due to electroweak corrections. The
  total is computed adding linearly the absolute values of all the
  contributions.}
\label{tab:deltamw}
\end{center}
\end{table}

\begin{table}[hbt]
\begin{center}
\begin{tabular}{|l|c|c|c|c|} 
\hline 
\multicolumn{5}{|c|}{$\gw$ Electroweak Correction Systematic Errors ($\MeVm$)} \\
\hline 
Uncertainty Source & $\enqq$ & $\mnqq$ & $\tnqq$ & $\qqqq$ \\
\hline
ISR                      & 2.0 & 2.0 & 2.0 & 2.0 \\
FSR                      & 1.0 & 1.0 & 2.0 & - \\
NF ${\mathcal{O}}(\alpha)$ & 2.0    & 2.0 & 2.0 & 2.0 \\
NL ${\mathcal{O}}(\alpha)$ & 4.0    & 4.0 & 4.0 & 4.0  \\
4-$f$ Background         & 2.0  & 1.0 & 6.0 & 1.0 \\   
\hline
Total & 11 & 10 & 16 & 9 \\
\hline
\end{tabular}
\caption{Summary of the systematic uncertainties on the $\W$ width due to electroweak corrections. The
  total is computed adding linearly the values of all the
  contributions.}
\label{tab:deltagw}
\end{center}
\end{table}

The total uncertainty per channel is conservatively computed summing linearly the values of the contributions. All the numbers have been rounded to 0.5~$\MeVm$.

Reference ~\cite{delphi4fgen} also reports a comparison of {\tt YFSWW} with the other completely independent Monte Carlo generator {\tt RacoonWW}~\cite{racoonww} which implements radiative corrections in the DPA. This study has not been directly used in the error estimation presented here due to the limitations in the treatment of non-collinear radiation in {\tt RacoonWW}. However, this study does provide additional confidence in the validity of the {\tt YFSWW} calculation.

As can be seen, the uncertainty on the $\W$ mass associated with the electroweak radiative corrections is found to be less than 10~$\MeVm$.

\subsection{LEP Collision Energy}

\label{sec:systebeam}

The average \LEP\ collision energy is evaluated at 15 minute intervals of running or after significant changes in the beam energy. The measured centre-of-mass energy is imposed as a constraint in the kinematic fit, and hence the relative error on the collision energy translates to approximately the same fractional error on the $\W$ mass determination. The effect of the uncertainty on the $\W$ width determination is negligible.

The beam energy is estimated using the \LEP\ energy model, discussed in section~\ref{sec:lep} based on 16 NMR probes in dipole magnets around the \LEP\ ring calibrated with the RDP technique. The compatibility of three cross-check methods with this determination was used to determine a set of small energy offsets. The relative size of this offset was energy dependent,  rising to a maximum of $1.6 \times 10^{-5}$ at 207~$\GeV$ centre-of-mass energy.

The \LEP\ energy working group also assessed the uncertainties in the collision energies and supplied these in the form of a 10$\times$10 correlation matrix. 
The uncertainties increase as the collision energy increases, due to the fact that higher energies are further from the RDP normalisation region.
The errors are given in  table~\ref{tab:ebcombres}. 
At 183~$\GeV$ centre-of-mass energy the uncertainty
on the collision energy is 20.3~$\MeV$. This rises to 23.7~$\MeV$ at 202~$\GeV$. For the
energy points at values of 205 and 207~$\GeV$, taken in the year 2000, 
there is an additional uncertainty due to the `Bending Field Spreading' 
strategy, in which the corrector magnets were powered in a coherent manner
to increase the overall dipole field and thus the \LEP\ energy \cite{lepener} . This leads to a larger error for the year 2000.
For the energy points at 161 and 172~$\GeV$, taken in the year 1996, there
is also a small increase in the error, compared to 183~$\GeV$, due to
increased uncertainties in the NMR calibration for this year.

\begin{table}
\begin{center}
\begin{tabular}{|l|cccccccccc|} \hline
             & \multicolumn{10}{|c|}{ $\sqs$ \ Nominal [$\GeV$] } \\ \cline{2-11}
             & 161  & 172  & 183  & 189  & 192  & 196  & 200  & 202  & 205   & 207   \\ \cline{2-11}
E$_{\rm{cm}}$ Error  [$\MeV$]  &  25.4 &  27.4 & 20.3 & 21.6 & 21.6 & 23.2 & 23.7 &  23.7 &  36.9 &  41.7 \\ \hline
\end{tabular}
\end{center}
\caption[]{Uncertainties on the \LEP\ energies for the different centre-of-mass energy points.
}
\label{tab:ebcombres}
\end{table}

The mean energy difference between the electron and positron beams is less than 4~$\MeV$ at all energies and hence the effect on the $\W$ mass or width determination is negligible. The momentum spread of the electrons or positrons in a bunch gives rise to a variation in the centre-of-mass energy of the collisions and boost of the centre of mass frame with respect to the laboratory frame. The spreads in centre-of-mass collision energies have been evaluated by the LEP energy working group \cite{lepener} and range from 144 to 265~$\MeV$. The corresponding effects for the $\W$ mass and width analyses are negligible.

\subsection{Aspect Ratio}

The aspect ratio is defined as the ratio of the length to the width of the detector. As all the sub-detectors of DELPHI are aligned with respect to the Vertex Detector, the knowledge of the aspect ratio is limited by the precision to which the position and dimensions of the Vertex Detector can be measured. The effect of a mismeasurement of the aspect ratio is to introduce a bias on the measurement of the polar angle, $\theta$. As the $\W$ boson production polar angle is not isotropic but forward peaked, a mismeasurement of the aspect ratio would result in a small bias on the average opening angle of the $\W$ decay products, and hence induce a small bias on the reconstructed $\W$ mass.  

The correspondence of hits in the overlapping silicon modules is sensitive to a misalignment of the Vertex Detector. In fact the study of these overlaps constitutes an essential part of the procedure for the alignment of the Vertex Detector. From this study, discussed further in \cite{radreturnDELPHI}, it is concluded that a reasonable estimate of the aspect ratio uncertainty is $3\times10^{-4}$. Such a bias would result in a shift in $\W$ mass below 1~$\MeVm$ for the semi-leptonic channel, and of 2~$\MeVm$ for the fully-hadronic one. The effect on the $\W$ width is negligible.

\subsection{Background Description}

The background events for the $\W$-pair selection are from four-fermion or hadronic two fermion processes.
 
The four-fermion background uncertainty is studied and described in the
electroweak corrections uncertainties (section~\ref{sec:radcorr}) and in the jet description studies (section~\ref{sec:jetdesc}) parts of this paper. 

The dominant source of background to $\W$ pair
production, both in the semi-leptonic and in the fully-hadronic channel,
is from $\Z \rightarrow q\bar{q}(\gamma)$ events.  

In the semi-leptonic channel the 2-fermion background is relatively small with the main uncertainty in its rate arising from the discrepancy
between data and simulation in the rate of misidentification of
energetic photons (from radiative return to the $\Z$ peak events) as electrons. 
This misidentification is mainly due to the electron-positron conversion of photons and the spurious associations of forward vertex detectors hits to an electromagnetic cluster in the calorimeter. A data-simulation comparison shows that a
10\% fluctuation of the background is possible without significantly
degrading the agreement between the data and simulation. The theory uncertainty on the 2-fermion cross-section is  generally small, 
in the worst case at the 2\% level~\cite{2f-therr}.
 
In the fully-hadronic channel the 2-fermion background is more
important, and the major contribution to the uncertainty is 
from the four-jet final state production mechanism. The study
performed in~\cite{wwxsec} has shown that the maximal difference in
the estimated 2-fermion background rate is 10\%  coming from changing from
{\tt PYTHIA} to {\tt HERWIG} as the hadronisation model, with the {\tt
  ARIADNE} model giving intermediate results. The effect on the $\W$ mass is 13~$\MeVm$ at $\sqs = 189~\GeV$, and 4~$\MeVm$ at $\sqs = 206.5~\GeV$, while the effect on the $\W$ width is 40~$\MeVm$ over the whole range of centre-of-mass energies.

In summary, applying a variation of $\pm10\%$ on the $\Z \rightarrow q\bar{q}(\gamma)$ event rate is used to provide an estimate of the systematic uncertainty on the background level for both the semi-leptonic and fully-hadronic channel mass and width measurements. This variation also covers any discrepancies seen in the data and simulation comparison plots shown in this paper.

The importance of the background event mass distribution has also been investigated. In the semi-leptonic analyses the mass distribution taken from the simulation has been replaced with a constant level and half of the variation in the result has been taken as a systematic. In the fully-hadronic channel this systematic was assessed by changing the generator used for the background between {\tt PYTHIA}, {\tt HERWIG} and {\tt ARIADNE}. 

The background level and background shape uncertainties were added in quadrature and the resulting errors are reported in tables \ref{tab:systmw189}, \ref{tab:systmw205} and \ref{tab:systgw} below.

\subsection{Bose-Einstein Correlations}
\label{sec:bec}

Correlations between final state hadronic particles are dominated by 
Bose-Einstein Correlations (BEC), a quantum mechanical effect 
which enhances the production of identical bosons close in phase space.
The net effect is that multiplets of identical bosons are produced with
smaller energy-momentum differences than non-identical ones.

BEC for particles produced from the same $\W$ boson affect the 
normal fragmentation and are therefore treated implicitly in the 
fragmentation uncertainties which are constrained by the large 
amount of $\Z$-data. BEC for pairs of particles coming from different $\W$s 
cannot be constrained or safely predicted by the information from 
single hadronically decaying vector bosons. 

A dedicated and model-independent measurement of the BEC effect  
was performed by the \DELPHI\ collaboration 
in~\cite{dbec} while other \LEP\ experiments have
made similar measurements~\cite{lbec}. Comparing these results
with Monte Carlo models constitutes the only way to estimate potential 
systematic uncertainties from BEC. The \LUBOEI\ model BE$_{32}$~\cite{luboei}
was found to give the largest shift in the measured value of $\mw$  
for a given amount of BEC. Other models give smaller shifts and some
models predict no appreciable BEC shifts at all. 
It was decided not to apply any corrections due to BEC and evaluate 
the systematic error as the largest predicted
shift consistent with the \DELPHI\ data. 
The predicted shift plus one standard deviation of its error is used as the estimator of the systematic error. 

The \DELPHI\ result for BEC is a 2.4 standard deviation evidence for BEC 
between different $\W$s and a correlation strength, $\Lambda$, which can be 
compared to the BE$_{32}$ prediction at the same effective correlation length scale:

\begin{equation} 
\Lambda_{\rm data}\Big/{\Lambda_{\rm BE_{32}}} = 0.55 \pm 0.20 
(\rm Stat.) \pm 0.11 (\rm Syst.).
\end{equation}

The predicted mass shift, BEC inside $\W$s only $-$ BEC inside and between $\W$s,  using BE$_{32}$ (with model parameters ${\rm PARJ(92)}=1.35$ and ${\rm PARJ(93)}=0.34$) is 40$\pm$10~$\MeVm$ for the standard mass analysis, 33$\pm$11~$\MeVm$ for the cone jet mass reconstruction analysis and $-17 \pm 20~\MeVm$ for the $\W$ width analysis.
The observed mass shift in BE$_{32}$ is linear in the observed
correlation, $\Lambda_{\rm BE_{32}}$. Applying the one standard deviation upper bound of the correlation parameter this translates into a systematic error of 31~$\MeVm$ from BEC for the standard analysis and 26~$\MeVm$  for the cone analysis. A systematic error of 20~$\MeVm$ is applied for the $\W$ width. The mass and width shifts were evaluated with the simulation model over the full range of centre-of-mass energies and no energy dependence was observed. The shifts reported are the average values. Conservatively, these errors are applied as symmetric uncertainties.

The combined \DELPHI\ BEC measurements of the correlation strength and effective correlation length scale suggest that the between-$\W$ BEC occur with an effective correlation length scale which is larger that the one
predicted by BE$_{32}$. If this is the case, the number of pairs 
effectively affected by the BEC is reduced and also the effect 
per pair is diminished. Furthermore, the other \LEP\ experiments have reported smaller values of $\Lambda_{\rm data}\Big/{\Lambda_{\rm BE_{32}}}$ than that observed by \DELPHI. Hence the systematic uncertainties applied in this analysis are considered conservative.

\subsection{Colour Reconnection}
\label{sec:cr}

In the reaction $e^+ e^- \rightarrow \WW \rightarrow
(q_1\bar{q}_2)(q_3\bar{q}_4)$ the hadronisation models used for this
analysis treat the colour singlets $q_1\bar{q}_2$ and $q_3\bar{q}_4$
coming from each $\W$ boson independently. However, interconnection effects
between the products of the two $\W$ bosons may be expected since the lifetime of the $\W$ bosons ($\tau_{\W} \simeq \hbar/\gw \simeq 0.1~\mbox{fm}/\it{c}$) is an order of magnitude smaller than the typical hadronisation times. 

The exchange of coloured gluons between partons from hadronic systems
from different $\W$ bosons can induce the so-called colour reconnection
(CR) effect in the development of the parton shower. This
effect can in principle distort the properties of the final hadronic
system and therefore affect the $\W$ mass measurement, if not properly accounted 
for in the simulation.

At perturbative level the effects are expected to be small~\cite{phLEP2},
and the impact on the reconstructed $\W$ mass has been evaluated to be at most $5~\MeVm$. However, CR effects can be large at hadronisation level, due to the large numbers of soft gluons sharing the space-time region. These effects have been studied by introducing CR effects into hadronisation models and comparing with \DELPHI\ data and are reported in \cite{delphicr}.

The most studied model, and the one used for the evaluation of the
systematic uncertainty on the $\W$ mass and width measurement, is the
Sj\"{o}strand-Khoze ``Type 1'' model (\SKI)~\cite{perturb}. This
model of CR is based on the Lund string fragmentation phenomenology:
the strings are considered as colour flux tubes with some volume, and
reconnection occurs when these tubes overlap. The probability of
reconnection in an event,$P_{reco}$, is parameterised by the value $\kappa$,
according to the volume of overlap between the two strings
$V_{overlap}$:
\begin{equation}
  \label{eq:sk1prob}
  P_{reco} = 1-e^{- \kappa V_{overlap}}.
\end{equation}
The parameter $\kappa$ determines the reconnection probability. By
comparing the data with the model predictions evaluated at several
$\kappa$ values it is possible to determine the value most consistent
with the data and extract the corresponding reconnection probability. 

Another model has been developed by the same authors (\SKII') and also implemented in \PYTHIA\ but is found to predict a smaller shift on the reconstructed $\W$ mass than \SKI\ for the same reconnection probability.

Further CR models are available in the \HERWIG\ and \ARIADNE\ Monte Carlo programs. In \ARIADNE, which implements an adapted version of the Gustafson-H\"akkinen model \cite{gh}, the model used \cite{ar2} allows for reconnections between partons originating in the same $\W$ boson, or from different $\W$ bosons if they have an energy smaller than the width of the $\W$ boson. The mass shift from CR is evaluated from the difference between the shift when the reconnections are made only in the same $\W$ boson and when the full reconnections are made. In the standard \DELPHI\ analysis, the shift was found to be $11 \pm 11$~$\MeVm$.

In \HERWIG\ the partons are reconnected, with a reconnection probability of 1/9, if the reconnection results in a smaller total cluster mass. The shift in the reconstructed $\W$ mass at 189~$\GeV$ centre-of-mass energy was found to be $29 \pm 7~\MeVm$, the same shift as obtained from a $\kappa$ value of 0.29 in the \SKI\ model.

\DELPHI\ has performed two analyses to compare these simulation models with data which are described in detail in~\cite{delphicr}.

The first one is based on the measurement of the particle flow between
the jets in a four jets $\WW$ event. On a subsample of strictly four-jet events two regions can be defined, the region between jets from the same $\W$ (called inside-$\W$ regions) and the region between jets from different $\W$ bosons (called between-$\W$ regions). The
ratio $R$ of the particle fluxes in the inside-$\W$ and between-$\W$ regions
(limiting the analysis to the central part of these regions) is an observable sensitive to  CR effects. The comparison of the flux measured in
real data with the prediction of the \SKI\ model as a function of
$\kappa$ allows the value to be determined which is most consistent with data, and
its uncertainty.

The second method used exploits the observation  that in the direct reconstruction analysis of the $\W$ mass, different
$\W$ mass estimators have different sensitivities to CR effects. 
As discussed in section~\ref{sec:cone} removing particles from the inter-jet regions reduces the sensitivity to CR effects and hence can be used to measure the CR effect. The correlation between the measurement of the mass shift (using the standard or cone jet reconstruction techniques) and the measurement of the mass from these techniques is only $11\%$.

From the combination of these two analyses and in the framework of the \SKI\ model, the value of the $\kappa$ parameter most compatible with the data is found to be~\cite{delphicr}:

\[ \kappa =2.2\pm^{2.5}_{1.3}. \]

\begin{figure}[htp]
  \begin{tabular}{c}
    \epsfig{file=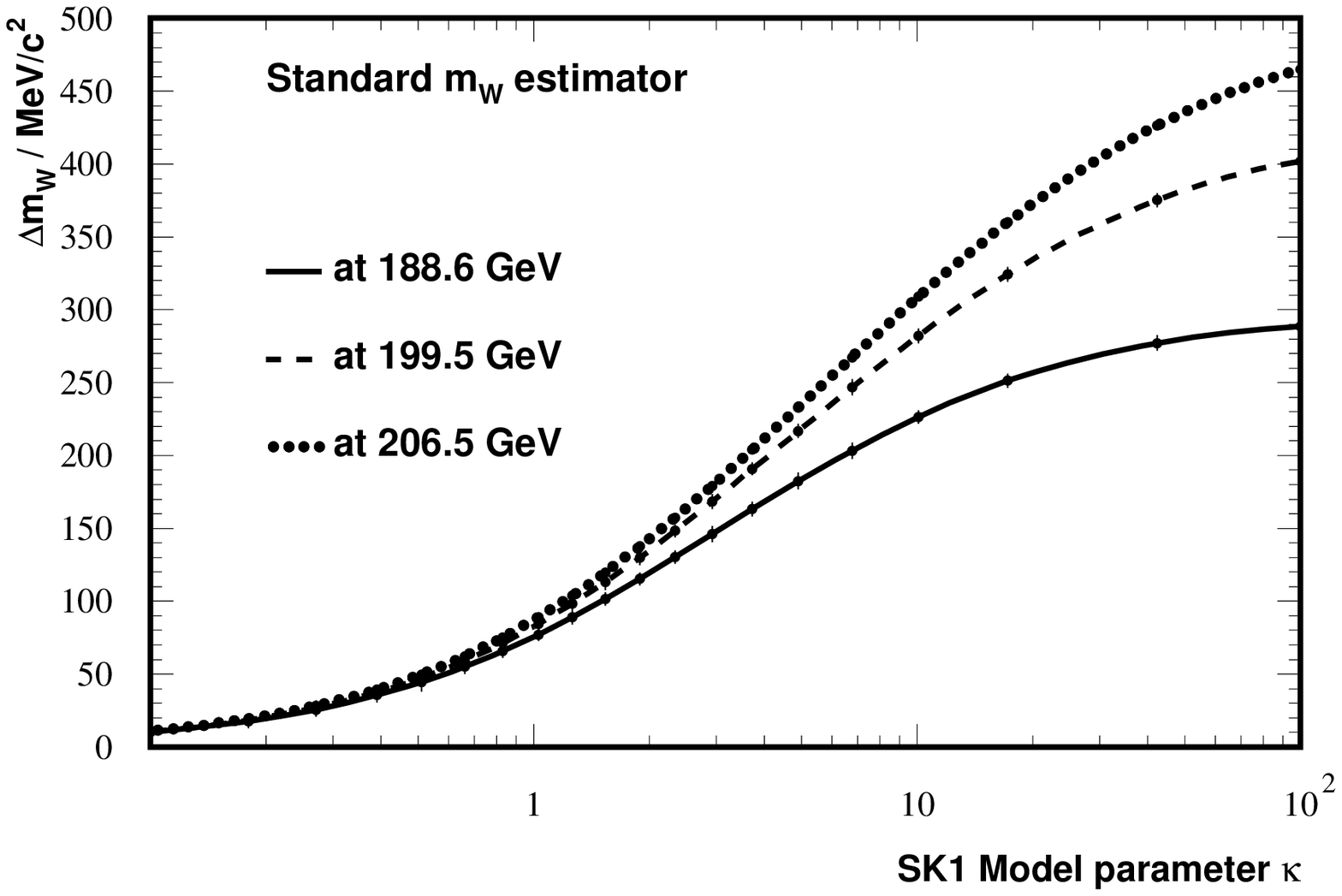,width=\textwidth} \\
    \epsfig{file=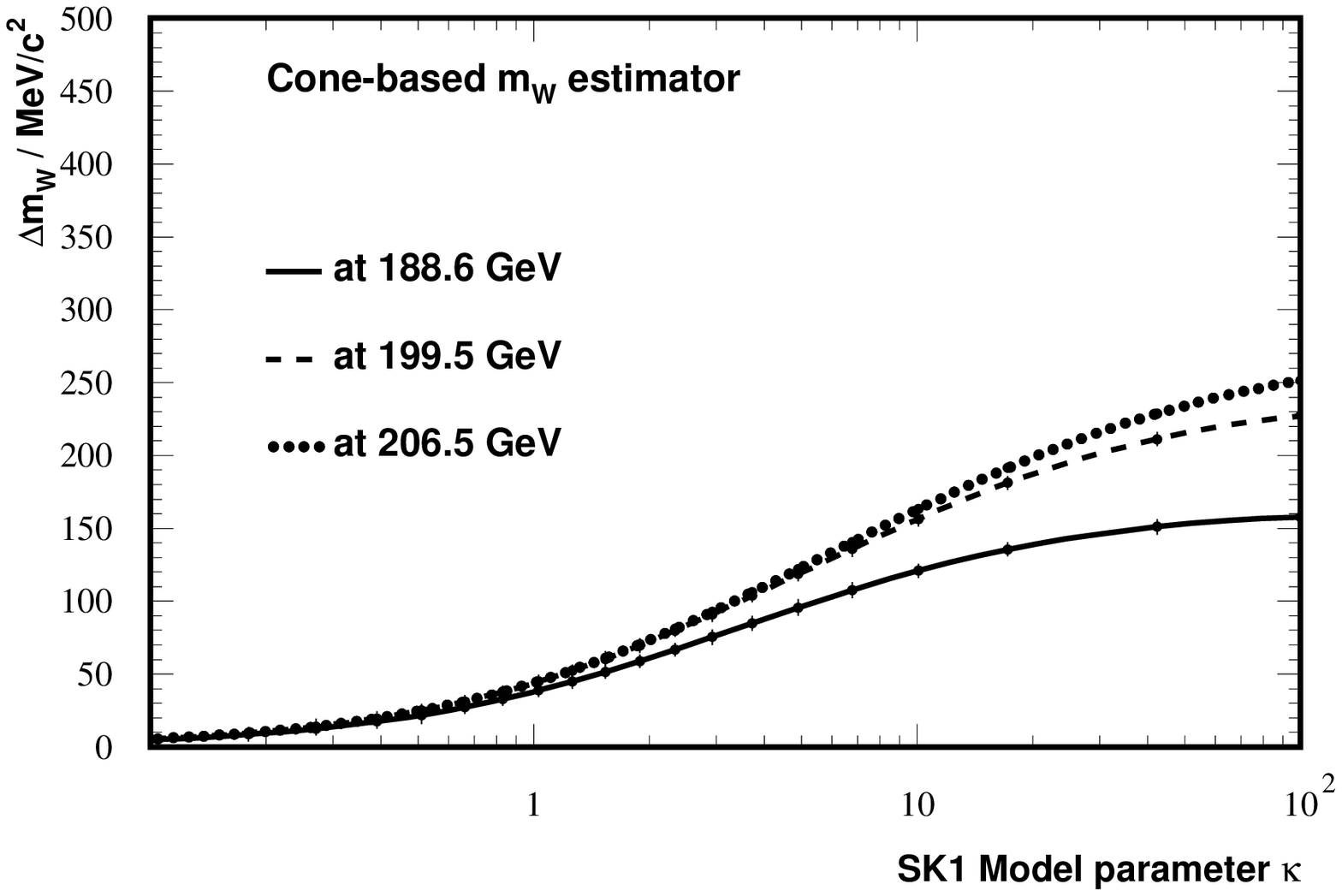,width=\textwidth}
  \end{tabular}
  \caption{$\W$ mass shift caused by the colour reconnection effect as described in the \SKI\ model plotted as a function of the model parameter $\kappa$ which controls the fraction of reconnected events. The upper plot is for the standard $\W$ mass analysis and the lower plot when the cone jet reconstruction technique is applied.}
\label{fig:sk1mass}
\end{figure}

\begin{figure}[htbp]
\epsfig{file=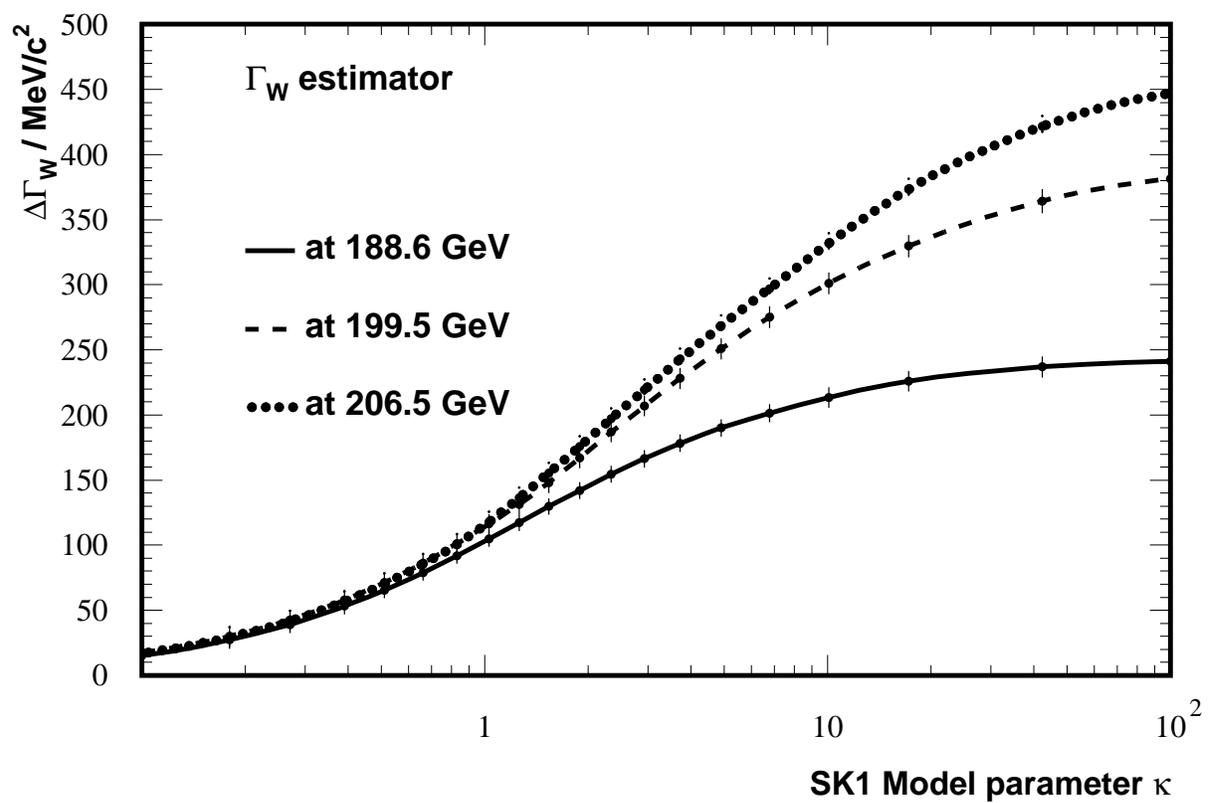,width=\textwidth}
  \caption{$\W$ width shift caused by the colour reconnection effect as described in the \SKI\ model plotted as a function of the model parameter $\kappa$ which controls the fraction of reconnected events.}
\label{fig:sk1width}
\end{figure}

The CR shift in the reconstructed $\W$ mass as a function of the \SKI\ $\kappa$ parameter is provided as figure~\ref{fig:sk1mass},  the results of the standard and cone jet reconstruction techniques are indicated. Figure~\ref{fig:sk1width} shows the CR shift for the $\W$ width reconstruction analysis. 
 
The systematic uncertainty on the $\W$ mass and width is calculated using the one standard deviation upper bound of $\kappa$ of 4.7. As reported above, this systematic error is considerably larger than that which would be evaluated from the $\ARIADNE$ or $\HERWIG$ CR models. Furthermore, this value of $\kappa$ is larger than that reported by the other \LEP\ experiments \cite{lepcr}. The CR $\W$ mass shift is dependent on the centre-of-mass energy in the \SKI\ model as shown in figures~\ref{fig:sk1mass} and \ref{fig:sk1width}. However, we prefer not to rely on the centre-of-mass energy evolution of the \SKI\ CR shift (leading to a change in relative weights when averaging the results from different centre-of-mass energies) and instead choose to quote the systematic errors at 200~$\GeV$ (close to the average centre-of-mass energy of the data). In light of the significant range of CR effect estimates no correction is made to the $\W$ mass or width results and for simplicity a symmetric systematic uncertainty is applied. The corresponding systematics uncertainties on the $\W$ mass are $212~\MeVm$ (standard), $116~\MeVm$ (cone jet reconstruction) and $247~\MeVm$ for the $\W$ width analysis. 

 \begin{table}[ht]
 \begin{center}
  \begin{tabular}{|l|ccc|c|}     \hline
\multicolumn{5}{|c|}{$\mw$  Systematic Errors ($\MeVm$) at 189 $\GeV$} \\ \hline
Sources of Systematic Error       & $\enqq$ & $\mnqq$ & $\tnqq$ & $\qqqq$ \\ \hline \hline
 Statistical Error on Calibration &  12     & 10      & 15      &  4  \\
 Lepton Corrections               &  24     & 16      & -       &  -  \\
 Jet Corrections                  &  18     & 15      & 19      & 24  \\
 Fragmentation                    &  10     & 10      & 13      & 12  \\
 Electroweak Corrections           &   9     &  4      & 5       & 5   \\
 Background                       &   5     &  1      &  12     & 17    \\
\hline
 LEP Energy                       &   9    &  9      &  9      &  9   \\
\hline
 Bose-Einstein Correlations       & -       & -       & -       &  31/26    \\ 
 Colour Reconnection              & -       & -       & -       &  212/116 \\
\hline
\end{tabular}
  \caption{Contributions to the systematic error on the $\W$ mass measurement for data taken at a nominal centre-of-mass energy of 189~$\GeV$. Where two uncertainties are reported in the $\qqqq$ analysis column the first corresponds to the standard analysis and the second to the cone jet reconstruction analysis. }
  \label{tab:systmw189}
 \end{center}
\end{table}

\begin{table}[ht]
 \begin{center}
  \begin{tabular}{|l|ccc|c|}     \hline
\multicolumn{5}{|c|}{$\mw$  Systematic Errors ($\MeVm$) at 205 $\GeV$} \\ \hline
Sources of Systematic Error       & $\enqq$ & $\mnqq$ & $\tnqq$ & $\qqqq$ \\ \hline 
 Statistical Error on Calibration &  15     & 10      & 17      &   4  \\
 Lepton Corrections               &  25     & 21      &    -    &   -  \\
 Jet Corrections                  &  26     & 21      & 33      & 28   \\
 Fragmentation                    &  10     & 10      & 13      & 12  \\
 Electroweak Corrections          &   9     &  4      &  5      &  5  \\
 Background                       &   4     &  6      & 19      &  5  \\
\hline
 LEP Energy                       &  15     &  15     &  15     & 15   \\
\hline
 Bose-Einstein Correlations       & -       & -       & -       & 31/26 \\ 
 Colour Reconnection              & -       & -       & -       & 212/116  \\ \hline
\end{tabular}
  \caption{Contributions to the systematic error on the $\W$ mass measurement for data taken at a nominal centre-of-mass energy of 205~$\GeV$. Where two uncertainties are reported in the $\qqqq$ analysis column the first corresponds to the standard analysis and the second to the cone jet reconstruction analysis.}
  \label{tab:systmw205}
 \end{center}
\end{table}

\begin{table}[ht]
 \begin{center}
  \begin{tabular}{|l|c|c|}     \hline
\multicolumn{3}{|c|}{$\gw$  Systematic Errors ($\MeVm$) at 205 $\GeV$} \\ \hline
Sources of Systematic Error       & $\lnqq$ & $\qqqq$ \\ \hline 
 Statistical Error on Calibration &  15     &   9 \\
 Lepton Corrections               &  48     & -  \\
 Jet Corrections                  &  38     & 169 \\
 Fragmentation                    &  29     &   8  \\
 Electroweak Corrections          &  11     &   9  \\
 Background                       &  43     &  51  \\
\hline
 Bose-Einstein Correlations       & -   & 20   \\ 
 Colour Reconnection              & -   & 247  \\ \hline
\end{tabular}
  \caption{Contributions to the systematic error on the $\W$ width measurement for data taken at a nominal centre-of-mass energy of 205~$\GeV$.}
  \label{tab:systgw}
 \end{center}
\end{table}

\section{Results}
\label{sec:res}
\label{sec:comb}

The results of the analyses and the final combinations of these results are presented in this section. The results are obtained at a range of nominal centre-of-mass energies and in the four event selection channels. Combined results are obtained from an average of these results and also an average with the previously published \DELPHI\ data \cite{delpaper161,delpaper172} that have not been reanalysed in this paper.  

Subdividing the results by data-taking years and nominal centre-of-mass energies enables a proper treatment of the correlated systematic uncertainty from the \LEP\ collision energy and other dependences on the centre-of-mass energy or
data-taking period.  A detailed breakdown of the sources of systematic uncertainty, as shown in tables~\ref{tab:systmw189},\ref{tab:systmw205} and \ref{tab:systgw}, is provided for each result and the correlations specified. 

The combination is performed and the evaluation of the components of
the total error assessed using the Best Linear Unbiased Estimate (BLUE) technique \cite{lyons}.

\subsection{$\W$ Mass}

  The $\W$ mass is extracted separately in the analyses designed to select the $\enqq$, $\mnqq$ and $\tnqq$ decay channels. The values obtained are given in table~\ref{tab:qqlvResult} for the analysed centre-of-mass collision energies. The semi-leptonic channel analysis results are combined into a single $\lnqq$ value for each year of data taking. When performing these combinations the following sources of systematic uncertainty are taken as fully-correlated between lepton channels and between years: electroweak corrections, fragmentation, jet corrections, lepton corrections, background. The \LEP\ energy measurement correlations are taken from the matrix supplied in \cite{lepener}. The simulation calibration statistics are taken as uncorrelated. 

   The $\W$ mass is also obtained from the $\qqqq$ channel using both the standard and cone jet reconstruction technique. The results obtained from these analyses are given in table~\ref{tab:qqqqResult}.

   In addition to the analyses presented in this paper, measurements of the $\W$ mass have also been made using the data collected in 1996.

\begin{table}[htb]
\begin{center}
\begin{tabular}{|l|l|l|c|}
\hline
Year & Energy & Channel & $\mw\ \GeVm$ \\ 
\hline
1996 &  172  & $\enqq$ & 80.450 $\pm$ 0.870(Stat.) $\pm$ 0.085(Syst.) $\pm$ 0.013(LEP)\\
1996 &  172  & $\mnqq$ & 80.560 $\pm$ 0.760(Stat.) $\pm$ 0.062(Syst.) $\pm$ 0.013(LEP)\\
1996 &  172  & $\lnqq$ & 80.510 $\pm$ 0.570(Stat.) $\pm$ 0.051(Syst.) $\pm$ 0.013(LEP)\\
\hline
1997 &  183  & $\enqq$ & 80.852 $\pm$ 0.411(Stat.) $\pm$ 0.034(Syst.) $\pm$ 0.009(LEP)\\
     &       & $\mnqq$ & 80.573 $\pm$ 0.331(Stat.) $\pm$ 0.024(Syst.) $\pm$ 0.009(LEP)\\
     &       & $\tnqq$ & 80.233 $\pm$ 0.396(Stat.) $\pm$ 0.025(Syst.) $\pm$ 0.009(LEP)\\
     &       & $\lnqq$ & 80.548 $\pm$ 0.216(Stat.) $\pm$ 0.024(Syst.) $\pm$ 0.009(LEP)\\
\hline
1998 &  189  & $\enqq$ & 79.848 $\pm$ 0.275(Stat.) $\pm$ 0.035(Syst.) $\pm$ 0.009(LEP)\\
1998 &       & $\mnqq$ & 80.238 $\pm$ 0.195(Stat.) $\pm$ 0.026(Syst.) $\pm$ 0.009(LEP)\\
1998 &       & $\tnqq$ & 80.055 $\pm$ 0.288(Stat.) $\pm$ 0.030(Syst.) $\pm$ 0.009(LEP)\\
1998 &       & $\lnqq$ & 80.096 $\pm$ 0.139(Stat.) $\pm$ 0.026(Syst.) $\pm$ 0.009(LEP)\\
\hline
1999 &  192  & $\enqq$ & 80.025 $\pm$ 0.789(Stat.) $\pm$ 0.036(Syst.) $\pm$ 0.009(LEP)\\
     &       & $\mnqq$ & 80.604 $\pm$ 0.467(Stat.) $\pm$ 0.028(Syst.) $\pm$ 0.009(LEP)\\
     &       & $\tnqq$ & 80.161 $\pm$ 0.664(Stat.) $\pm$ 0.033(Syst.) $\pm$ 0.009(LEP)\\
     &  196  & $\enqq$ & 80.391 $\pm$ 0.349(Stat.) $\pm$ 0.037(Syst.) $\pm$ 0.010(LEP)\\
     &       & $\mnqq$ & 80.024 $\pm$ 0.270(Stat.) $\pm$ 0.031(Syst.) $\pm$ 0.010(LEP)\\
     &       & $\tnqq$ & 80.269 $\pm$ 0.417(Stat.) $\pm$ 0.036(Syst.) $\pm$ 0.010(LEP)\\
     &  200  & $\enqq$ & 80.383 $\pm$ 0.365(Stat.) $\pm$ 0.037(Syst.) $\pm$ 0.010(LEP)\\
     &       & $\mnqq$ & 80.374 $\pm$ 0.282(Stat.) $\pm$ 0.032(Syst.) $\pm$ 0.010(LEP)\\
     &       & $\tnqq$ & 80.197 $\pm$ 0.438(Stat.) $\pm$ 0.040(Syst.) $\pm$ 0.010(LEP)\\
     &  202  & $\enqq$ & 80.193 $\pm$ 0.453(Stat.) $\pm$ 0.039(Syst.) $\pm$ 0.010(LEP)\\
     &       & $\mnqq$ & 80.120 $\pm$ 0.341(Stat.) $\pm$ 0.033(Syst.) $\pm$ 0.010(LEP)\\
     &       & $\tnqq$ & 81.399 $\pm$ 0.574(Stat.) $\pm$ 0.042(Syst.) $\pm$ 0.010(LEP)\\
     & 192-202 
             & $\lnqq$ & 80.296 $\pm$ 0.113(Stat.) $\pm$ 0.030(Syst.) $\pm$ 0.009(LEP)\\
\hline
2000 &  206  & $\enqq$ & 80.814 $\pm$ 0.267(Stat.) $\pm$ 0.040(Syst.) $\pm$ 0.016(LEP)\\ 
     &       & $\mnqq$ & 80.340 $\pm$ 0.193(Stat.) $\pm$ 0.032(Syst.) $\pm$ 0.016(LEP)\\
     &       & $\tnqq$ & 80.701 $\pm$ 0.272(Stat.) $\pm$ 0.042(Syst.) $\pm$ 0.016(LEP)\\
     &       & $\lnqq$ & 80.551 $\pm$ 0.136(Stat.) $\pm$ 0.034(Syst.) $\pm$ 0.016(LEP)\\
\hline
\end{tabular}
\vspace{0.2cm}
\caption{Measured $\W$ mass (in $\GeVm$) from the semi-leptonic decay channel analyses with the nominal centre-of-mass energies (in $\GeV$) of each data sample indicated. The values marked $\lnqq$ are the combined values of the three semi-leptonic channel analyses. The values obtained from the data recorded in 1996 and analysed in \cite{delpaper172} are also included.}
\label{tab:qqlvResult}
\end{center}
\end{table}

\begin{table}[htb]
\begin{center}
\begin{tabular}{|l|l|l|c|}
\hline
Year & Energy & Analysis & $\mw\ \GeVm$ \\
\hline
1996 &  172 & std  & 79.900 $\pm$ 0.590(Stat.) $\pm$ 0.050(Syst.) $\pm$ 0.214(FSI) $\pm$ 0.013(LEP)\\
\hline
1997 &  183 & std  & 80.137 $\pm$ 0.185(Stat.) $\pm$ 0.046(Syst.) $\pm$ 0.214(FSI) $\pm$ 0.009(LEP)\\
     &      & cone & 80.100 $\pm$ 0.191(Stat.) $\pm$ 0.046(Syst.) $\pm$ 0.119(FSI) $\pm$ 0.009(LEP)\\
\hline
1998 &  189 & std  & 80.519 $\pm$ 0.107(Stat.) $\pm$ 0.032(Syst.) $\pm$ 0.214(FSI) $\pm$ 0.009(LEP)\\
     &      & cone & 80.533 $\pm$ 0.119(Stat.) $\pm$ 0.032(Syst.) $\pm$ 0.119(FSI) $\pm$ 0.009(LEP)\\
\hline
1999 &  192 & std  & 80.711 $\pm$ 0.281(Stat.) $\pm$ 0.032(Syst.) $\pm$ 0.214(FSI) $\pm$ 0.009(LEP)\\
     &      & cone & 81.076 $\pm$ 0.294(Stat.) $\pm$ 0.032(Syst.) $\pm$ 0.119(FSI) $\pm$ 0.009(LEP)\\ 
     &  196 & std  & 80.248 $\pm$ 0.159(Stat.) $\pm$ 0.032(Syst.) $\pm$ 0.214(FSI) $\pm$ 0.010(LEP)\\
     &      & cone & 80.240 $\pm$ 0.192(Stat.) $\pm$ 0.032(Syst.) $\pm$ 0.119(FSI) $\pm$ 0.010(LEP)\\
     &  200 & std  & 80.274 $\pm$ 0.149(Stat.) $\pm$ 0.032(Syst.) $\pm$ 0.214(FSI) $\pm$ 0.010(LEP)\\
     &      & cone & 80.227 $\pm$ 0.164(Stat.) $\pm$ 0.032(Syst.) $\pm$ 0.119(FSI) $\pm$ 0.010(LEP)\\
     &  202 & std  & 80.537 $\pm$ 0.199(Stat.) $\pm$ 0.031(Syst.) $\pm$ 0.214(FSI) $\pm$ 0.010(LEP)\\
     &      & cone & 80.248 $\pm$ 0.231(Stat.) $\pm$ 0.031(Syst.) $\pm$ 0.119(FSI) $\pm$ 0.010(LEP)\\
     & 192-202
            & std  & 80.365 $\pm$ 0.090(Stat.) $\pm$ 0.032(Syst.) $\pm$ 0.214(FSI) $\pm$ 0.010(LEP)\\
     &      & cone & 80.339 $\pm$ 0.103(Stat.) $\pm$ 0.032(Syst.) $\pm$ 0.119(FSI) $\pm$ 0.010(LEP)\\
\hline
2000 &  206 & std  & 80.318 $\pm$ 0.092(Stat.) $\pm$ 0.032(Syst.) $\pm$ 0.214(FSI) $\pm$ 0.015(LEP)\\
     &      & cone & 80.171 $\pm$ 0.104(Stat.) $\pm$ 0.032(Syst.) $\pm$ 0.119(FSI) $\pm$ 0.015(LEP)\\
\hline
\end{tabular}
\vspace{0.2cm}
\caption{
Measured $\W$ mass (in $\GeVm$) from the fully-hadronic decay channel analysis with the nominal centre-of-mass energies (in $\GeV$) of each data sample indicated. Results are provided for both the standard (std) and cone jet reconstruction techniques applied. The value obtained from the data recorded in 1996 and analysed in \cite{delpaper172} is also included.}
\label{tab:qqqqResult}
\end{center}
\end{table}

\subsubsection{$\bfW$ Mass from the $\bfWW$ Cross-section}
\label{sec:xsec}

The $\DELPHI$ collaboration has measured the total \CCTHREE\ $\WW$ cross-section, as a function of centre-of-mass energy, using the full data sample collected by the collaboration during \LEP2\ operations \cite{wwxsec}. Assuming the validity of the cross-section dependence predicted by the Standard Model these measurements can be translated into a measurement of the $\W$ mass. Only the cross-section measurements close to the $\WW$ threshold have significant sensitivity to the $\W$ mass.

The Standard Model cross-section dependence on the $\W$ mass is obtained from the \WPHACT\ and \YFSWW\ generator setup, as discussed in section~\ref{sec:simul}, and cross-checked with the improved Born approximation calculation. The theoretical error on the total $\WW$ cross-section near threshold was estimated as $2\%$  decreasing with increasing collision energy to $0.5\%$ in the DPA-valid region \cite{YBGEN}, the corresponding error on the $\W$ mass is marked below as Theor. The sources of experimental systematic error have not been reevaluated and are as reported in \cite{delpaper161}, apart from use of the revised collision energy uncertainty.

From a $\chisq$ fit of the measured cross-sections at centre-of-mass energies of 161.31, 172.14 and 182.65~$\GeV$ the mass has been determined to be
\begin{eqnarray*}
   \mw & = & 80.448 \pm 0.434 (\rm Stat.) \pm 0.090 (\rm Syst.) \pm 0.043 (\rm Theor.) \pm 0.013 (\rm LEP)~\GeVm .
 \end{eqnarray*}

\subsubsection{$\W$ Mass from Direct Reconstruction at $\sqs = 172~\GeV$}
\label{sec:172}

For completeness, we also report here on the relatively small data sample (10~$\ipb$) recorded in 1996 at $\sqs = 172~\GeV$. This sample was analysed and $\W$ mass results published using the $\enqq$, $\mnqq$ and $\qqqq$ decay channels in \cite{delpaper172}. The $\qqqq$ analysis was performed using a standard analysis rather than a cone jet reconstruction based analysis.

This data sample has not been reprocessed, nor have $\W$ width results been produced with this sample.
The estimates of systematic uncertainties are retained from the original paper except for the uncertainties arising from colour reconnection and Bose-Einstein Correlations in the $\qqqq$ channel, where the errors reported above for the standard analysis are used, and the use of the final \LEP\ collision energy uncertainty.
The revised values are
\begin{eqnarray*}
   \mw & = & 80.51 \pm 0.57 (\rm Stat.) \pm 0.05 (\rm Syst.) \pm 0.01 (\rm LEP)~\GeVm ,
 \end{eqnarray*}

\noindent for the combined semi-leptonic channels, and
\begin{eqnarray*}
   \mw & = & 79.90 \pm 0.59 (\rm Stat.) \pm 0.05 (\rm Syst.)  \pm 0.21 (\rm FSI.) \pm 0.01 (\rm LEP)~\GeVm ,
 \end{eqnarray*}

\noindent for the fully-hadronic decay channel. These values have been included in tables~\ref{tab:qqlvResult} and \ref{tab:qqqqResult}.

\subsubsection{Combined Results}

The combinations of the results are performed, assuming that the following components of the error are fully-correlated between years (and energy points) and between the fully-hadronic and semi-leptonic channels: electroweak corrections, fragmentation and jet correction. The lepton-related detector systematic in the semi-leptonic channel is also assumed to be fully correlated between years. The colour reconnection and Bose-Einstein effect in the fully-hadronic channel is assumed to be fully correlated between years. The error arising from calibration statistics is uncorrelated between years in the semi-leptonic analysis, as it was determined from independent Monte Carlo simulation samples, but this error is correlated in the fully-hadronic channel as the values were obtained from an overall fit to the samples at all centre-of-mass energies. This error source is uncorrelated in the combination of the semi-leptonic and fully-hadronic channel. The background-related systematic is assumed to be fully correlated between years in both the fully-hadronic and semi-leptonic analyses but uncorrelated between the two channels. The LEP centre-of-mass energy uncertainty is, of course, fully correlated between  the semi-leptonic and fully-hadronic decay channels but is only partially correlated between years. The inter-year correlations were assessed by the LEP energy working group \cite{lepener} and this correlation matrix was applied when performing the combinations reported here.

The results from the semi-leptonic $\W$ mass analyses in each year of data taking (1996-2000) have been combined. The result for the analysis aimed at selecting events in the $\enqq$ decay channel is:
\begin{eqnarray*}
   \mw & = & 80.388\pm 0.133(\rm Stat.) \pm 0.036 (\rm Syst.) \pm 0.010 (\rm LEP)~\GeVm,
 \end{eqnarray*}
the combination has a $\chisq$ probability of $25\%$.

The result for the analysis aimed at selecting events in the $\mnqq$ decay channel is:
\begin{eqnarray*}
   \mw & = & 80.294 \pm 0.098(\rm Stat.) \pm 0.028 (\rm Syst.) \pm 0.010 (\rm LEP)~\GeVm,
 \end{eqnarray*}
the combination has a $\chisq$ probability of $96\%$.

The $\tnqq$ selection includes significant cross-talk from events in other decay channels (see table \ref{tab:evtsel}) and a result from the 1996 data is not available. The result for the analysis aimed at selecting events in the $\tnqq$ decay channel (in the years 1997-2000) is:
\begin{eqnarray*}
   \mw & = & 80.387 \pm 0.144(\rm Stat.) \pm 0.033 (\rm Syst.) \pm 0.010 (\rm LEP)~\GeVm,
 \end{eqnarray*}
the combination has a $\chisq$ probability of $56\%$.

The result for the combined semi-leptonic $\W$ mass analyses is:
\begin{eqnarray*}
   \mw & = & 80.339 \pm 0.069(\rm Stat.) \pm 0.029 (\rm Syst.) \pm 0.009 (\rm LEP)~\GeVm,
 \end{eqnarray*}
the combination has a $\chisq$ probability of $16\%$.

Similarly, the results on the $\W$ mass extracted from the fully-hadronic event analysis have also been combined. The value from 1996 uses the standard reconstruction technique; the results of the cone-jet reconstruction technique are used for the other data taking years (1997-2000). The combined result is:
\begin{eqnarray*}
   \mw & = & 80.311 \pm 0.059(\rm Stat.) \pm 0.032 (\rm Syst.) \pm 0.119 (\rm FSI) \pm 0.010 (\rm LEP)~\GeVm,
 \end{eqnarray*}
the combination also has a $\chisq$ probability of $16\%$.

The mass difference between the $\W$ boson mass measurements obtained from the fully-hadronic and semi-leptonic channels $\Delta\mw(\qqqq - \lnqq)$, has been determined. 
A significant non-zero value for $\Delta\mw$ could indicate that Bose-Einstein or colour reconnection effects are biasing the value of $\mw$ determined from $\qqqq$ events. 
Since $\Delta\mw$ is primarily of interest as a cross-check of the possible effects of final state interactions, the errors from CR and BEC are set to zero in its determination and the results of the standard reconstruction technique, rather than the FSI effect-reducing cone-jet reconstruction technique, are used for the $\qqqq$ analysis. 
The result provides no evidence for FSI effects: 
\begin{eqnarray*}
   \Delta\mw(\qqqq - \lnqq) & = 0.024 \pm 0.090~\GeVm,
 \end{eqnarray*}
the combination has a $\chisq$ probability of $20\%$.

The final \DELPHI\ result for the $\W$ mass for the full \LEP2\ data sample is obtained by combining the values obtained from the direct reconstruction method in the $\lnqq$ analysis and cone jet reconstruction technique $\qqqq$ analysis in each data taking year. 
The value obtained from the threshold cross-section is also included in this average. The combined result is:
\begin{eqnarray*}
   \mw & = & 80.336 \pm 0.055(\rm Stat.) \pm 0.028 (\rm Syst.)~ \pm 0.025 (\rm FSI) \pm 0.009 (\rm LEP)~\GeVm,
 \end{eqnarray*}
the combination has a $\chisq$ probability of $15\%$.

Although the statistical error in the  $\lnqq$ and $\qqqq$ channels is similar, owing to the large systematic error attributed to final state cross-talk effects the weight of the fully-hadronic channel results in this average is $21\%$. The weight of the threshold cross-section measurement of the $\W$ mass is only $2\%$ due to the small data sample collected at $161~\GeV$ centre-of-mass energy. The full error breakdown of the averages is provided in table~\ref{tab:combmass}.

\begin{table}[ht]
 \begin{center}
  \begin{tabular}{|l|r|r|r|}     \hline
                                  & $\lnqq$   & $\qqqq$ & All   \\ \hline 
 Value                            & 80.339    & 80.311  &  80.336 \\
 Statistical Error                &   .069    &   .059  &    .055  \\
 \hline
 Statistical Error on Calibration &   .003    &   .004  &    .002  \\
 Lepton Corrections               &   .015    &   -~~   &    .012  \\
 Jet Corrections                  &   .020    &   .026  &    .021     \\
 Fragmentation                    &   .011    &   .012  &    .011     \\
 Background                       &   .007    &   .013  &    .006     \\
 Threshold Systematics            &   -~~     &   -~~   &    .002 \\
 Electroweak Corrections          &   .006    &   .005  &    .006     \\
\hline
 LEP Energy                       &   .009    &   .010  &    .009     \\
\hline
 Bose-Einstein Correlations       &   -~~     &   .026  &    .005    \\ 
 Colour Reconnection              &   -~~     &   .116  &    .024     \\
\hline
\end{tabular}
  \caption{The final results (in $\GeVm$) of the $\W$ mass analyses and the breakdown of the uncertainty into its component categories. The $\lnqq$ and $\qqqq$ results use the values obtained in these analysis channels from the direct reconstruction method. The column marked `All' uses the full direct reconstruction analyses and the threshold cross-section measurement. The $\qqqq$ results are taken from the cone jet reconstruction analysis, for all data except 1996 where the standard analysis was used.}
  \label{tab:combmass}
 \end{center}
\end{table}

The \DELPHI\ measurement of the colour reconnection effect is reported in \cite{delphicr}. This measurement places relatively loose constraints on the size of the W mass uncertainty from CR effects, and thus leads to the small impact of the fully-hadronic mass in the \DELPHI\ average. For comparison the value of the combined \DELPHI\ W mass as a function of the CR uncertainty  is shown in table \ref{tab:crmasscomb}. All other errors, including that arising from Bose-Einstein correlations, have been kept constant in these results.  

\begin{table}[htb]
\begin{center}
\begin{tabular}{|l|l|l|}
\hline
CR $\MeVm$ &  $\kappa_{\SKI}$ & $\mw\ \GeVm$ \\
\hline
0 & 0.00 & 80.326 $\pm$ 0.045(Stat.) $\pm$ 0.028(Syst.) $\pm$ 0.013(FSI) $\pm$ 0.010(LEP)\\
20 & 0.40 & 80.326 $\pm$ 0.045(Stat.) $\pm$ 0.028(Syst.) $\pm$ 0.016(FSI) $\pm$ 0.010(LEP)\\
40 & 0.89 & 80.328 $\pm$ 0.046(Stat.) $\pm$ 0.028(Syst.) $\pm$ 0.021(FSI) $\pm$ 0.010(LEP)\\
60 & 1.51 & 80.330 $\pm$ 0.048(Stat.) $\pm$ 0.028(Syst.) $\pm$ 0.024(FSI) $\pm$ 0.010(LEP)\\
80 & 2.30 & 80.333 $\pm$ 0.051(Stat.) $\pm$ 0.028(Syst.) $\pm$ 0.026(FSI) $\pm$ 0.010(LEP)\\
100 & 3.36 & 80.335 $\pm$ 0.054(Stat.) $\pm$ 0.028(Syst.) $\pm$ 0.026(FSI) $\pm$ 0.009(LEP)\\

\hline
\end{tabular}
\caption{The combined \DELPHI\ W Mass value as a function of the uncertainty ascribed to colour reconnection effects in the fully-hadronic decay channel. The values of the $\kappa_{\SKI}$ parameter that give rise to this shift in the $\qqqq$ $\W$ mass at a centre-of-mass energy of 200~$\GeV$ are also given.
}
\label{tab:crmasscomb}
\end{center}
\end{table}

\subsection{$\W$ Width}

The $\W$ width has been measured from the semi-leptonic and the fully-hadronic decay channel events. As the analysis is less sensitive to the $\W$ width than the $\W$ mass, the width is extracted by performing a combined fit of the three semi-leptonic channels rather than from each channel individually. The results are given in table~\ref{tab:widthResult}. The correlations assumed for the combinations are identical to those reported above for the $\W$ mass.

\begin{table}[htb]
\begin{center}
\begin{tabular}{|c|c|c|l|}
\hline
Year & Energy          & Channel & \hspace{2cm} $\gw~\GeVm$ \\
\hline
 1997 &  183   & $\lnqq$ & 2.495 $\pm$ 0.590(Stat.) $\pm$ 0.069(Syst.) \\
      &        & $\qqqq$ & 2.572 $\pm$ 0.460(Stat.) $\pm$ 0.092(Syst.) $\pm$ 0.248(FSI) \\
\hline
 1998 &  189   & $\lnqq$ & 3.056 $\pm$ 0.401(Stat.) $\pm$ 0.071(Syst.) \\
      &        & $\qqqq$ & 2.337 $\pm$ 0.260(Stat.) $\pm$ 0.114(Syst.) $\pm$ 0.248(FSI) \\
\hline
 1999 &  192   & $\lnqq$ & 2.342 $\pm$ 0.953(Stat.) $\pm$ 0.071(Syst.)\\
      &        & $\qqqq$ & 2.390 $\pm$ 0.756(Stat.) $\pm$ 0.126(Syst.) $\pm$ 0.248(FSI)\\
      &  196   & $\lnqq$ & 1.805 $\pm$ 0.440(Stat.) $\pm$ 0.072(Syst.)\\
      &        & $\qqqq$ & 2.545 $\pm$ 0.508(Stat.) $\pm$ 0.142(Syst.) $\pm$ 0.248(FSI)\\
      &  200   & $\lnqq$ & 2.153 $\pm$ 0.477(Stat.) $\pm$ 0.073(Syst.)\\
      &        & $\qqqq$ & 2.210 $\pm$ 0.376(Stat.) $\pm$ 0.157(Syst.) $\pm$ 0.248(FSI)\\
      &  202   & $\lnqq$ & 1.707 $\pm$ 0.649(Stat.) $\pm$ 0.076(Syst.)\\
      &        & $\qqqq$ & 1.797 $\pm$ 0.488(Stat.) $\pm$ 0.165(Syst.) $\pm$ 0.248(FSI)\\
      & 192-202
               & $\lnqq$ & 1.950 $\pm$ 0.277(Stat.) $\pm$ 0.072(Syst.)\\ 
      &        & $\qqqq$ & 2.210 $\pm$ 0.243(Stat.) $\pm$ 0.152(Syst.) $\pm$ 0.248(FSI)\\
\hline
 2000 &  206   & $\lnqq$ & 2.814 $\pm$ 0.364(Stat.) $\pm$ 0.083(Syst.)\\
      &        & $\qqqq$ & 1.979 $\pm$ 0.225(Stat.) $\pm$ 0.183(Syst.) $\pm$ 0.248(FSI)\\
\hline
\end{tabular}
\vspace{0.2cm}
\caption{Measured $\W$ widths (in $\GeVm$) from the semi-leptonic decay and fully-hadronic decay channel analyses with the nominal centre-of-mass energies (in $\GeV$) of each data sample indicated.}
\label{tab:widthResult}
\end{center}
\end{table}

The results from the semi-leptonic $\W$ width analyses in each year of data taking (1997-2000) have been combined, the result obtained is:
\begin{eqnarray*}
   \gw & = & 2.452 \pm 0.184(\rm Stat.) \pm 0.073 (\rm Syst.)~\GeVm,
 \end{eqnarray*}
the combination has a $\chisq$ probability of $9\%$.

Similarly, the results on the $\W$ width extracted from the fully-hadronic event analysis have also been combined, the result obtained is:
\begin{eqnarray*}
   \gw & = &  2.237\pm 0.137(\rm Stat.) \pm 0.139 (\rm Syst.)~ \pm 0.248 (\rm FSI)~\GeVm,
 \end{eqnarray*}
the combination has a $\chisq$ probability of $62\%$.

The final \DELPHI\ result for the $\W$ width for the full \LEP2\ data sample is obtained by combining the values obtained from the direct reconstruction method in the $\lnqq$ analysis and $\qqqq$ analysis in each data taking year. The combined result is:
\begin{eqnarray*}
   \gw & = &  2.404 \pm 0.140(\rm Stat.) \pm 0.077 (\rm Syst.)~ \pm 0.065 (\rm FSI)~\GeVm,
 \end{eqnarray*}
the combination has a $\chisq$ probability of $27\%$.

Although the statistical error in the  $\lnqq$ and $\qqqq$ channels is similar, owing to the large systematic error attributed to final state cross-talk effects the weight of the fully-hadronic channel results in this average is $26\%$. The full error breakdown of the averages is provided in table~\ref{tab:combwidth}.

\begin{table}[ht]
 \begin{center}
  \begin{tabular}{|l|r|r|r|}     \hline
                                  & $\lnqq$ & $\qqqq$ & All   \\ \hline 
 Value                            & 2.452   &  2.237  & 2.404 \\
 Statistical Error                &  .184   &   .137  &  .140 \\
 \hline
 Statistical Error on Calibration &  .006   &   .009  &  .005    \\
 Lepton Corrections               &  .041   &   -~~   &  .030    \\
 Jet Corrections                  &  .036   &   .129  &  .059      \\
 Fragmentation                    &  .029   &   .008  &  .024      \\
 Electroweak Corrections           &  .011  &   .009  &  .010     \\
 Background                       &  .037   &   .051  &  .031      \\
\hline
 Bose-Einstein Correlations       &   -~~   &   .020  &  .005   \\ \hline
 Colour Reconnection              &   -~~   &   .247  &  .065      \\
\hline
\end{tabular}
  \caption{The final results (in $\GeVm$) of the $\W$ width analyses and the breakdown of the uncertainty into its component categories. The $\lnqq$ and $\qqqq$ results use the values obtained in these analysis channels from the direct reconstruction method. The column marked `All' provides the result from combining the measurements made in both channels.}
  \label{tab:combwidth}
 \end{center}
\end{table}

\section{Conclusions}

The mass and width of the $\W$ boson have been measured using the reconstructed masses in $\ee \ra \WW$ events decaying to $\qqqq$ and $\lnqq$ states. The $\W$ Mass was also extracted from the dependence of the $\WW$ cross-section close to the production threshold. The full \LEP2\ data sample of $660~\ipb$ collected by the \DELPHI\ experiment at centre-of-mass energies from 161 to 209~$\GeV$ has been used. The final results are:
 \begin{eqnarray*}
   \mw & = & 80.336 \pm 0.055(\rm Stat.) \pm 0.028 (\rm Syst.)~ \pm 0.025 (\rm FSI) \pm 0.009 (\rm LEP)~\GeVm,
 \end{eqnarray*}
\begin{eqnarray*}
   \gw & = &  2.404 \pm 0.140(\rm Stat.) \pm 0.077 (\rm Syst.)~ \pm 0.065 (\rm FSI)~\GeVm.
 \end{eqnarray*}

These results supersede the previously published \DELPHI\ results~\cite{delpaper161,delpaper172,delpaper183,delpaper189}.

\newpage

\subsection*{Acknowledgements}
\vskip 3 mm
We are greatly indebted to our technical collaborators, to the
members of the CERN-SL Division for the excellent performance of
the \LEP\ collider, and to the funding agencies for their
support in building and operating the \DELPHI\ detector.
We also wish to offer our thanks to the \LEP\ energy working group 
for their measurement of the \LEP\ collision energy which plays an 
important role in the analysis presented in this paper.\\
We acknowledge in particular the support of \\
Austrian Federal Ministry of Education, Science and Culture,
GZ 616.364/2-III/2a/98, \\
FNRS--FWO, Flanders Institute to encourage scientific and technological 
research in the industry (IWT) and Belgian Federal Office for Scientific,
Technical and Cultural affairs (OSTC), Belgium, \\
FINEP, CNPq, CAPES, FUJB and FAPERJ, Brazil, \\
Ministry of Education of the Czech Republic, project LC527, \\
Academy of Sciences of the Czech Republic, project AV0Z10100502, \\
Commission of the European Communities (DG XII), \\
Direction des Sciences de la Mati$\grave{\mbox{\rm e}}$re, CEA, France, \\
Bundesministerium f$\ddot{\mbox{\rm u}}$r Bildung, Wissenschaft, Forschung 
und Technologie, Germany,\\
General Secretariat for Research and Technology, Greece, \\
National Science Foundation (NWO) and Foundation for Research on Matter (FOM),
The Netherlands, \\
Norwegian Research Council,  \\
State Committee for Scientific Research, Poland, SPUB-M/CERN/PO3/DZ296/2000,
SPUB-M/CERN/PO3/DZ297/2000, 2P03B 104 19 and 2P03B 69 23(2002-2004),\\
FCT - Funda\c{c}\~ao para a Ci\^encia e Tecnologia, Portugal, \\
Vedecka grantova agentura MS SR, Slovakia, Nr. 95/5195/134, \\
Ministry of Science and Technology of the Republic of Slovenia, \\
CICYT, Spain, AEN99-0950 and AEN99-0761,  \\
The Swedish Research Council,      \\
Particle Physics and Astronomy Research Council, UK, \\
Department of Energy, USA, DE-FG02-01ER41155, \\
EEC RTN contract HPRN-CT-00292-2002. \\


\pagebreak

\pagebreak


\begin{thebibliography}{99}

\bibitem{delpaper161} DELPHI Collaboration, P.~Abreu {\it et al.}, 
        \Journal{\PLB}{397}{158}{1997} 

\bibitem{delpaper172} DELPHI Collaboration, P.~Abreu {\it et al.},
        \Journal{\EUR}{2}{581}{1998}

\bibitem{delpaper183} DELPHI Collaboration, P.~Abreu {\it et al.},
        \Journal{\PLB}{462}{410}{1999}

\bibitem{delpaper189} DELPHI Collaboration, P.~Abreu {\it et al.},
        \Journal{\PLB}{511}{159}{2001}

\bibitem{lepmw} ALEPH Collaboration, S.~Schael {\it et al.},
\Journal{\EUR}{47}{309}{2006};\\
L3 Collaboration, P.~Achard {\it et al.},
\Journal{\EUR}{45}{569}{2006};\\
OPAL Collaboration, G.~Abbiendi {\it et al.},
\Journal{\EUR}{45}{307}{2006}


\bibitem{hadmw} 
CDF Collaboration, D0 Collaboration and Tevatron Electroweak Working Group, V.M.~Abazov {\it et al.},
        \Journal{\PRD}{70}{092008}{2004} 

\bibitem{lepener}
LEP Energy Working Group, R.~Assmann {\it et al.},
\Journal{\EUR}{39}{253}{2005}

\bibitem{radreturn} ALEPH Collaboration, R.~Barate {\it et al.}, 
\Journal{\PLB}{464}{339}{1999};  \\
L3 Collaboration, P.~Achard {\it et al.}, 
\Journal{\PLB}{585}{42}{2004}; \\
OPAL Collaboration, G.~Abbiendi {\it et al.},
\Journal{\PLB}{604}{31}{2004}


\bibitem{radreturnDELPHI}
DELPHI Collaboration, J. Abdallah {\it et al.}, 
\Journal{\EUR}{46}{295}{2006}

\bibitem{delphi} DELPHI Collaboration, P.~Aarnio {\it et al.},
        \Journal{\NIMA}{303}{233}{1991} \\
                 DELPHI Collaboration, P.~Abreu {\it et al.}, 
        \Journal{\NIMA}{378}{57}{1996} 

\bibitem{vft} The DELPHI Silicon Tracker Group, P.~Chochula {\it et al.},
    \Journal{\NIMA}{412}{304}{1998}

\bibitem{STIC} S.~J.~Alvsvaag {\it et al.},
    \Journal{\NIMA}{425}{106}{1999}

\bibitem{delphi4fgen} A.~Ballestrero, R.~Chierici, F.~Cossutti and E.~Migliore,
\Journal{\CPC}{152}{175}{2003}  

\bibitem{wphact}
  E.~Accomando and A.~Ballestrero,
\Journal{\CPC} {99}{270}{1997} \\
  E.~Accomando, A.~Ballestrero and E.~Maina, 
\Journal{\CPC} {150}{166}{2003}  

\bibitem{pythia} T. Sj\"ostrand {\it et al.}, 
  \Journal{\CPC} {135}{238}{2001} 


\bibitem{tauola} S.~Jadach, Z.~Was, R.~Decker and J.H.~Kuehn, 
\Journal{\CPC} {76}{361}{1993}  

\bibitem{yfsww} S.~Jadach, W.~Placzek, M.~Skrzypek, B.~F.~L.~Ward and Z.~Was,
        \Journal{\PLB}{417}{326}{1998}; \\
  S.~Jadach, W.~Placzek, M.~Skrzypek, B.~F.~L.~Ward and Z.~Was, 
\Journal{\CPC} {140}{432}{2001}  

\bibitem{photos} E.~Barberio and Z.~Was, 
   \Journal{\CPC} {79}{291}{1994}

\bibitem{ariadne} L.~L\"{o}nnblad, 
   \Journal{\CPC}{71}{15}{1992}  

\bibitem{herwig} G.~Corcella {\it et al.}, 
JHEP {\bf 0101} (2001) 010

\bibitem{deltune} DELPHI Collaboration, P.~Abreu {\it et al.},
        \Journal{\ZPC}{73}{11}{1996}

\bibitem{kk} S.~Jadach, B.~F.~L.~Ward and Z.~Was, \Journal{\CPC}{130}{260}{2000}
\bibitem{luclus} T. Sj\"ostrand,
        {\it PYTHIA 5.7 and JETSET 7.4: Physics and manual}, 
        CERN-TH-7112-93-REV (1995)

\bibitem{neural} Code kindly provided by J.~Schwindling and B.~Mansoulie

\bibitem{SPRIM} P.~Abreu {\it et al.}, 
        \Journal{\NIMA}{427}{487}{1999} 

\bibitem{aabtag} G.~Borisov, 
        \Journal{\NIMA}{417}{384}{1998}; \\
                 DELPHI Collaboration, P.~Abreu {\it et al.}, 
        \Journal{\EUR}{10}{415}{1999}

\bibitem{DURHAM} S.~Catani, Yu.L.~Dokshitzer, M.~Olsson, G.~Turnock and B.R.~Webber,
        \Journal{\PLB}{269}{432}{1991}; \\
        N. Brown, W. Stirling, 
        \Journal{\ZPC}{53}{629}{1992} 

\bibitem{CAMJET} Yu.L.~Dokshitzer, G.D.~Leder, S.~Moretti, B.R.~Webber,
        JHEP {\bf{9708}} (1997) 001

\bibitem{DICLUS} L.~L\"onnblad, 
        \Journal{\ZPC}{58}{471}{1993} 


\bibitem{delphicr}
DELPHI Collaboration,  J.~Abdallah {\it et al.},
\Journal{\EUR}{51}{249}{2007}

\bibitem{lepcr}
L3 Collaboration,  P.~Achard {\it et al.}, 
\Journal{\PLB}{561}{202}{2003}; \\
OPAL Collaboration, G.~Abbiendi {\it et al.},
\Journal{\EUR}{45}{291}{2006}; \\
ALEPH Collaboration, S.~Schael {\it et al.},
\Journal{\EUR}{47}{309}{2006}

\bibitem{pdg}
Particle Data Group, S. Eidelman {\it et al}.
\Journal{\PLB}{592}{1}{2004}

\bibitem{brem} H.~A.~Bethe and W.~Heitler, Proc. Roy. Soc. {\bf A146}(1934) 83
          
\bibitem{wmasssys} S.~Jadach, W.~Placzek, M.~Skrzypek, B.~F.~L.~Ward
  and Z.~Was, Phys. Lett. {\bf B523} (2001) 117

\bibitem{eweak} F.~Cossutti,
\Journal{\EUR}{44}{383}{2005}

\bibitem{KC}
A.P.Chapovsky and V.A.Khoze, \Journal{\EUR}{9}{449}{1999}

          
\bibitem{racoonww} A. Denner, S. Dittmaier, M. Roth and D. Wackeroth, \Journal{\NUCB}{560}{33}{1999}; \\
A. Denner, S. Dittmaier, M. Roth and D. Wackeroth, \Journal{\NUCB}{587}{67}{2000}
       

\bibitem{2f-therr} F.~Boudjema, B.~Mele {\it et al.}, {\it Standard Model
    Process}, Physics at LEP2, eds. G.~Altarelli, T.~Sj\"{o}strand and
    F.~Zwirner, CERN 96-01 (1996) Vol. 1, 207
  
\bibitem{wwxsec} DELPHI Collaboration, J.~Abdallah {\it et al.},  
\Journal{\EUR}{34}{127}{2004} 

\bibitem{dbec}
DELPHI Collaboration, J. ~Abdallah {\it et al.}, \Journal{\EUR}{44}{161}{2005}

\bibitem{lbec}
ALEPH Collaboration, S.~Schael {\it{et al.}},
\Journal{\PLB}{606}{265}{2005}; \\
OPAL Collaboration, G.~Abbiendi {\it{et al.}},
\Journal{\EUR}{36}{297}{2004}; \\
L3 Collaboration, P. Achard {\it{et al.}},
\Journal{\PLB}{547}{139}{2002} 

\bibitem{luboei} L.~L\"onnblad and T.~Sj\"ostrand,
        \Journal{\EUR}{2}{165}{1998} 

\bibitem{phLEP2}
V. Khoze {\it et al.}, {\it Colour Reconnection}, Physics at LEP2, eds. G. Altarelli, T.~Sj\"{o}strand and F.~Zwirner, CERN 96-01 (1996) Vol.1, 191 
   

\bibitem{perturb} T.~Sj\"ostrand and V.~Khoze,
        \Journal{\ZPC}{62}{281}{1994};\\
        T. Sj\"ostrand and V. Khoze,
        \Journal{\PRL}{72}{28}{1994}

\bibitem{gh}
G. Gustafson and J. H\"akkinen, 
     \Journal{\ZPC}{64}{659}{1994}

\bibitem{ar2} 
L.~L\"{o}nnblad,
     \Journal{\ZPC}{70}{107}{1996}  
    
\bibitem{lyons}
 L.~Lyons, D.~Gibaut and P.~Clifford, \Journal{\NIMA}{270}{110}{1988}

\bibitem{YBGEN} 
{\it{LEP2 Monte Carlo Workshop : Report of the Working Groups on Precision Calculations for LEP2 Physics}}
eds. G. Passarino, R. Pittau, S. Jadach, CERN-2000-009 (2000)


\end{thebibliography}
\end{document}